# Diverse Behaviors in Non-Uniform Chiral and Non-Chiral Swarmalators

Steven Ceron[1,2], Kevin O'Keeffe[3], Kirstin Petersen[4*]

[1] Sibley School of Mechanical and Aerospace Engineering, Cornell University; Ithaca, NY 14853, USA.
[2] Computer Science and Artificial Intelligence Lab, Massachusetts Institute of Technology, Cambridge, MA 02139, USA.
[3] Senseable City Lab, Massachusetts Institute of Technology, Cambridge, MA 02139, USA.
[4] Electrical and Computer Engineering, Cornell University; 136 Hoy Road, Ithaca, NY 14853, USA.
*Correspondence to: kirstin@cornell.edu

**Abstract:** We study the emergent behaviors of a population of swarming coupled oscillators, dubbed 'swarmalators'. Previous work considered the simplest, idealized case: identical swarmalators with global coupling. Here we expand this work by adding more realistic features: local coupling, non-identical natural frequencies, and chirality. This more realistic model generates a variety of new behaviors including lattices of vortices, beating clusters, and interacting phase waves. Similar behaviors are found across natural and artificial micro-scale collective systems, including social slime mold, spermatozoa vortex arrays, and Quincke rollers. Our results indicate a wide range of future use cases, both to aid characterization and understanding of natural swarms, and to design complex interactions in collective systems from soft and active matter to micro-robotics.

# INTRODUCTION

Synchronization (self-organization in time) and swarming (self-organization in space) are present in many natural and artificial systems. Synchronization occurs in flashing fireflies[1], firing heart cells[2], spiking neurons[3,4], and chorusing frogs[5]. Swarming, on the other hand, is seen in cell collectives that migrate in response to external signals[6,7], flocks of birds that seamlessly change collective flight direction[8], and schools of fish that coalesce and move together to the collective's advantage[9].

A combination of synchronization and swarming also occurs in diverse contexts, from biological micro-scale collectives[10,11] and chemical micromotors[12,13], to magnetic domain walls[14] and robotic drones[15]. Despite its ubiquity, a theoretical understanding of this interplay between synchronization and swarming is lacking. Synchronization research has, broadly speaking, focused on oscillators which may synchronize in time, but not move around in space[16-19]. Swarming research has done the reverse; it has studied units moving through space[20,21] that synchronize spatially dependent variables like orientation, but not internal phase variables in time. As such, the interplay of synchronization and swarming defines a new kind of collective dynamics about which little is known.

The first steps to explore this domain were taken a few years ago by Igoshin et al.[22], Tanaka et al.[23,24], and Levis et al.[25,26]; they derived models of chemotactic oscillators[22-24] and revolving agents[25-26] and produced diverse behaviors. Later, O'Keeffe et al[27] proposed a generalized Kuramoto model of 'swarmalators' (short for "swarming oscillators") whose states have been seen in natural[6,10,11,28] and artificial[13,29-32] micro-scale collectives and is now being extended and implemented on robot swarms[33-38].

This paper continues swarmalator research by introducing a new swarmalator model that facilitates diverse emergent behaviors useful to researchers in fields ranging from bio-inspired swarm robotics to microrobotics to biology; our motivation is to comprehensively map out the space of emergent behaviors the model generates. In O'Keeffe et al.[27], for simplicity's sake, the swarmalators were identical, non-chiral, and globally coupled. Here, we relax these idealizations and consider swarmalators in a two-dimensional world which are non-identical (swarmalators have different natural frequencies), chiral (swarmalators have inherent clockwise and counterclockwise circular orbits), and locally coupled (swarmalators can only couple to motion and phase of neighbors within a given radius). Each of these relaxations enhances the model's descriptive power of real systems. (1) Non-identical natural frequencies are representative of systems where agents have differing internal parameters and/or may be noisy. The behaviors of micro-robots, for example, vary due to manufacturing tolerances. (2) The chirality in revolving swarmalators enables our model to relate the internal and spatial parameters while mimicking realistic systems where the direction of revolution may be opposite. Following the previous example of micro-robots, their orientation (and thus motion patterns) may flip randomly when deposited on a water surface. (3) Local coupling ensures that this model is descriptive of distributed systems, where agent sensors or reactive response have a limited range of view. Again, micro-robots are a good example, as their physical interactions drop off rapidly with distance. The inclusion of these realistic features enable many new behaviors that to date have not been found in other swarmalator models including interacting phase waves, organized arrays of vortices, concentric phase self-organization, radial oscillation. Many of these behaviors qualitatively resemble those displayed by vortex arrays of sperm[10], the flocking patterns of Quinke rollers[30,31], the various life stages of slime mold[6], spatiotemporal waves of cellular self-organization related to embryology[39-41], and the radial oscillation of phototactic micromotors[12]; the most remarkable states are summarized in Supplementary Fig. 1 and paired with their qualitatively similar real-world counterparts in Supplementary Table 1. Additional states shown in this paper, such as bouncing and revolving clusters, do not to our knowledge have any natural counterparts, but could be useful as inspiration for an applied context. For example, artificial swarmalator systems such as aerial or marine drone collectives could use periodic orbits to maximize surveillance[42] or information-sharing between agents that do not always remain close to each other [43]. Emergent behaviors could also be designed for colloids towards high-precision medicine[44]. We hope this work will inspire new studies on swarmalators; and be useful in the characterization and control of natural and artificial collective systems.

**RESULTS**

**The Model**
Swarmalators have a spatial position $x_i$ and an internal phase $\theta_i$ which evolve according to Equations (1) and (2).

$$\dot{x}_i = v_i + \frac{1}{N}\sum_{j \neq i}^{N}\left[\frac{x_j - x_i}{|x_j - x_i|}(A + J\cos(\theta_j - \theta_i - Q_{\dot{x}})) - B\frac{x_j - x_i}{|x_j - x_i|^2}\right] \quad (1)$$

$$\dot{\theta}_i = \omega_i + \frac{K}{N}\sum_{j \neq i}^{N}\frac{\sin(\theta_j - \theta_i - Q_{\dot{\theta}})}{|x_j - x_i|} \quad (2)$$

As per Equation (1), each oscillator $i$ has an inherent velocity ($v_i$), a spatial attraction to all other agents defined by a unit vector between agents' positions and a positive coefficient ($A = 1$) which ensures that agents do not dissipate infinitely, a spatial phase interaction coefficient ($J \in [-1,1]$) which enables agents with similar phases to move towards each other when $J$ is positive and move away from similarly phased agents when $J$ is negative, and a global repulsive term defined by a power law and a positive coefficient ($B = 1$) which ensures that agents do not aggregate at a single point in space. In Equation (2), each agent follows the Kuramoto model where there is a natural frequency ($\omega_i$), a phase coupling coefficient ($K$), and an inverse dependence on the distance between agents.

In the original model, all (uncoupled) swarmalators moved in the same direction with $|v_i| = v_0$, and $v_0$ was set to zero. Here, we modify this by setting $v_i = c_i n_i$, so that now the swarmalator model can be generalized to instances where there is a direct mapping between an agent's internal state (phase) and its orientation along a circular orbit.

$$c_i = \omega_i R_i \quad (3)$$

$$n_i = \begin{bmatrix} \cos\left(\theta_i + \frac{\pi}{2}\right) \\ \sin\left(\theta_i + \frac{\pi}{2}\right) \end{bmatrix} \quad (4)$$

Here, $c_i$ is dependent on an agent's natural frequency ($\omega_i$) and its radius of revolution in real space ($R_i$), and holds a value of -1, 1, or 0 throughout our study. $n_i$ is a vector pointing in the direction orthogonal to the angle denoted by $\theta_i$, within the global coordinate system. When $v_i = 0$, an agent has no inherent motion and its phase is an internal state; however, when $v_i \neq 0$, an agent follows periodic circular orbits in real space with some inherent radius of revolution and its phase is its angular position about its orbit within the global coordinate system. Our new definition of $v_i$ enables us to model swarmalator collectives where each agent's phase is either an internal state ($c_i = 0$) or it has a mapping to its position within real space ($c_i = -1$ or $c_i = 1$). $|c_i|$ is the inherent speed of each agent, and for most of this study we only consider collectives where all agents share the same $|c_i|$. When an agent has no inherent motion in real space, $c_i = 0$, $v_i$ goes away and, throughout the collective, there is no inherent motion (no circular orbiting). $\omega_i$ is never equal to zero in the cases we present; however, past studies have explored the emergent behaviors of swarmalator collectives when $\omega_i = 0$. If there is inherent motion, meaning that $c_i > 0$, then $\omega_i > 0$, this means that the agent's inherent circular orbit within real space is counterclockwise (CCW). If $c_i < 0$, then $\omega_i < 0$, meaning that its inherent circular orbit is clockwise (CW). The nature of our model is similar to recent work by Togashi[45], which maps an agent's internal state / phase to its shape (radius), which affects how it interacts with its neighbors and thus the emergent collective behaviors.

We also include new phase offset terms, $Q_{\dot{x}}$ and $Q_{\dot{\theta}}$, defined in Equations (5) and (6), which enable 'frequency coupling'. The motivation here is to increase the attraction between agents with opposing signs for their natural frequency; this enables more realistic emergent behaviors reminiscent of hydrodynamically and mechanically coupled systems[10]. Supplementary Discussion 1 and Supplementary Fig. 2 further discuss the phase offset terms and their effect on the collective's oscillatory behavior.

$$Q_{\dot{x}} = \frac{\pi}{2}\left|\frac{\omega_j}{|\omega_j|} - \frac{\omega_i}{|\omega_i|}\right| \quad (5)$$

$$Q_{\dot{\theta}} = \frac{\pi}{4}\left|\frac{\omega_j}{|\omega_j|} - \frac{\omega_i}{|\omega_i|}\right| \quad (6)$$

Finally, we consider swarmalators with several different cases of natural frequencies ω:
1. Single frequency ($F1$): $\omega_i = 1$ for all swarmalators.
2. Two frequencies ($F2$): Exactly half of the swarmalators have $\omega_i = 1$ and the other half have $\omega_i = -1$.
3. Single uniform distribution ($F3$): All swarmalators have their natural frequency randomly selected from a single uniform distribution, such that $\omega_i \sim U(1, \Omega)$.
4. Double uniform distribution ($F4$): Exactly half of the swarmalators have their natural frequency randomly selected from one uniform distribution ($\omega_i \sim U(1, \Omega)$) and the second half have their natural frequency selected from another uniform distribution ($\omega_i \sim U(-\Omega, -1)$).

Throughout these definitions and the remainder of the text, $U(X, Y)$ defines a uniform distribution on the interval [X,Y] and $\Omega = 3$ for most of the study; we present additional results in the supplementary material to showcase the collective behaviors when $\Omega$ is higher or lower. The different cases / distributions of natural frequencies are referred to as $F1$, $F2$, $F3$, and $F4$. The natural frequency distributions F1 and F2 are representative of idealized cases where the collective either shares a single value or is split in half between two equal and opposite values. F3 and F4 represent more disordered cases, common in real-world noisy scenarios. More details on the natural frequency distributions used for this study can be found in Supplementary Discussion 2 and Supplementary Fig. 3.

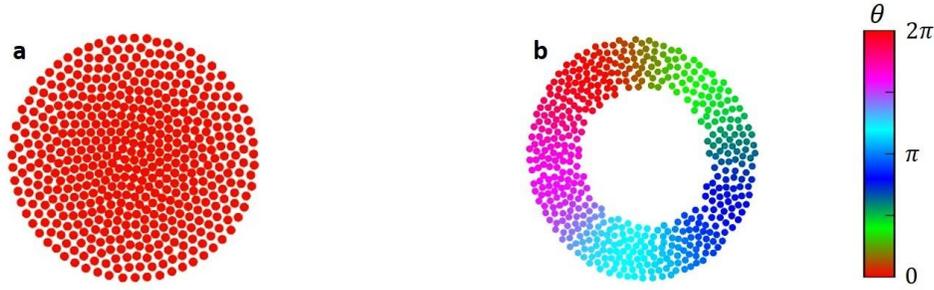

**Fig. 1. Space-phase order parameter.** Two of the five emergent behaviors in the original swarmalator study. **(a)** Static sync ($S \approx 0$). **(b)** Static phase wave ($S \approx 1$). The color bar represents each agent's phase between 0 and $2\pi$; the same color representation is used throughout the work.

To catalog the various collective states of our model, we use several order parameters, some of which were introduced in Ref. 27. The first is,

$$S_\pm = \frac{1}{N} \sum_{j=1}^{N} e^{i(\phi_j \pm \theta_j)} \tag{7}$$

Here, $\phi_j$ is an agent $j$'s angular position with respect to the collective centroid ($\phi_j = \begin{pmatrix} y_j \\ x_j \end{pmatrix}$), where $x_j$ and $y_j$ are its spatial coordinate positions with respect to the collective centroid, and $\theta_j$ is its phase; this measures a kind of circumferential 'space-phase order' in the system. When there is no correlation between the angular position and phase, as in Fig. 1a, $S_\pm = 0$; when there is maximal correlation, as in Fig. 1b, $S_\pm = 1$. Throughout our work, we will refer to $\max(S_+, S_-)$ as $S$ and use heat maps of the order parameter in $K - J$ parameter space to draw out the approximate locations of the different emergent states. We also show several plots of the order parameters $S_+$, $S_-$, $Z$, $\gamma$, and $\beta$ (defined in the Methods section), which aid in explaining key differences between the states.

**Non-Chiral Swarmalators**
We begin our exploration with the simplest case of non-chiral swarmalators with no frequency coupling: $c_i = 0$ for all agents, $Q_{\dot{x}} = 0$, and $Q_{\dot{\theta}} = 0$. We numerically simulated Equations (1) and (2) using an Euler method with stepsize $dt = 0.1$ for $T = 1000$ time units at which steady states were achieved. We scanned over the state space $J \in [-1,1]$ and $K \in [-1,2]$. We use the same numerical scheme and parameter regions in $K$ and $J$ space throughout our work.

Numerical simulations revealed a zoo of convergent self-organization behaviors; some stationary, some dynamic. Some of the most interesting are depicted in Fig. 2, the full set are reviewed and shown in Supplementary Discussions 3-4, Supplementary Figs. 4-12, and Supplementary Movie 1. Here, we discuss in detail the properties of each of the most interesting states when the natural frequency distributions are $F1$ and $F2$; however, we strongly recommend viewing Supplementary Movie 1 to have a clear picture of the states in mind before reading this description (we also recommend this for the other sections in the paper). Results regarding the emergent collective behaviors when the system has distributions $F3$ and $F4$ are described in the supplementary material.

The heat map in Fig. 2a shows the general locations of the emergent collective behaviors within $K - J$ parameter space when the natural frequency distribution is $F1$, while Fig. 2b refers to systems where the distribution is $F2$. We note that static async (Fig. 2c) is found through much of negative $K -$ space and phase waves (static, active, and splintered) appear when $K$ is negative and $J$ is positive. This is clear from the bright triangular region in Fig. 2a which indicates there is high circumferential phase organization. Note that $S$ shows very little difference between the region of static phase waves and splintered phase waves, but dips slightly at low $K$ and high $J$ where the active phase wave emerges. As opposed to previous studies where the static phase wave remained static in both space and phase, the phase wave shown in Fig. 2d has a static annulus formation and a circumferentially traveling phase wave. When there are several natural frequency groups, however, the number of phase waves change, and they become significantly more dynamic. Fig. 2e shows two interacting phase waves, each composed of agents with the same natural frequency; the two annulus formations oscillate in place as each of their phase waves travels around the circular border and interacts with the opposing group's phase wave. The two phase waves shown here are concentric, but it is worth noting that the collective can also settle on an inter-locked ring formation where either of the natural frequency groups can be closer to the collective centroid. Regardless of whether the phase waves are concentric or inter-locked, they remain circumferentially ordered by phase, which enables a high $S$ value along negative $K$ and positive $J$, as shown in Fig. 2b.

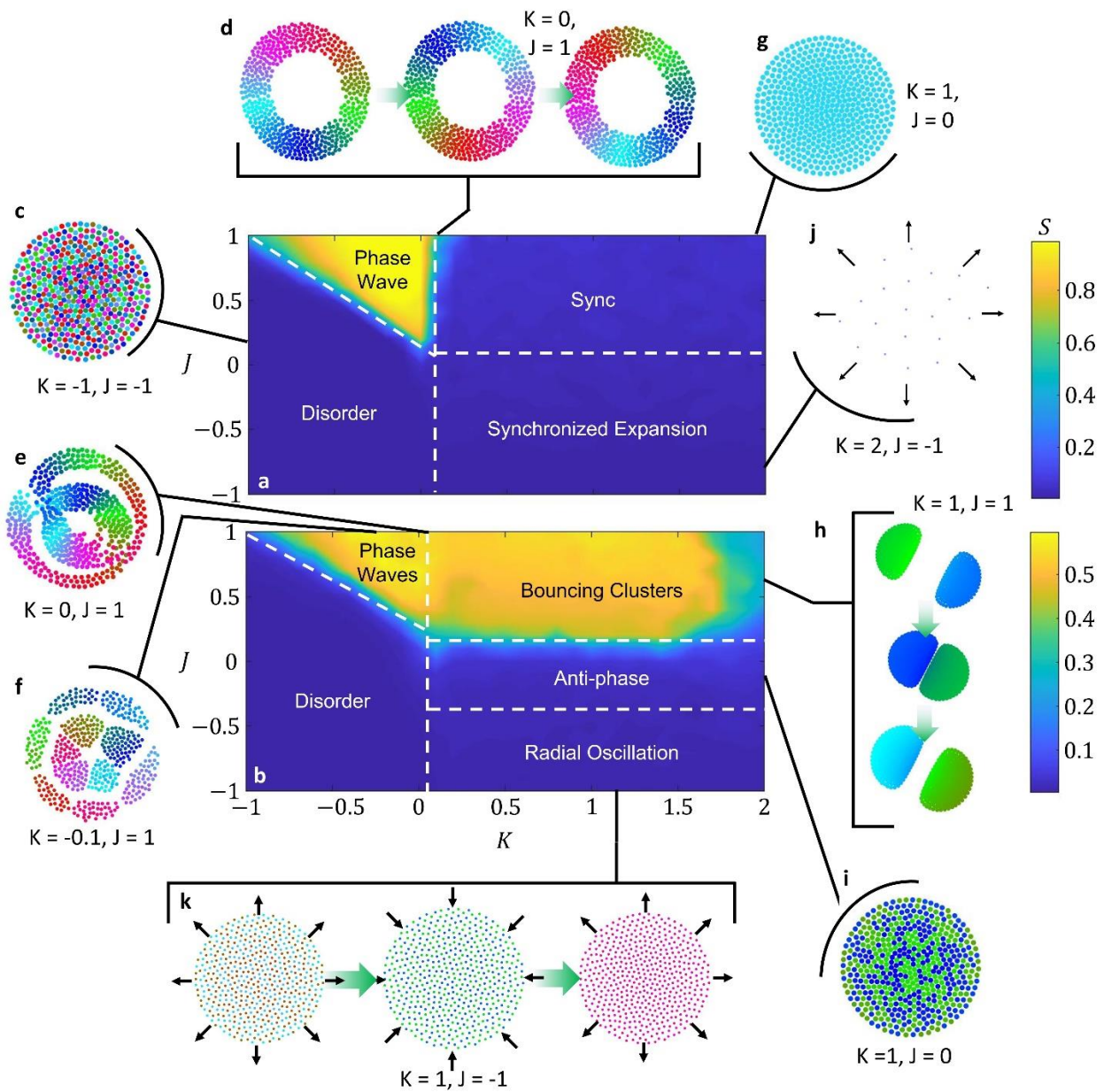

**Fig. 2. Non-Chiral swarmalators with no natural frequency spread.** Heat maps of $S$ across $K - J$ parameter space are shown for test cases with natural frequency distributions **(a)** $F1$ and **(b)** $F2$. **(c)** Static async / disorder. **(d)** Phase wave traveling circumferentially across static agents. **(e)** Double interacting phase waves. **(f)** Double interacting splintered phase waves. **(g)** Static sync. **(h)** Bouncing clusters. **(i)** Static anti-phase. **(j)** Expanding synchronized collective. **(k)** Periodic radial oscillation.

At $K \approx -0.1, J = 1$, the collective forms a single splintered phase wave when all agents share the same natural frequency, and then display concentric splintered phase waves that rotate about the collective centroid in opposing directions (Fig. 2f). Fig. 3 shows that in these cases there is high $S$ while all other calculated order parameters ($Z$, $\gamma$, and $\beta$) remain low. Supplementary Fig. 8 and Supplementary Movie 2 showcase the emergent behaviors when there are two, three, four, or five natural frequency values within the collective. In each of these cases, the collective is evenly split, such that an equal number of agents have each of the natural frequency values, and we find that, with the same $K$ and $J$ values, the number of concentric splintered phase wave formations is correlated to the number of natural frequencies present. When there are more than three natural frequency groups, however, the spacing between the groups is not clear and there is no clear direction of rotation in each layer of the formation.

We also observe the static sync state when there is only one natural frequency (Fig. 2g); however, the collective behavior is significantly more dynamic when the natural frequency distribution is $F2$ because two clusters form and synchronize within themselves, but do not synchronize with the opposing cluster (Fig. 2h). Their inability to synchronize causes the collective to enter a bouncing cluster state where each natural frequency group oscillates between attraction and repulsion with the other group, and the two produce symmetric, periodic oscillations about their mutual centroid. The oscillatory behavior is due to the clusters having equal and opposite natural frequencies; this means one group moves about the phase unit circle in the CW direction while the other moves in the CCW direction. Since both groups have the same absolute natural frequency value, they move about the phase unit circle at the same rate, and share the same phase / are offset by $\pi$ two times per phase cycle. This means the two groups oscillate between having no phase difference and the maximum phase difference and produce periodic attraction and repulsion to each other. Notice in Fig. 2b and Fig. 3 that $S$ remains high around low positive $K$, but dips slightly at very high $K$. There is relatively high spatial-phase order for bouncing clusters since the two groups independently synchronize and remain symmetric about an axis running through the collective centroid; Fig. 3b shows that $\beta$ increases as a result of the cluster separation and this allows us to pinpoint the onset of the bouncing cluster behavior when $J$ is positive. At high $K$, the collective eventually transitions from the bouncing cluster state closer to a static sync state, which lowers $S$ because there is no longer circumferential phase organization. The bouncing cluster behavior becomes even more interesting when there are more than two natural frequency groups, a greater number of bouncing clusters form, but their behavior is no longer symmetric about a single axis. Supplementary Fig. 9 and Supplementary Movie 3 show and characterize bouncing cluster collectives when there are more than two natural frequency groups.

At high $K$ and $J = 0$, collectives with the natural frequency distribution $F2$ create circular formations and enter static anti-phase states (Fig. 2i); agents synchronize within their own natural frequency group because of the high phase coupling factor, but remain offset by phase from the opposing group. Since $J = 0$ ensures that there is no spatial attraction between like-phased agents, there is a fairly uniform distribution of agents from both groups across the formation. The formation is circular because $A = 1$, which enables all agents to attract to each other through a unit vector model, regardless of their phase or distance to each other; the power law model for repulsion ensures that they do not converge to a single point in real space. Once the collective aggregates into a stable circular shape, agents remain fixed and their interaction distance, which still affects their phase coupling behavior, does not change. Some effects of this can be picked out at lower $K$: agents are unable to synchronize and as a result phase waves are seen traveling radially. Agents in the center couple more easily to all other agents, than those close to the boundary; as a result, they are able to lead the phase waves within each natural frequency group. As $K$ gets closer to 2, agents begin to freeze their phase behavior; the two natural frequency groups synchronize within their own group, but maintain a constant phase offset with the opposing group; $K$ is high enough that the two groups phase lock.

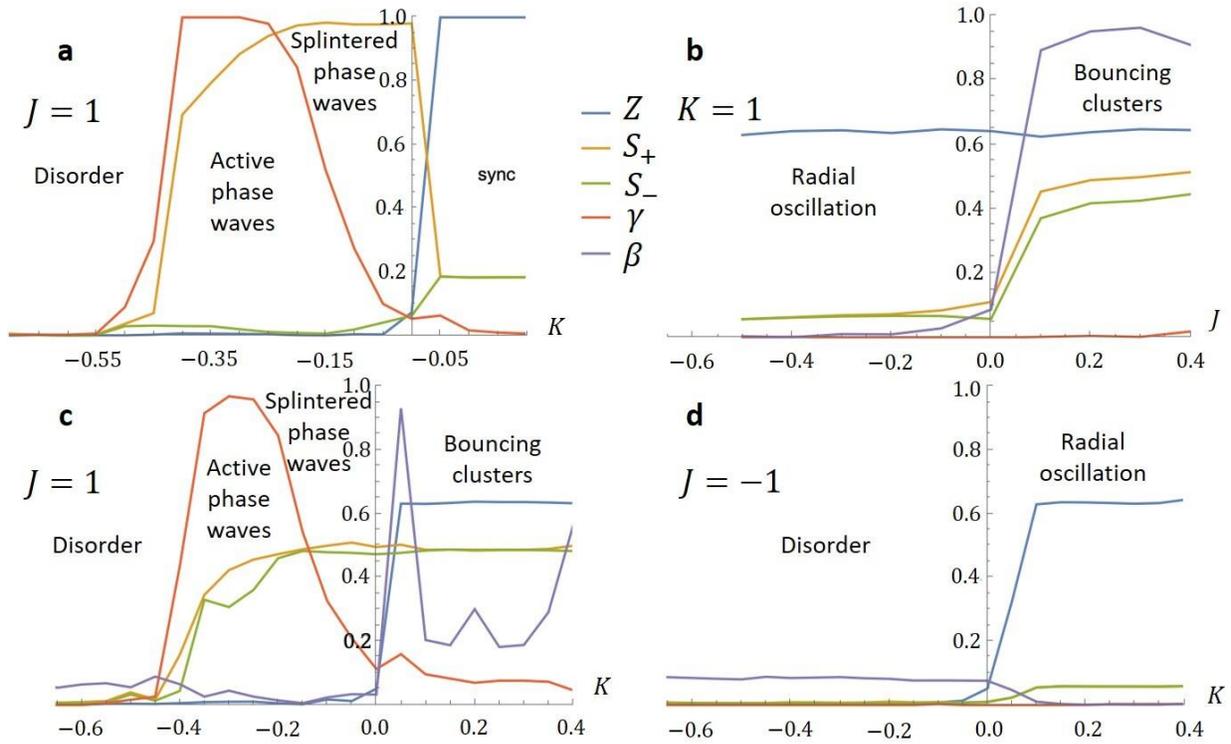

**Fig. 3. Order parameters for non-chiral swarmalators.** The order parameters $Z$, $S_+$, $S_-$, $\gamma$, and $\beta$ are plotted for non-chiral swarmalators when the natural frequency distribution is $F2$. (a) $J = 1$. (b) $K = 1$. (c) $J = 1$. (d) $J = -1$.

Finally, we observe a synchronized state in collectives with a single natural frequency that drives the collective to a circular formation that then proceeds to expand ($K > 0, J < 0$), shown in Fig. 2j. In this case, agents synchronize and then repel from similarly phased agents; since all share the same phase, the repulsion leads to uniform expansion. Supplementary Discussion 4 and Supplementary Fig. 11 address the expansion behavior for various values of $J$ and find that at $J = -1$, the collective expands indefinitely. The same region of $K - J$ parameter space for a natural frequency distribution of $F2$ reveals a radial oscillation state, where the collective evenly distributes agents from both natural frequency groups across a circular formation and proceeds to periodically expand and contract (Fig. 2k). The phase interaction behavior is similar to when there are bouncing clusters; the two groups oscillate between attraction and repulsion, but instead of splitting into multiple clusters, agents coalesce in one so that they are close to agents from the opposing group. As the two groups' phase difference decreases, agents repel each other and the collective expands, but as their phase difference reaches its maximum, attraction occurs and the group contracts. Because agents are close to other agents with a natural frequency of opposite sign, the order parameters $S$ and $\beta$ remain low while $Z$ goes to mid-range, as shown in Fig. 3d.

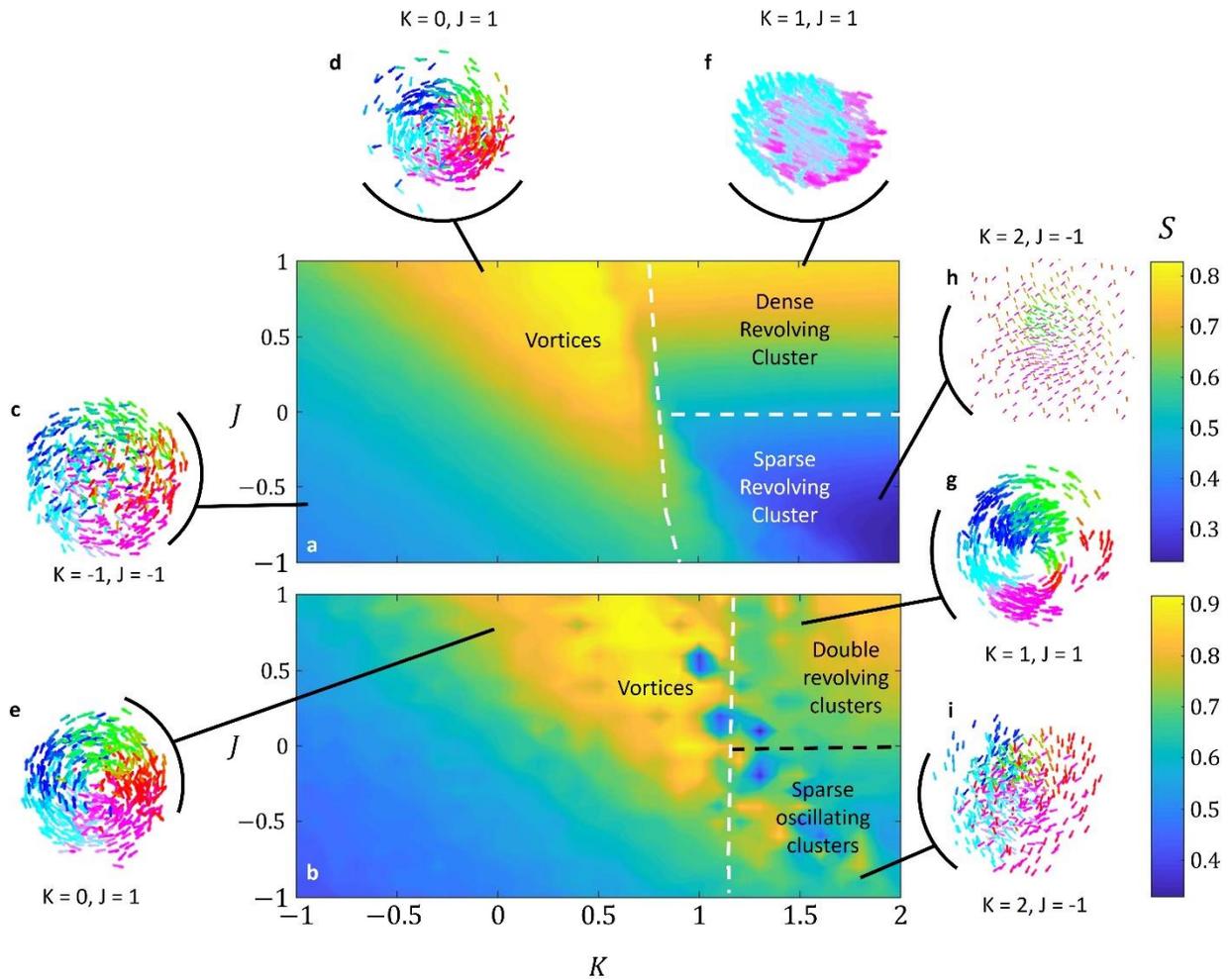

**Fig. 4. Revolving swarmalators with a natural frequency spread.** Heat maps of $S$ across $K - J$ parameter space shown for test cases with a natural frequency distribution of **(a)** $F3$ and **(b)** $F4$. **(c)** Disordered vortex. **(d-e)** Vortex. **(f-g)** Revolving clusters. **(h-i)** Sparse revolving clusters.

**Revolving Swarmalators**

Revolving swarmalators' motion and phase coupling behavior is defined by Equations (1) and (2) when $c_i \neq 0$ for all agents and $Q_{\dot{x}}, Q_{\dot{\theta}} = 0$; each agent's natural frequency is defined according to one of the four natural frequency distributions ($F1 - F4$). An overview of the revolving swarmalators' behaviors when there is no frequency coupling is further reviewed in Supplementary Discussions 5-6, shown in Supplementary Movie 6, and characterized in Supplementary Figs. 13-16. Results regarding the emergent collective behaviors when the natural frequency distributions are $F1$ and $F2$ are reviewed in Supplementary Discussion 5.

Revolving collectives composed of agents revolving in the same direction, but with a spread of revolution radii between $\frac{1}{3}$ and 1 are summarized in Fig. 4. Both heat maps demonstrate relatively high circumferential spatial phase order across the $K - J$ parameter space and especially high $S$ triangular regions spanning positive and negative $K$ and $J$. Disordered phase waves / vortices are also found with a natural frequency spread (Fig. 4c and Fig. 5) at low $K$; ordered phases waves lie in the triangular regions. As opposed to most previous phase wave formations where an annulus formed, the frequency spread enables agents to revolve at different inherent radii and form a vortex formation like the one shown in Figs. 4d-e. However, the vortex in Fig. 4d looks marginally more organized since all agents revolve in the same direction.

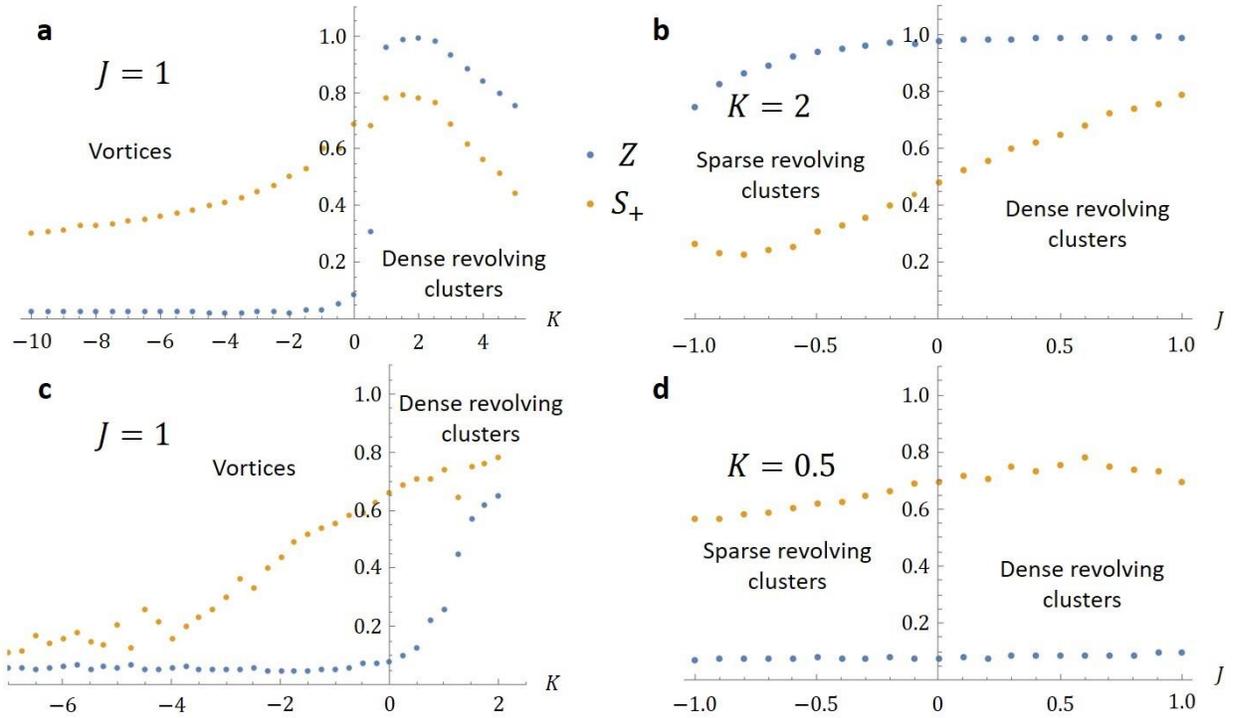

**Fig. 5. Order parameters for revolving swarmalators.** The order parameters $Z$ and $S_+$ are plotted for chiral swarmalators when the natural frequency distribution is $F2$ (top row) and $F4$ (bottom row). (a) $J = 1$, F2 frequency distribution. The dense revolving clusters exist for K > 0 as indicated by $R, S_+ > 0$. When $K < 0$, the vortex is born in which $R = 0, S_+ > 0$ (b) $K = 2$, F2. Here there is no sharp distinction between the two states, as indicated by both $S_+, R$ varying smoothly as $J$ increases from -1 to 1. Instead, there is a cross over from sparsity $R < 1$ to density $R \approx 1$ (the density in space maps onto a density in phase coherence) (c) $J = 1$. F4. The picture is the same as in panel (a); (d) $K = 0.5$, F4, the picture is the same as in (d).

The single and multiple synchronized revolving clusters are also found at high $K$ and $J$ (Figs. 4f-g); the main difference is that the agents are more tightly pack when the natural frequency distribution is $F3$ than when it is $F4$. Fig. 4f shows a tight cluster revolving and almost reaching phase synchrony, while Fig. 4g shows two more elongated clusters that begin to revolve about each other and never reach phase synchrony. This general revolving cluster behavior persists through negative $J$ as well, the main difference is that the clusters become much sparser (Figs. 4h-i ). The $S$ value also dips at high $K$ and low $J$ because the synchronized clusters become so sparse that they occupy wide angular regions of the circular trajectory so the circumferential phase organization decreases.

**Frequency-Coupled Chiral Swarmalators**
Frequency-coupled chiral swarmalators (FCCS) affect each other's motions similar to the way counter revolving agents might locally affect each other through physical interactions in a real-world setting. In Equations (1) and (2), $Q_{\dot{x}}$ and $Q_{\dot{\theta}}$ are defined by Equations (5) and (6); therefore, the natural frequency sign difference between two agents plays an important role in how they will affect each other's motion and phase coupling. Because the spatial and temporal interactions are altered by the sign difference between agents' natural frequencies, we only show results for when the natural frequency distributions are $F2$ and $F4$; the emergent behaviors of FCCS when the natural frequency distributions are $F1$ and $F3$ are the same as non-frequency-coupled revolving swarmalators since here all agent have the same sign for their natural frequencies and there is no chirality. The emergent behaviors of FCCS are summarized in Fig. 6 and further discussed and shown in Supplementary Discussion 6, Supplementary Figs. 17-18, and Supplementary Movie 7.

When the collective is split into two equal and opposite natural frequencies, behaviors emerge that appear similar to those from the regular chiral swarmalators; increased repulsion between agents with opposite natural frequency signs, however, makes the collective split into different clusters and minimize the trajectory intersection along several regions of the $K-J$ parameter space, shown in Figs. 6a,b. This is not always the case; for example, when the natural frequency distribution is $F2$ and there is low $K$ and $J$, an annulus forms with agents from both natural frequency groups moving close to each other (Fig. 6c). The same region of $K-J$ parameter space when the natural frequency distribution is $F4$ yields high mixing and high disorder (Fig. 6d); the disorder most likely occurs because of the large spread of natural frequencies. The bright rectangular region in the lower half of Fig. 6b indicates a single vortex emerging with high mixing between agents with opposite signs of natural frequency (Fig. 6e).

The first demonstration of high repulsion between counter-revolving agents is shown in Figs. 6f-h. Figs. 6f-g show two interacting phase waves that intersect each other when the natural frequency distribution is $F2$; the non-frequency-coupled swarmalators also exhibit interacting phase waves, but are sometimes difficult to distinguish since they are concentric. The phase wave intersection is caused by the phase shift terms $Q_{\dot{x}}$ and $Q_{\dot{\theta}}$ which enable agents with opposite revolving directions to repel similar phases. At very low $K$, the phase waves become counter-rotating ellipses, which decreases the circumferential phase organization of the whole collective and lowers $S$ (upper left region of Fig. 6a); when $K$ is close to zero, the collective maintains circular double phase waves. When there is a frequency spread, the collective forms two vortices that intersect similarly to the double phase waves; in addition to the vortices, however, there are loose agents around the intersection region that have a high natural frequency and are unable to join the organized formations.

Double revolving clusters are also observed when the natural frequency distribution is $F2$ (Fig. 6k-l); however, FCCS inhibit trajectory intersection for high $K$ and $J$. Rather than follow the same circular trajectory in opposite directions, each cluster follows a fairly circular trajectory, but repels away from the opposing cluster as soon as they begin to intersect. This behavior is much less evident at low $J$ because the clusters become sparser so they can intersect more without being close to other agents.

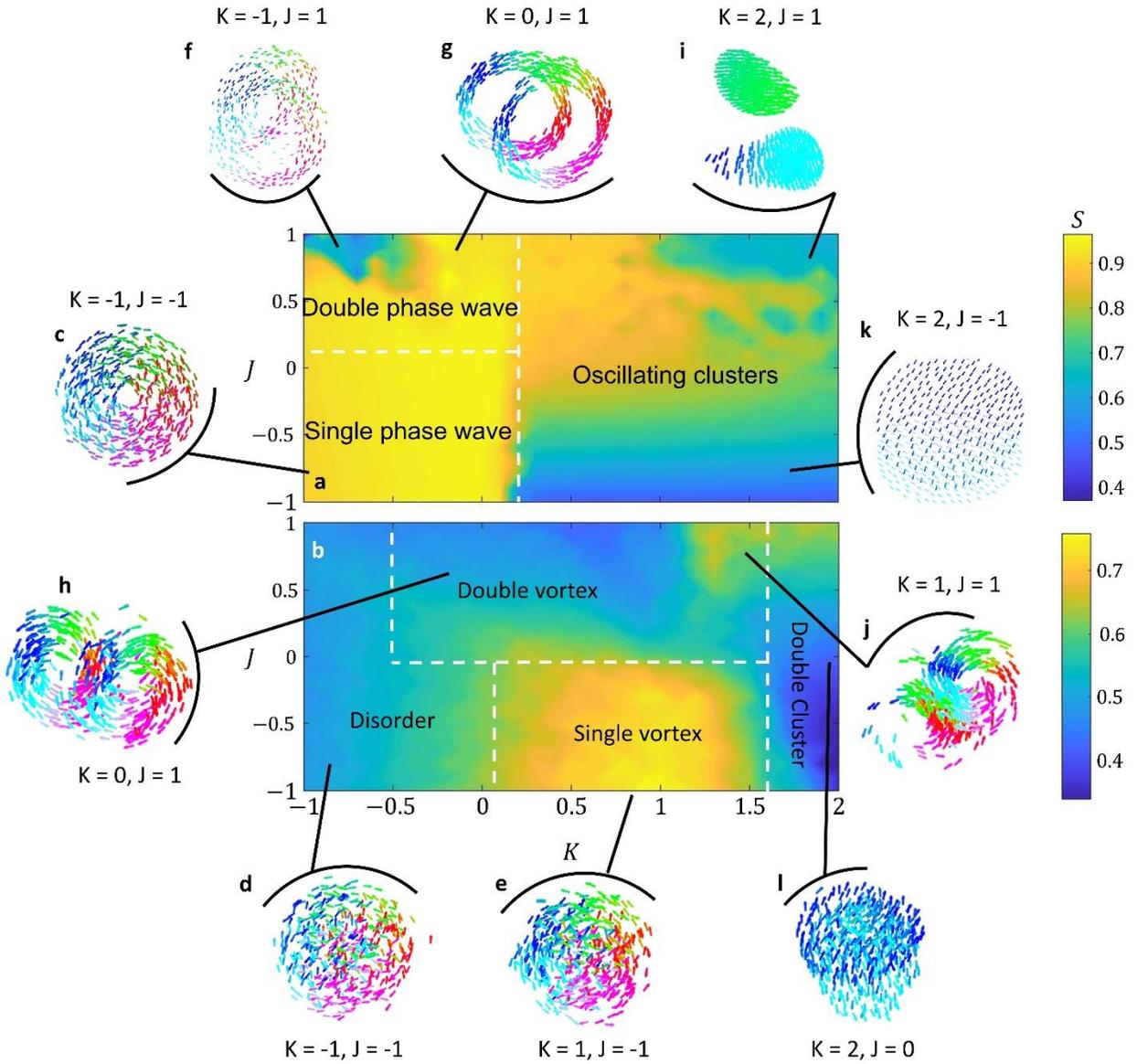

**Fig. 6. Frequency-coupled chiral swarmalators.** Heat maps of $S$ across $K - J$ parameter space are shown for natural frequency distributions **(a)** $F2$ and **(b)** $F4$. **(c)** Phase wave. **(d-e)** Disordered vortices. **(f-g)** Double phase waves. **(h)** Double vortex. **(i)** Dense revolving clusters. **(j)** Vortex. **(k-l)** Revolving clusters.

### Fine-Grained Analysis

Our analysis so far has been at the global level, the order parameters in Figs. 3 and 5 being single numbers for the entire collective. We supplement this with a finer grained analysis by computing three correlation functions: the velocity auto-correlation function, the position-position correlation function, and the phase-phase correlation function:

$$C_{v,v}(t,\tau) = \langle \frac{1}{N} \sum_i v_i(t) \cdot v_i(t-\tau) \rangle \tag{8}$$

$$C_{x,x}(r) \coloneqq g(r) \coloneqq \langle \frac{1}{N(N-1)} \sum_{i,j} \delta(r - r_{ij}) \rangle \tag{9}$$

$$C_{\theta,\theta}(r) := \langle \frac{1}{N(N-1)} \sum_{i,j} \hat{n}_i \cdot \hat{n}_j \delta(r_{ij} - r) \rangle \tag{10}$$

Throughout Equations (8) – (10), $\langle \cdot \rangle$ denotes the ensemble average.

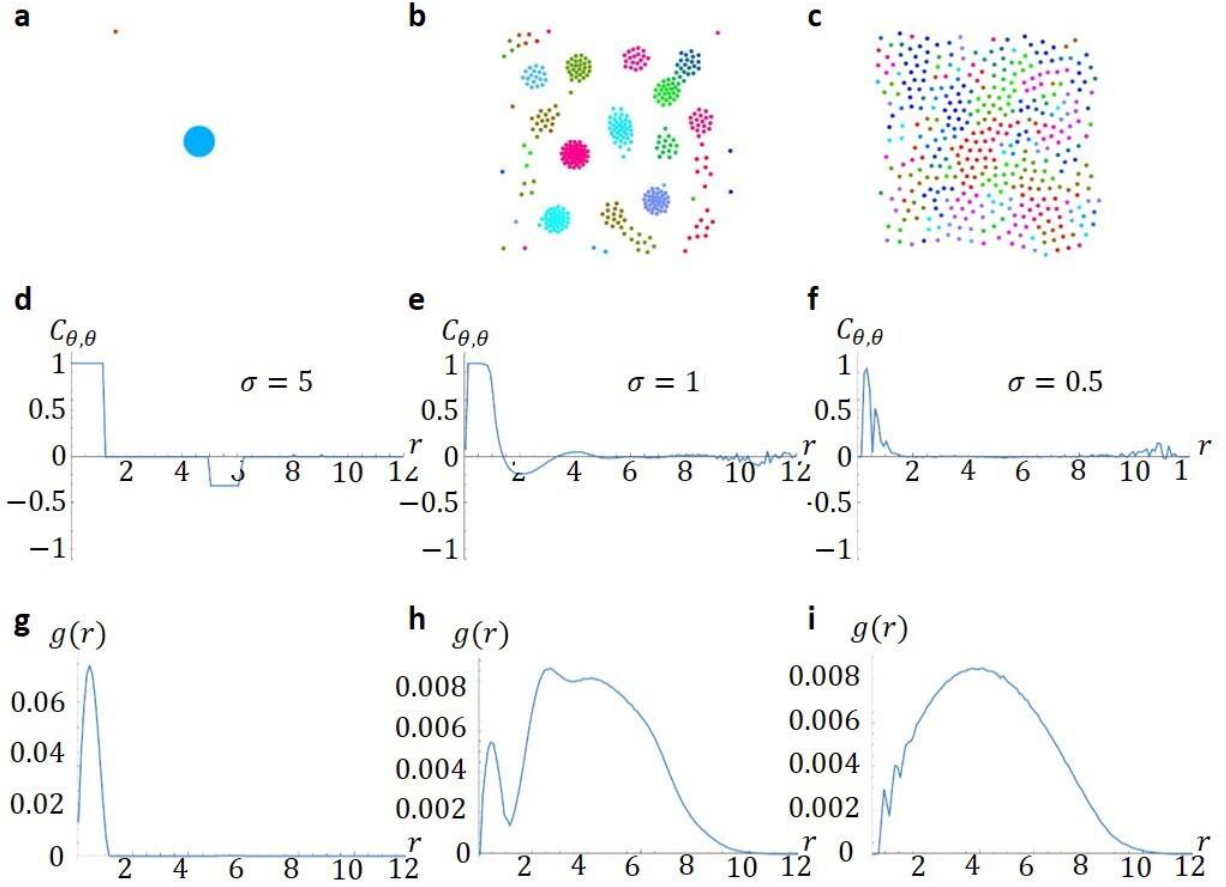

**Fig. 7. Correlation Functions.** (a-c) Scatter plots of state. (d-f) Phase-phase correlation function for different coupling ranges of $\sigma$. (g-h) $g(r)$ for different coupling ranges of $\sigma$. Simulation parameters: $(J, K, d, \Omega) = (1, 1, 0.05, 0)$, $(dt, T, N) = (0.25, 500, 300)$ with initial position drawn uniformly at random in a box of length $L = 4$ and phases drawn uniformly at random from $[-\pi, \pi]$ (d-i) Each data point is the average of 500 simulations

Since we have presented a large number of collective states, it is infeasible to calculate the above correlation functions for each state. Instead, we focus on just the sync state and the phase wave / vortex state, which in a sense are the primitives for the other states; for example, the bouncing cluster state is composed of two sync states, and double phase wave is composed of two phase waves.

The local coupling instance of the chiral swarmalator model is defined by Equations (15)-(17): agents only sense other agents within a set radius ($\sigma$). Supplementary Discussions 8, 9, and 10 and Supplementary Movies 8-12 present a comprehensive analysis of the correlation functions for the sync and phase wave states under various effects, such as dynamical noise $d$ and coupling of finite range $\sigma > 0$. We present the most interesting effects for the sync state here; the phenomena are virtually the same for the phase wave / vortex state (Supplementary Discussions 9 and 10). Supplementary Figs. 20-27 show many of the emergent behaviors for parameters of the local coupling instance, while Supplementary Figs. 28-35 explore how the collective's phase coherence, kinetic energy, and linear momentum fluctuate when there are different natural frequency distributions, a spectrum of $J$ and $K$ values, and a spectrum of $\sigma$ ranging from 0.2 to 2. In Supplementary Figs. 36-39, we also quantify how many clusters appear, the collectives' normalized speed, and the collectives' average normalized angular momentum at different values of $K$ and $J$ for a range $\sigma$ between 0.2 and 2.

Fig. 7 shows how the sync state deforms under local coupling. For long range $\sigma = 5$, a single sync cluster is realized in Fig. 7a. The phase-phase correlation function $C_{\theta,\theta}(r) \approx 1$ over this radius, then drops to zero (Fig. 7d). Similarly, the position-position correlation function $g(r)$ has a single peak and drops to zero beyond $r \approx 2$ consistent with a single cluster. For larger noise $d > 0, \Omega > 0$, the sync cluster develops a slight phase gradient reflected in a dip in $C_{\theta,\theta}(r)$ (Supplementary Figs. 54 and 55). For intermediary coupling $\sigma = 3$, multiple clusters appear (b) with little inter-cluster phase correlation; $C_{\theta,\theta}(r)$ drops to zero for $r \approx 3$. $g(r)$ develops two new, blurred peaks, which is consistent to a state with multiple clusters (Fig. 7h). As before, adding noise and a frequency spread $d > 0, \Omega > 0$ leads to an intra-cluster phase gradient. Finally for short range coupling $\sigma = 0.5$, a single crystal-like state emerges (Fig. 7c) with length scale $r \approx 10$, as evidenced by the domain of $g(r)$ in Fig. 7i, and local phase coherence, as evidenced by $C_{\theta,\theta}(r)$ having three descending peaks over a length scale $r \approx 2$ (Fig. 7f). This state bifurcates into an incoherent gas-like state when $d > 0$ and $\Omega > 0$.

The same single / multiple cluster and gas-like transition under increasingly local coupling is observed for vortices, both non-chiral (Supplementary Fig. 57) and chiral (Supplementary Fig. 59).

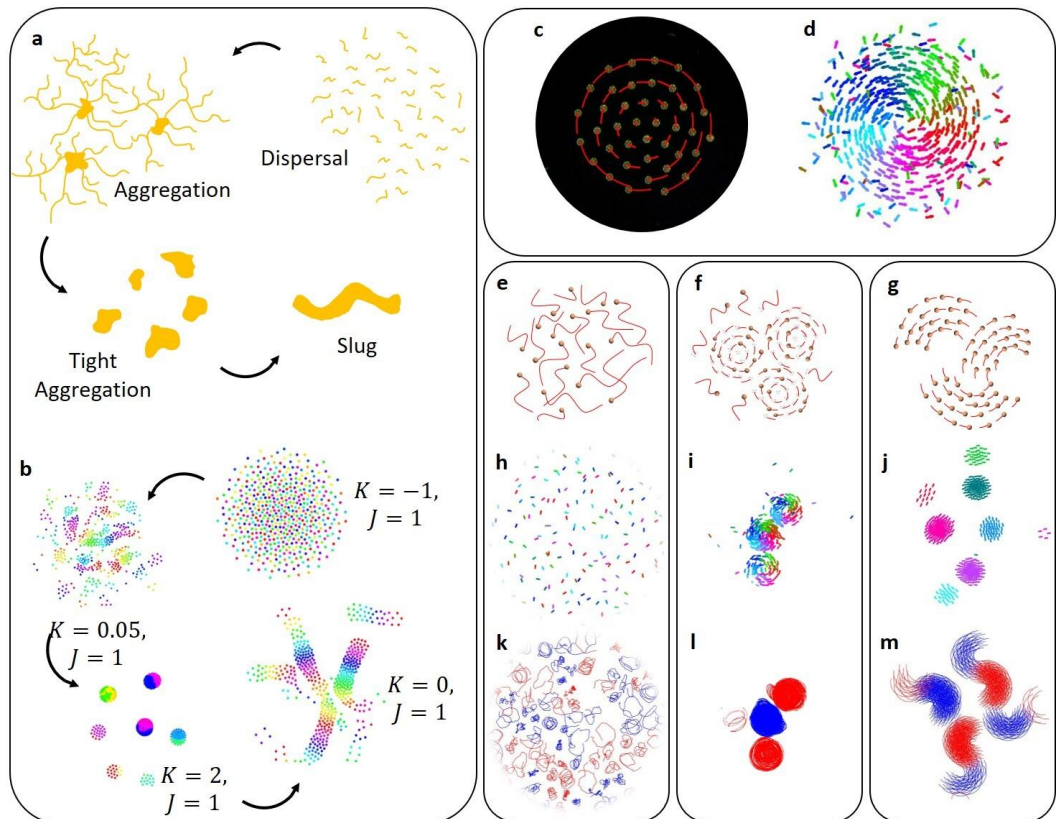

**Fig. 8. Swarmalators resembling real swarms.** (a) Graphical representation of slime mold life stages. (b) Stages from (a) reproduced by non-chiral swarmalators when $\sigma = 1.4$. (c) Emergent vortex-like formation of micro-disks used in Gardi et al.[50]. (d) Emergent single vortex-like formation qualitatively resembling the one shown in (c). (e-m) FCCS resembling Quincke rollers. (e-g) Graphical representation of Quincke rollers exhibiting three distinct states: gas-like, multiple vortices, and flocking vortices. (h-j) Swarmalator behaviors that closely match the behaviors seen in Refs. 30 and 31; the natural frequency distributions are (h) $F4$, (i) $F4$, and (j) $F2$; the parameters here are (h) $\sigma = 0.8, K = -1, J = -1$, (i) $\sigma = 1.6, K = 0, J = 1$, (j) $\sigma = 0.8, K = 1, J = 1$, (k-m) Trajectories of swarmalator collectives shown in (h-j). Blue trajectories correspond to agents with a negative natural frequency and red trajectories to a positive natural frequency.

**Real World Parallels**

The swarmalator research field is still building momentum. While the models demonstrate rich and exciting spatiotemporal patterns[33-38], direct use-cases continue to be elusive, earning them a 'toy-model' reputation[46]. To indicate the potential of our model to break this barrier, we use this section to discuss parallels between emergent behaviors in swarmalators and a range of natural and engineered swarms. Figs. 8 and 9 show how locally coupled swarmalators qualitatively resemble emergent behaviors in slime mold, vortex arrays of spermatozoa, rotating magnetic colloids, and Quincke rollers, for example. We show additional examples in the Supplementary Material and Supplementary Movies 13-17.

Social slime mold has various life stages that consist of aggregation, collective motion, and dispersal[6,47]. Here, we demonstrate that locally coupled non-chiral swarmalators can qualitatively reproduce these four general life-stages of the multi-cellular amoebae. Fig. 8a shows the various life stages for Dictyostelium Discoideum where thousands of cells begin to aggregate through starvation; as the starvation period continues, prolonged aggregation enables tight clusters that can then combine to form a plasmodium or "slug". This formation enables cells to move together even though there is no central brain within the collective. The slug moves around the environment until it finds nutrients and the collective disperses. Fig. 8b shows the swarmalators' version of the general behaviors graphically represented in Fig. 8a; Supplementary Movie 16 shows the transitions between these different behaviors. The non-chiral swarmalators in Fig. 8b change $K$ and $J$ and have the natural frequency distribution $F2$ and $\sigma = 1.4$. The sparse clusters when $K = 0.05, J = 1$ result from agents that partially synchronize within small groups; when $K = 2, J = 1$, the collectives form tighter clusters. The slug stage is achieved by setting $K = 0, J = 1$; this turns off phase coupling while enabling attraction between agents with a similar phase. A set of phase wave ribbons form instead of the annulus since agents can only sense locally. We have not done an exhaustive exploration of state transitions, but we did find that clusters are harder to form when there is high dispersion. Although phase behavior is not the same as what is exhibited in the real social slime mold, we believe this holds great promise for developing complex behaviors with oscillator-like artificial swarms. The non-chiral instance of our model can also be modified as described in Equations (16) and (17) to capture the radial phase waves seen in aggregations of embryonic genetics oscillators[41]; this is explored in greater detail in Supplementary Figs. 60-62 and Supplementary Movie 17.

Spinning magnetic microrobots like those shown in Refs. 48-51 and by Yan et al.[29] produce rotating formations; an image of a rotating microrobot collective[51] is shown in Fig. 8c, and we can create similar vortices (Fig. 8d) with the regular chiral swarmalators. Note that the rotating crystal-like behavior demonstrated in each of these works is not accurately reproduced with the swarmalators because agents are free to move out of their circular trajectory over time, whereas in the real system, the particles in the rotating magnetic colloids are physically enclosed in the crystal-like formation. The microrobots in Gardi et al.[51] are able to move freely, therefore this system is more adequate for realizing swarmalator-like vortices. Indeed, Supplementary Fig. 63 shows that when the collective has an exponential distribution of natural frequencies between zero and one, the speed and angular velocity about the collective centroid are comparable between the physical experiment and the swarmalator simulations. Following these comparisons, it follows that the general revolving behavior of the swarmalators could be further tuned to study the behavior of other revolving collectives, like those shown by Han et al.[30], Zhang et al.[31], Grzybowski et al.[48,49], and Wang et al.[50].

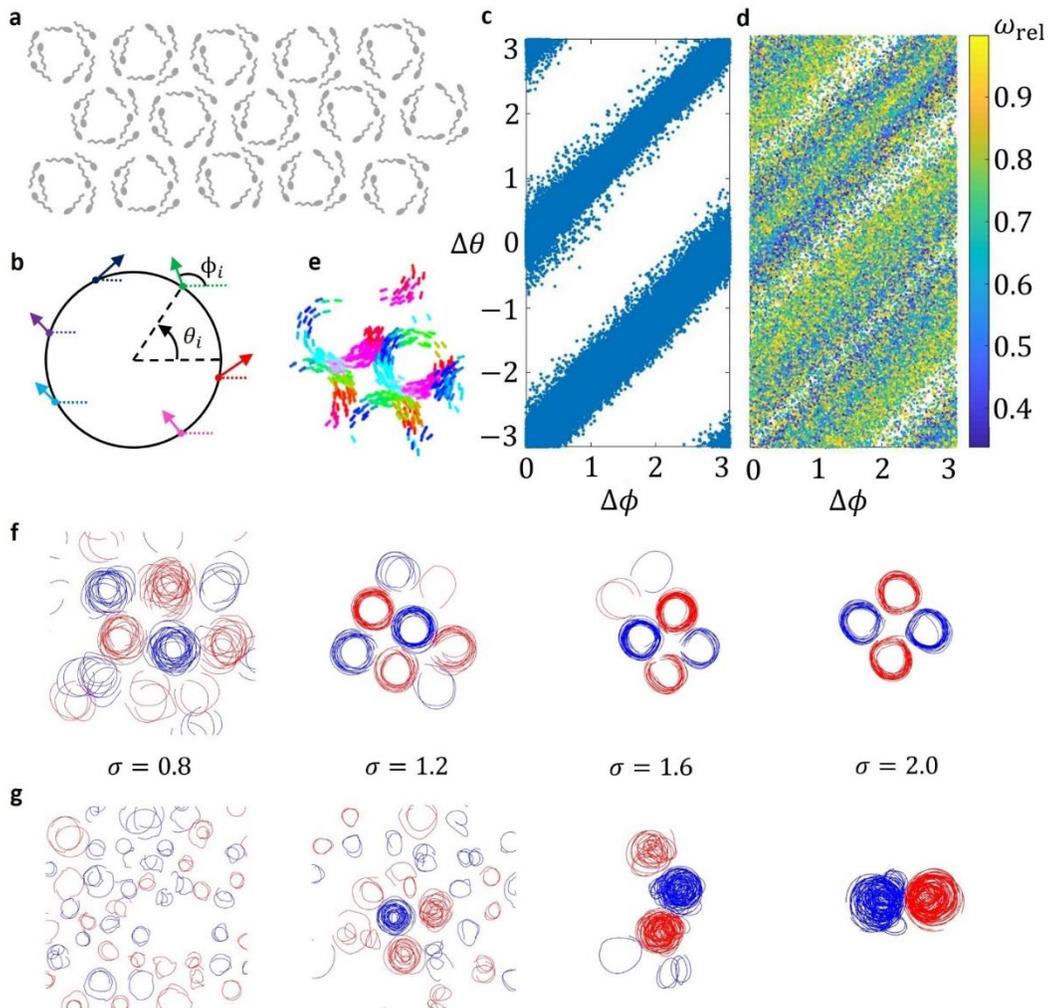

**Fig. 9. FCCS resembling spermatozoa. (a)** Graphical representation of sperm vortex arrays. **(b)** Head orientation vs. position along circular trajectory. **(c)** Plot of head orientation vs. angular position when the natural frequency distribution is $F2$. **(d)** Scatter plot of phase (angular position) vs. head orientation color coded by the relative natural frequency; this corresponds to when the natural frequency distribution is $F4$. The result is similar to what is shown in Fig. 2 of Riedel et al.[10] **(e)** Snapshot of swarmalators demonstrating this behavior when $K = 0, J = 1, \sigma = 1.2$, and the natural frequency distribution is $F2$. **(f)** Trajectories of swarmalators self-organizing into arrays of vortices when the natural frequency distribution is $F2$. **(g)** Trajectories of swarmalators self-organizing into several adjacent vortices when the natural frequency distribution is $F4$.

We also qualitatively compare the formations of the FCCS to Quincke roller collectives[30,31], which exhibit different formations including gas-like, single vortex, multiple vortices, and flocking vortices; these states are graphically depicted in Fig. 8e-g. Figs. 8h-j show sample emergent formations and the corresponding trajectories of swarmalators exhibiting those states when we vary values for $K, J$, and the natural frequency distribution. Each of the trajectory snapshots are chosen because they qualitatively compare to formations shown in Refs. 30 and 31. There is a vast literature on the emergent behaviors of Quincke rollers with many more behaviors that could be worth imitating using the swarmalator models. Finally, Fig. 9 highlights some of the similarities between spermatozoa vortex arrays and the swarmalators when multiple phase waves emerge. Fig. 9a shows a graphical representation of the spermatozoa vortex arrays shown by Reidel et al.[10], where the vortices are tightly packed. In their paper, the authors provide a histogram of the sperm head orientation versus position about their circular trajectory; this concept is depicted in Fig. 9b where each dot represents an agent, its color the phase ($\theta_i$) or position about the circular trajectory, and the orientation is $\phi_i$. We show two similar plots (Fig. 9c,d) and find that the general trend remains similar for swarmalators, at low and high $\sigma$ as shown in

Supplementary Fig. 64. This is powerful because we can use this model to replicate some of the emergent collective behaviors of a natural system without considering the specific hydrodynamic coupling mechanisms that occurs in the real system. Fig. 9e shows similar packing of the ring-like formations; although, as shown by the trajectories in Fig. 9f, the swarmalator vortices are in square grids across all $\sigma$ rather than hexagonal grids like those depicted in Fig. 9a. The FCCS have a strong attraction between agents with differing natural frequency when their phase is offset by $\pi$, so agents from opposing natural frequency groups will have a stronger attraction than agents from the same group. We observe some of the same behavior when the natural frequency distribution is $F4$; the vortex-like formation is not as organized when the natural frequency distribution was $F2$ but vortices with opposing chirality form next to each other because of the attraction between phase offsets of $\pi$.

**DISCUSSION**

We explored non-chiral and chiral swarmalators across various natural frequency distributions and found a zoo of emergent behaviors. Non-chiral swarmalators have no inherent motion, yet their temporal behavior drives them towards dynamic behaviors like radial oscillation, synchronized expansion, bouncing clusters, static anti-phase states, concentric phase self-organization, and many variations of phase waves, sync, and partial sync states. Chiral swarmalators couple spatial and temporal behavior by adding an inherent circular motion to each agent; the emergent behaviors follow similar trends to the non-chiral swarmalators in the $K - J$ parameter space, but the circular motion also enables higher levels of spatial-phase order even when there is little or no phase coupling. Some of the most interesting emergent behaviors include single and multiple vortices and phase waves, along with dense and sparse revolving clusters. Throughout the non-chiral and chiral swarmalators, global and local coupling illuminates the effect of distance on the emergent self-organization occurring for different values of $K$ and $J$. The emergent behaviors qualitatively mimic a wide variety of collective systems including slime mold, spermatozoa, magnetic colloids, and Quincke roller collectives.

Past work on swarming systems includes several models similar to the Viscek model[20], which demonstrates that collectives can transition from disordered states to flocks by modifying individual agents' behavior according to their neighbors' motion and orientation. Modified Viscek models defined the agents' motion as inherent revolutions; this led to a breadth of states including revolving flocking patterns[25,26], vortices, and active foams[21]. These models address emergent swarming states in the presence of noise and agent-to-agent coupling; however, if inherent motion is removed, many of the behaviors cease to exist. Our swarmalator model enables us to explore resulting emergent behaviors when the inherent motion is turned on or off. Its strength therefore lies in the fact that it can represent inherently mobile agents where the phase is correlated to orientation, as was the case in the previously mentioned works[20,21,25,26], as well as collectives where there is no inherent motion and the phase is an internal property. To the best of our knowledge, no other active matter model is this concise and offers such a diverse set of emergent collective behaviors. Indeed, many may be finely tuned to closely mimic the behavior of some real world collectives[51-53]; however, with this new model we can switch between wildly different behaviors by switching just a few global parameters. As shown in the summarizing lists in Supplementary Fig. 1 and Supplementary Table 1, our model covers a wide variety of behaviors that qualitatively resemble many real-world collectives[5,10-12,31,40,47,49,50,53,56-60] and are only partially discovered in other active matter models[20, 24, 25-27, 36, 45].

Given the universality of phase dependence and circular motion, this model may be used to advance studies across fields, from inspiring new theoretical advances in active matter to modeling frameworks in developmental and behavioral biology. These minimalistic coordination schemes may also support macro-scale robot swarms where constituents travel in periodic orbits that exceed their range of communication[43,51], multi-robot systems acting in the human space where more advanced perception and reasoning presents security concerns[52], and microrobot swarms in biomedical applications which are fundamentally limited in both sensing and computation power[53-55]. We hope this encourages a line of fundamental studies on the emergent behaviors that result from an interdependence between time-domain and spatial-domain-specific parameters.

**METHODS**

**General Simulation Parameters**
Numerical studies were run on Matlab using Euler integration, 500 agents, a time step size of $dt = 0.1$, and a final time step of $t_f = 1000$. Simulations for the $S$ order heat maps were run for 10 trials.

**Order Parameters**
The order parameter $\gamma$ represents the fraction of swarmalators in a collective that have completed at least one cycle of phase and position about the collective centroid after the transient period.

The Kuramoto order parameter $Z$ gives a measure of the overall phase coherence / degree of synchrony. The parameter is defined by the following equation.

$$Z = \frac{1}{N} \sum_{j=1}^{N} e^{i(\theta_j)} \tag{11}$$

The order parameter $\beta$ is the difference in the average position of swarmalators with a positive natural frequency and those with a negative natural frequency, $n_{\omega > 0}$ is the number of swarmalators with a positive natural frequency, $n_{\omega < 0}$ is the number of swarmalators with a negative natural frequency, and $x_{\omega > 0}$ and $x_{\omega > 0}$ are the respective positions of the swarmalators:

$$\beta = \left| \frac{1}{n_{\omega > 0}} \sum_{j=1}^{n_{\omega > 0}} x_{\omega > 0} - \frac{1}{n_{\omega < 0}} \sum_{j=1}^{n_{\omega < 0}} x_{\omega < 0} \right| \tag{12}$$

**Local Coupling Model**
The local coupling model uses step functions to determine whether agents affect each other's motion and phase behavior.

$$\dot{x}_i = v_i + \frac{1}{N} \sum_{j \neq i}^{N} \left[ \left( \frac{x_j - x_i}{|x_j - x_i|} \left( A + J \cos(\theta_j - \theta_i - Q_{\dot{x}}) \right) \right. \right. \tag{13}$$
$$\left. \left. - B \frac{x_j - x_i}{|x_j - x_i|^2} \right) H(\sigma - |x_j - x_i|) \right] + \xi_i(t)$$

$$\dot{\theta}_i = \omega_i + \frac{K}{N} \sum_{j \neq i}^{N} \left( \frac{\sin(\theta_j - \theta_i - Q_{\dot{\theta}})}{|x_j - x_i|} \right) H(\sigma - |x_j - x_i|) + \eta_i(t) \tag{14}$$

$$H(\sigma - |x_j - x_i|) = \begin{cases} 1, & \sigma - |x_j - x_i| > 0 \\ 0, & \sigma - |x_j - x_i| \leq 0 \end{cases} \quad (15)$$

We can study the effect of white noise on the emergent collective behaviors by adding in the terms $\xi_i$ and $\eta_i$. The white noise terms are $\xi_i$ and $\eta_i$ and are generated from the same distribution. The maximum distance for coupling is defined by the variable $\sigma$.

**Model Variation**

The non-chiral instance of the model can be slightly modified to demonstrate behavior very similar to embryonic genetic oscillators.

$$\dot{x}_i = v_i + \frac{1}{N}\sum_{j \neq i}^{N}\left(\frac{x_j - x_i}{|x_j - x_i|^2}(A + J\cos(\theta_j - \theta_i)) - B\frac{x_j - x_i}{|x_j - x_i|^4}\right) + \xi_i(t) \quad (16)$$

$$\dot{\theta}_i = \omega_i + \frac{K}{N}\sum_{j \neq i}^{N}\sin(\theta_j - \theta_i - \alpha) + \eta_i(t) \quad (17)$$

Here, we include a constant phase offset term $\alpha$ and modify the distance relation between swarmalators; $v_i$ is equal to zero for all agents and noise is included in the model.

**Acknowledgments:** We thank Gaurav Gardi for fruitful discussions on possible venues of application for the chiral swarmalators and further analysis that can be done for these swarming systems. We also thank Nils Napp for valuable feedback on how to improve the clarity of this work.

**Funding:** S. C. and K. P. thank the National Science Foundation Graduate Research Fellowship, the National Science Foundation grant 2042411, and the Packard Foundation Fellowship for Science and Engineering.

**Author contributions:** S. C., K. O. K, and K. P. designed the study; S. C. contructed the model, S. C. and K. O. K. performed data processing and analysis; S. C. wrote the manuscript; all authors discussed the results and contributed to the editing of the manuscript; K. P. supervised the research.

**Competing interests:** Authors declare that they have no competing interests.

Supplementary Information for

**Diverse Behaviors in Non-Uniform Chiral and Non-Chiral Swarmalators**

**The PDF file includes:**

Supplementary Table 1. Summary of emergent states in this work and other works.

Discussion 1. System oscillation characterization.
Discussion 2. Natural frequency and revolution radius distributions.
Discussion 3. Non-chiral swarmalators with a frequency spread.
Discussion 4. Overview and characterization of non-chiral swarmalators.
Discussion 5. Chiral swarmalators with discrete sets of natural frequencies.
Discussion 6. Overview and characterization of chiral swarmalators.
Discussion 7. Order Parameter.

Discussion 8. Analysis of sync state.
Discussion 9. Analysis of Phase waves / vortices.
Discussion 10. Analysis of Revolving Swarmalators.

**Other Supplementary Material for this manuscript includes the following:**

**Supplementary Table 1. Summary of emergent states in this work and other works.**

*A refers to the work in this paper

| No. | States | Real-world systems that exhibit similar states | Models, by reference, that demonstrate similar states ||||||||
|---|---|---|---|---|---|---|---|---|---|---|
| | | | A | 20 | 24 | 25 | 26 | 27 | 36 | 45 |
| 1 | Static Sync | Spermatozoa[10], bristlebots[56], Janus particles[57], chemical micromotors[58, 59], magnetic domain walls[60] | ■ | | ■ | | | ■ | | ■ |
| 2 | Static Async | Quincke rollers[31], bristlebots[56] | ■ | | | | | ■ | | |
| 3 | Phase Wave | Ferromagnetic colloids[61], vinegar eels[11], Japanese tree frogs[5] | ■ | | ■ | | | ■ | | |
| 4 | Splintered Phase Wave | ~ | ■ | | | | | ■ | | |
| 5 | Active Phase Wave | Mosh pit[62], Colloidal rollers[31], Spermatozoa[10], Bristlebots[56] | ■ | | | | | ■ | | |
| 6 | Linear Phase Wave | Slime mold[47] | ■ | | ■ | | | | | |
| 7 | Multiple Bouncing Clusters | ~ | ■ | | ■ | | | | ■ | |
| 8 | Concentric Phase Self-Organization | Embryonic genetic oscillators[40] | ■ | | | | | | | ■ |
| 9 | Radial Oscillation | Phototactic micromotors[12] | ■ | | | | | | | |
| 10 | Synchronized Expansion | Ferromagnetic microrobots[53] | ■ | | | | | | | |
| 11 | Anti-Phase State | ~ | ■ | | | | | | | |
| 12 | Vortex | Mosh pit[62], ferromagnetic microrobots[49, 50, 53, 54] | ■ | | | | | | | |
| 13 | Gas-like | Ferromagnetic microrobots[53], Quincke rollers[31], slime mold[47] | ■ | ■ | | ■ | ■ | | | |
| 14 | Dense Revolving Cluster | Bristlebots[56] | ■ | | | | ■ | | | |
| 15 | Multiple Revolving Cluster | Bristlebots[56], Quincke rollers[31] | ■ | ■ | | | ■ | | | |
| 16 | Sparse Revolving Cluster | Quincke rollers[31] | ■ | | | | ■ | | | |
| 17 | Multiple Vortices | Spermatozoa[10] | ■ | | | | | | | |
| 18 | Elongated Vortices | ~ | ■ | | | | | | | |

**Discussion 1. System oscillation characterization**

This discussion reviews the oscillatory swarmalator behavior in attraction and repulsion across various phase interaction parameter spaces. The plots in Supplementary Fig. 2 show why oscillation in attraction and repulsion occurs as the agents' phases vary. Fig. S1a shows a heat map of the spatial phase interaction term in Equation (1) from the main text: when the two agents' phases are used as coordinates to plot a point in $\theta_i - \theta_j$ space, we can determine whether that point lies along attraction or repulsion regions. If the point lies close to the line $\theta_j = \theta_i$, then it lies within the yellow region which indicates a positive value for the spatial phase interaction, which enables attraction. When the point lies close to either of the lines $\theta_j = \theta_i - \pi$ or $\theta_j = \theta_i - \pi$, then the spatial phase interaction is negative, which enables repulsion. If agents vary their phases at the same rate, they will maintain the same phase difference over time and according to the phase distribution they will maintain a constant amount of attraction or repulsion. However, with natural frequency distributions the phase difference changes over time and the ($\theta_i,\theta_j$) point moves through attraction and repulsion regions.

  Supplementary Fig. 2b shows the effect of a phase offset on the spatial phase interaction term. Here, the space is defined by the phase difference changing between $-2\pi$ and $2\pi$ and a fixed phase offset ($Q_{\dot{x}}$) also changing between 0 and $2\pi$; the trend shown continues over larger ranges. In Supplementary Fig. 2a there were two bands of repulsion and one band of attraction; however, the phase offset enables two additional bands, three bands of repulsion and two bands of attraction. The more complex space means that agents are more prone to switch between attraction and repulsion when their phase difference changes over time. Supplementary Fig. 2c shows the spatial phase interaction value when $Q_{\dot{x}}$ is defined by Equation (4) from the main text. Here, the three possible values for $Q_{\dot{x}}$ are $-\pi$, 0, and $\pi$.

  The frequency dependence is relevant when modeling physical systems, where agents with the same direction of motion will tend to align and repel agents that are revolving in the opposing direction. The three lines show the phase shifts along positive and negative phase differences that occur with frequency coupling. Future swarmalator studies with different natural frequency distributions may find the particular spatial phase interaction space that leads to especially useful static or dynamic formations. Further research along these lines may find that macro-scale robot collectives can be made to move in specific ways and self-organize into any formation by simply communicating some phase value to surrounding agents and reacting to the neighbors' phase through a programmable spatial phase interaction.

## Discussion 2. Natural frequency distributions.

This discussion reviews the methods used for generating the four natural frequency distributions used in this study and how they relate to the inherent orbiting radius when $c_i$ is equal to 1 or $-1$.

- Frequency distribution $F1$: The collective shares the same natural frequency ($\omega = 1$), $c_i = 1$ and $R_i = 1$.
- Frequency distribution $F2$: The collective is split into two natural frequency groups; half of the collective's natural frequency is $\omega = 1$ and the second half has $\omega = -1$. Here, the first half has $c_i = 1, R_i = 1$ and the other half with $c_i = -1, R_i = 1$.
- Frequency distribution $F3$: There is a spread of natural frequencies across only positive values so that the whole collective has $\omega \sim U(1,3)$, $c_i = 1$, and $P(R) \sim R^{\wedge}(-2)$.
- Frequency distribution $F4$: There is a spread of natural frequencies across only positive values so that one half of the swarmalators have $\omega \sim U(1,3)$ and the second half have $\omega \sim U(-3,-1)$. If the swarmalators have inherent circular orbits, then first half of the collective has $c_i = 1$ and the second half has $c_i = -1$, and both halves have $P(R) \sim R^{\wedge}(-2)$.

For chiral swarmalators and frequency-coupled chiral swarmalators (FCCS) this ensures that agents have a negative or positive sign to their natural frequency and an inherent angular velocity in their circular orbit which determines their direction of motion. For revolving swarmalators $R_i$ is a physically relevant term since it defines an agent's revolution radius.

The four natural frequency distributions ($F1$, $F2$, $F3$, and $F4$) are depicted in Supplementary Fig. 3.

**Discussion 3. Non-chiral swarmalators with a natural frequency spread**

This discussion presents some of the most interesting emergent collective behaviors when non-chiral swarmalators have a natural frequency spread. Supplementary Figs. 4a-b show that disorder lies throughout much of the $K-J$ parameter space and an example formation is shown in Supplementary Fig. 4c. At higher values of $K$ and $J$, however, clusters of agents with lower $|\omega|$ begin to synchronize while agents with higher $|\omega_i|$ remain asynchronous and at the outer edges of the collective. Supplementary Fig. 4d shows the first instance of partial synchronized collectives with a positive natural frequency spread; at $K=1$, the collective begins to couple and initially a small, synchronized cluster forms in the center of a ring of asynchronous agents, much like the one in Supplementary Fig. 4e. Over time, agents from the asynchronous group begin to couple more closely with the cluster and therefore move towards the center. The agents joining the center cluster have a higher natural frequency; therefore, the group does not synchronize and an elongated phase wave forms along an axis. Agents from the outer boundary coalesce at both ends of the phase wave, slowly stepping into phase with the surrounding agents.

Supplementary Fig. 4e shows synchronization; Similar behaviors occur when the natural frequency spread is across positive and negative values, except two sync clusters form since agents with the same sign of natural frequency can synchronize more easily. As shown in the sequential images in Supplementary Fig. 4f, two phase waves periodically bounce from each other's boundary because of the oscillatory attraction and repulsion and are surrounded by a cloud of agents with higher natural frequencies. The mechanisms of phase interaction here are like those underlying the behavior in main text Fig. 2h, but the spatial oscillation amplitude remains much lower because of the many agents along the border that are unable to join one of the synchronized clusters. Another partial synchronization state emerges at high $K$ when $J=0$ for both natural frequency distributions; agents coalesce into a circular formation and many of them synchronize (Supplementary Figs. 4g-h).

Finally, a state emerges at high $K$ and low $J$; agents self-organize concentrically by phase and an asynchronous cluster forms at the center. When $J<0$, agents with a high $|\omega_i|$ spatially attract since their phase difference is large much of the time, thus a cluster forms. Conversely, agents with a lower $|\omega_i|$ synchronize more easily and repel each other because of the low phase difference; this drives them to form a sparse cloud around the center cluster (Supplementary Figs. 4i-j). A greater fraction of the collective forms the sparse cloud in Supplementary Fig. 4i than in Supplementary Fig. 4j because all agents have natural frequencies distributed across positive values, which enables a greater fraction to have low $|\omega_i|$ and synchronize more easily. In Supplementary Movie 1 there is slight radial oscillation; however, we classify these as separate states because (1) the radial oscillation is almost insignificant, (2) much of the collective remains clustered and is not expanding, and (3) the perimeter cloud of agents has a noticeable concentric phase wave traveling from the center out that is not as obvious in the radially oscillating collective from main text Fig. 2j.

**Discussion 4. Overview and characterization of non-chiral swarmalators**

This discussion reviews the collective behaviors exhibited by non-chiral swarmalators with no frequency coupling. The behaviors achieved by swarmalators when $c, Q_{\dot{x}}, Q_{\dot{\theta}} = 0$ for integer values of $K$ and $J$ are shown for the four natural frequency distributions in Supplementary Fig. 5. The formations shown here have only scratched the surface of what emerges when the coefficients within $\dot{x}_i$ and $\dot{\theta}_i$ are varied. The formations shown in Supplementary Fig. 6 correspond to swarmalators with discrete natural frequencies; Supplementary Figs. 6a-g help the reader visualize each of the corresponding behavioral characterizations that are relevant in Supplementary Figs. 6h-n.

    The first formation shown and characterized in Supplementary Figs. 6a and 6h is the interacting phase waves that results from half of the collective having $\omega_i = -1$ and the other half $\omega_i = 1$. The mean speed and $S$ order are plotted for the range of $K$ when $J = 1$. Throughout most of the cases in which $K > 0$, the collective holds a stable mean speed between 0.1 and 0.15; however, around the point where two interacting phase waves emerge ($K \approx 0$), the collective increases its mean speed because of the increased oscillation that occurs when the agents spread out along rings and are each closer to agents from the opposite natural frequency group. When $K, J > 0$, two clusters form that oscillate between attraction and repulsion; this enables the entire collective to be in constant motion, and after each natural frequency group synchronizes within itself, the speed remains constant ($K \approx 0.3$). At these higher values of $K$, the cluster formation ensures that each agent's motion is affected mainly by two values at any point in time: the average phase within its own natural frequency group and the average phase of the opposite group. When the ring formations emerge, the global spatial-phase order decreases because of the presence of two natural frequency groups, but the concurrent interaction between so many different phases enables each agent to oscillate in place more quickly.

    In Supplementary Fig. 6i the local coherence measures the average phase coherence of each agent with neighbors that are within a distance of 0.15. The static anti-phase locked state occurs when $J = 0$; when $K < 0$, the local coherence remains at a very small value and then spikes up as soon as it is positive. Another increase in phase coherence occurs when $K > 1.5$; here, the agents synchronize with most other members within the same natural frequency group.

    Supplementary Fig. 6j shows the amplitude of the distance between the centroids of the two natural frequency groups throughout $K - J$ space. The amplitude is highest when two clusters form and "bounce" off each other's circumference ($K, J > 0$). At high values of $K$, ($K \approx 2$) the whole collective begins to synchronize which halts the bouncing cluster behavior and forms the sync state that has been observed in earlier works[27]. We have run longer simulations with lower values of $K$ for this natural frequency distribution and have found that the bouncing clusters continue oscillating about their common centroid with the same amplitude. We demonstrate that clustering also occurs when there are more than two natural frequency groups. A collective of 500 agents with random phases are evenly divided among the following five natural frequencies: $-1, -0.5, 0, 0.5,$ and $1$. When $K, J = 1$, five independently synchronous clusters form that oscillate between attraction and repulsion with the surrounding clusters. This behavior is shown in Supplementary Movie 2 and the five separate clusters in Supplementary Fig. 9a. Supplementary Figs. 9b-f show the oscillation between each natural frequency group's centroid as it oscillates between attraction and repulsion with each of the others. Since there are more than two groups, the oscillation is noisy as the frequency groups in the center are squeezed out of the visible range of the axis along which the groups align. We also show images of collectives with three, four, five, and nine natural frequency groups in Supplementary Fig. 9g-j and display each of their bouncing behaviors in Supplementary Movie 2. We believe this emergent behavior could be especially relevant to robot collectives at the macro scale that can separate and cluster into distinct groups simply by changing their natural frequencies.

The expanding and contracting collectives are analyzed in Supplementary Fig. 6k where the collective's radii is recorded for the peak minimum and maximum as it oscillates. When $K < 0, J = -1$, agents cannot synchronize and as a result there is minimal repulsion across the collective since few agents share the same phase. As soon as $K > 0$, the collective increases its degree of synchrony within each natural frequency group, and agents repel each other when they share the same phase. Agents with $\omega_i = -1$ are essentially traveling about the phase unit circle in the clockwise direction while the agents with $\omega_i = 1$ are traveling in the counterclockwise direction. Since both groups are traveling at the same rate about the phase unit circle in opposite directions, they exhibit the same phase two times for each phase oscillation cycle. The repulsion between the two groups is highest when the whole collective shares the same phase, which results in the maximum point for the radial oscillation cycle. The greatest attraction occurs when the phases of the two natural frequency groups are at opposite ends of the phase unit circle; this results in the minimum collective radius for the radial oscillation cycle.

Supplementary Fig. 6e shows the phase wave behavior that results when the natural frequency distribution is $F1$ and $K = 0, J = 1$. This is very similar to what was shown in 2017 by O'Keeffe et al[27]. with the difference that although each agent remains essentially unmoving once the phase wave formation is reached, the whole collective oscillates in place so that a uniform phase wave oscillates circumferentially about the annulus formation. In the original swarmalator study, the collective was shown either in a static phase wave state where the agents' motion and phase remained static, or in an active phase wave state where their phase and motion were dynamic. The phase wave state achieved in our study results from the fact that the natural frequency for all agents is much higher than what is needed for phase-coupling-induced motion given by the spatial phase coupling term in $\dot{x}_i$. A lower natural frequency would similarly produce the active phase wave state. Supplementary Fig. 6l shows the spatial-phase order throughout the collective for different values of $K$ and $J$; these results allow us to get a closer look at the behavior of the order parameter $S$ throughout $K - J$ parameter space than what is shown in the main text.

The collective reaches another state where degree of synchrony increases with distance from the collective's centroid when $F1$ and $K = 0, J = -1$, as shown in Supplementary Figs. 6f and 5m. As opposed to exponential-like and step-like behavior in degree of synchrony as a function of increasing distance from the collective centroid seen for collectives with natural frequency distributions $F3$ and $F4$, the plot shown in Supplementary Fig. 6m shows a very linear behavior. The linear behavior is the same when $K = 0.5, 1, 1.5$, and 2, and reaches a maximum degree of synchrony value of ~0.2. From Supplementary Fig. 6f we can distinctly see a ring of agents with similar phases lining the circular boundary of the collective.

Supplementary Fig. 6g shows an expanding collective when $F1$ and $K = 2, J = -1$; since $J$ is negative, agents with similar phases will repel and expand the radius of the collective. Supplementary Fig. 6n shows the rate of expansion and the final radius (after 500 time steps) for different values of $K$ when $J = -1$. After $K$ reaches a high enough value to synchronize agents' phases, radial repulsion begins, and the collective maintains the same rate of expansion and final radius after 500 time steps. Supplementary Fig. 9 explores the expansion behavior across 1000 time steps for different values of $J$; this plot demonstrates that when $0 > J > -1$, the collective expands and reaches a steady collective radius at which it remains. When $J = -1$ the collective continues expanding and does not seem to stop; this is reasonable since the phase coupling-induced motion in $\dot{x}_i$ is defined by a unit vector and the coefficient for global attraction ($A$) has a value of 1. When $J = -1$ and all agents are synchronized, the unit vector term in $\dot{x}_i$ is cancelled out and only the power law repulsion term remains active; thus, all agents repel each other slightly and the collective expands indefinitely.

When the natural frequency distribution is $F4$, the collective has a uniform distribution of natural frequencies across positive and negative values. This ensures that agents with the same natural frequency sign will not synchronize as easily as when the natural frequency distribution is $F2$. The emergent behaviors for collectives with natural frequency distributions are shown in Supplementary Figs. 7a-d and characterized in Supplementary Figs. 7e-h. The first emergent collective behavior presented for this natural frequency distribution is shown in Supplementary Fig. 7a ($K = 2, J = -1$), where a cluster of phase-disordered agents forms at the center and is surrounded by rings of agents with greater phase synchrony. Supplementary Fig. 7e explores the phase synchrony behavior as a function of distance from the collective's centroid for different positive values of $K$. At $K = 0, 0.5$, and $1$, the degree of synchrony remains below $0.05$ across all distances from the centroid. Once $K \approx 1.5$, there is a jump in the degree of synchrony at about $0.7 R_{max}$ and which increases with distance from the collective centroid ($R_{max}$ is the collective radius). These periodic rings of higher synchrony occurs because of the spread in natural frequency; agents with a higher $|\omega_i|$ will have a harder time synchronizing their phase with their neighbors. Since $J$ is negative in these cases, agents with greater phase differences will tend to attract to each other and cluster closer to the collective centroid. Agents with a lower $|\omega_i|$ will synchronize more easily with other agents that share the same sign for their natural frequency; this means that these agents will repel each other more and move towards the collective's boundary where they will have the greatest distance from similarly phased agents. Agents with positive and negative natural frequencies will have high phase differences which enables greater attraction. As a result, although the degree of synchrony increases the collective boundary, it does not surpass $0.1$ because agents with differing phases attract and sometimes occupy the same annulus regions used for the measurements in this plot.

Supplementary Fig. 7b ($K = 2, J = 1$) shows two clusters of agents form close to the collective centroid while a ring of phase-disordered agents encapsulate the clusters. Fig. 5f shows how the degree of synchrony for the whole collective and within each natural frequency group increases with $K$. This shows that higher values of $K$ are needed to fully synchronize each group and the whole collective. Similar bouncing cluster behavior is also observed for $K, J > 0$; however, the clusters do not repel each other at such great distances as was the case when $F2$ since there is greater phase disorder resulting from the spread of natural frequencies.

When $F3$, the collective has a uniform distribution between $0$ and $1$ for the agents' natural frequencies. Supplementary Fig. 7c shows the collective almost forming a sync state when $K = 2, J = 1$; however, a tight ring of asynchronous agents encapsulate the center cluster. The outer agents have higher natural frequency which makes it more difficult for them to synchronize with the center agents. Since $J$ is positive, agents with similar phases attract to each other; therefore, the outer agents tend to radially oscillate as their phase difference with the center cluster increases or decreases. In Supplementary Fig. 7g, we show that the collective synchrony increases with $K$, but only reaches about $0.6$ because of the spread in natural frequency.

When $K = 2, J = -1$ (Supplementary Fig. 7d), the collective exhibits a more organized version of what emerges at the same values of $K$ and $J$ when the natural frequency is spread across negative and positive values. A tight cluster of phase disordered agents forms at the center while rings of more coherent agents encapsulate it. The agents along the boundary have a lower natural frequency than those in the cluster and thus synchronize more easily. Supplementary Fig. 7h shows how the coherence within concentric rings increases with the distance from the collective's centroid. Starting at about $K = 1.5$ the degree of synchrony begins to rise at a distance of $\sim 0.4$ from the collective centroid. Since all agents share the same sign for their natural frequencies, there are lower phase differences between agents with a lower $|\omega|$ which enables greater spacing between the phase-ordered agents and thus a larger region of phase-ordered agents than when $F2$. Supplementary Fig. 10 and Supplementary Movie 4 shows a dynamic plot of the agents' distance from the collective centroid as a function of phase, and highlights the concentric phase self-organization behavior for different $J$.

**Discussion 5. Chiral swarmalators with discrete sets of natural frequencies**

This discussion reviews the chiral collective's behavior when all agents revolve in the same direction at the same rate and when half revolve clockwise (CW) and the other half counterclockwise (CCW) at the same rate. The emergent formations across $K$ and $J$ in the four natural frequency distributions is shown in Supplementary Fig. 13; however, only the upper left and lower right quadrants of the figure relate to this discussion. A summary of the most interesting behaviors and the $S$ order heat maps across $K-J$ parameter space is shown in Supplementary Fig. 14. Supplementary Figs. 14a-b shows that $S$ remains relatively high even when $K$ and $J$ are negative; a very disordered phase wave forms along this region (Supplementary Fig. 14c) with high spatial-phase order at the outer boundary and low order towards the center. The circular motion and negative $K$ drives agents to follow circular paths that will intersect where the phase difference is greatest. Agents on opposite sides of the collective, revolving in the same direction, minimize the distance between each other when they are offset by $\pi$ along their circular trajectories, and results in the emergent formation in Supplementary Fig. 14c.

Disordered phase waves also exist at low $J$ and high $K$; Supplementary Fig. 14d shows a phase wave along an axis that results from a rotating ellipse formation that steadily increases its major axis over time. In this case, similar phases spatially attract, but temporally repel; instead of moving towards synchrony, agents anti-couple and become asynchronous. This enables the formation to slowly stretch out over time. Similar elliptical formations are observed when the collective has two natural frequency groups; however, the two ellipses that emerge rotate about a common center in opposite directions and maintain relatively constant major axes over time. The equal and opposite natural frequencies inhibit perfect anti-coupling, so the collective maintains a steady rotating formation over time (Supplementary Movie 6).

Like the non-chiral swarmalators, distinct phase waves emerge when $K = 0$ (Supplementary Figs. 14e-g). With two natural frequency groups, $K = 0$, $J = 1$ enables agents to separate into concentric, counter-rotating groups that each oscillate between being on the inside and being on the outside (Supplementary Fig. 14f). This is driven by their spatial repulsion and their attempt to each settle at the same revolution radius. A phase wave also forms when there is no spatial or temporal coupling (Supplementary Fig. 14g), the collective forms an annulus because of the agents' inherent circular motion with equal radii, their spatial attraction through the unit vector model, and repulsion through a power law model. All agents maintain the same radius of revolution, but switch between the inner and outer boundaries of the annulus.

Sync states are also observed at positive $K$ and $J$ when the collective revolves in the same direction (Supplementary Fig. 14h); and a high $S$ is maintained because the collective maintains high circumferential phase order about its time-average centroid. Similar behavior is observed when there are two natural frequency groups; two clusters independently synchronize and revolve along the same circular trajectory in opposite directions (Supplementary Fig. 14i). Contrary to the bouncing clusters observed for the non-chiral swarmalators, these collectives maintain a steady trajectory and pass through each other two times per phase cycle because of their inherent circular motion. Their inherent motion practically inhibits synchronization since agents from opposite natural frequency groups have a harder time clustering. Finally, expanding synchronized collectives are also found for chiral swarmalators when the collective has uniform revolution behavior (Supplementary Fig. 14j) and sparse revolving clusters when there are two natural frequency groups (Supplementary Fig. 14k). The sparse revolving clusters are very similar to the dense revolving clusters, except that there is much higher neighbor spacing. Each cluster size also remains relatively constant, as opposed to constant expansion (Supplementary Fig. 14j) or radial oscillation (main text Fig. 2k), because of the inherent circular motion that inhibits much of the phase interaction effects between the two groups.

## Discussion 6. Overview and characterization of chiral swarmalators

This discussion reviews the collective behaviors of regular and frequency-coupled chiral swarmalators. The behaviors of collectives composed of chiral swarmalators with no natural frequency coupling ($c = 1, Q_{\dot{x}}, Q_{\dot{\theta}} = 0$) when $K$ and $J$ hold integer values and different natural frequency distributions are shown in Supplementary Fig. 14. Supplementary Figs. 15a-e show many of the emergent behaviors when the collective has discrete natural frequencies and each behavior is characterized in Supplementary Figs. 15f-j.

Supplementary Fig. 15a highlights chiral swarmalators' emergent behaviors when there is no phase coupling and only global attraction enabled by a unit vector model and repulsion through a power law. The phase wave formation results from the fact that all agents have the same angular velocity and thus the same radius of revolution; half have an inherent clockwise motion and the other half a counterclockwise motion. Since $K, J = 0$, there is no phase coupling behavior and agents simply revolve about the same centroid. The plot of $S$ in Supplementary Fig. 15f shows a clearer view of the spatial-phase order when $J = 0$ than the heat map shown in the main text. As opposed to non-chiral swarmalators, which maintain a low spatial-phase order when the spatial phase coupling term is zero in $\dot{x}_i$, the chiral swarmalators naturally self-organize into a phase wave formation with an even distribution of phases across the collective. The high value for $S$ when $K > -0.9$ is a direct result of each agent's inherent circular motion and each collective starting with a uniform distribution of phases, meaning the agents' orientations are spread across $0$ and $2\pi$. Even when agents anti-couple ($K < 0$), the inherent circular motion ensures that a single spinning vortex will form. We also note that the trend in inner radius of the collective decreases with $K$, this results from the fact that as $K$ decreases, agents will tend to phase couple more easily with other agents with whom they have a phase difference closer to $\pi$; this will drive the collective to lower the inner radius since this is the point at which agents with a phase difference of $\pi$ will be closest to each other.

The two concentric ring formations shown in Supplementary Fig. 15b happen when $F2$ and $K = 0, J = 1$. Each ring is formed by a natural frequency group and the two are composed of agents revolving in opposite directions. As shown in Supplementary Fig. 15g, the average radius of each frequency group oscillates over time, and this behavior persists even when the collectives run for longer simulation times. The radial oscillation occurs because each natural frequency group has equal and opposite angular velocities that are paired with the phase behavior. As agents minimize their phase difference with agents from the other natural frequency group, which happens twice for each phase cycle, the attraction between these agents maximizes so that a single ring appears. Since the phase difference between agents is continuously increasing or decreasing, the two ring formations quickly separate back into concentric formations. Supplementary Fig. 15g shows that the two groups oscillate between being outside and inside. This most likely occurs because each group has a natural revolution radius of 1 and it is easier for each agent to continue traveling along the direction closest to their instantaneous direction of travel when the attraction between the two groups begins to decrease. This means that agents from the outer ring, which were moving towards the center as their attraction with agents from the other natural frequency group grew, will continue traveling towards the collective centroid once the attraction begins to decrease. It is interesting to note that while the frequency of radial oscillation remains almost constant throughout time, the amplitude changes a good amount; this could be due to slightly non-uniform distributions in the agents' $x, y$, and $\theta$ values at the beginning of the simulation. The slightly non-uniform distributions could drive each group to reach variable maximum and minimum radii; however, the average radius remains the same over time and equal to each group's inherent revolution radius.

The dense clusters observed for non-chiral swarmalators are also seen for positive values of $K$ and $J$ with the chiral swaramalators (Supplementary Fig. 15c). The general revolving cluster behavior, however, persists throughout negative $J$ as well for the chiral swarmalators when $F2$ as shown by the heat map of maximum distance between the centroid of the two natural frequency groups in Supplementary Fig. 15h. The densest clusters are formed at high $K$ and $J$ since this enables agents from the same natural frequency group to synchronize and cluster with similarly phased agents. As $J$ decreases, each natural frequency group cluster becomes sparser and forms an elliptical-like formation that oscillates in attraction and repulsion to the other cluster. When $J < 0$, lower $J$ enables agents within the same natural frequency group to increase their neighbor spacing since they synchronize their phases and repel each other more strongly. The inherent circular motion of the agents, however, induces clustering since they share the same phase within each natural frequency group. Agents from opposite natural frequency groups still undergo oscillatory spatial attraction and repulsion two times per phase cycle (revolution cycle). This induces agents from opposite natural frequency groups to have high attraction and repulsion through each revolution; an elliptical-like formation enables agents from each group to balance the time-varying attraction / repulsion forces that drive each agent to (1) maintain its inherent circular motion, (2) increase its spacing from similarly phased agents, and (3) move towards or away from agents in the opposing natural frequency group (Supplementary Fig. 15d). Supplementary Fig. 15i shows that the major axis of each natural frequency group increases as $J$ becomes lower, which agrees with the statements above.

The image shown for Supplementary Fig. 15e shows the collective with $F1, K = -1, J = 1$ at the end of a simulation; during the simulation the collective starts out in a phase wave formation and over time begins to elongate into a rotating ellipse until the collective forms an ellipse with a very small minor axis (essentially a line) as shown in the image. The behavior repeats itself across different values of $K$ and the ellipse's major axis increases dramatically as $J$ approaches $-1$, as shown in Supplementary Fig. 15j.

When $F4$, half of the collective has an inherent circular motion in the clockwise direction, and the other half the counterclockwise direction; agents have an uniform distribution of revolution radii between 0 and 1. Supplementary Figs. 16a-e show some of the most interesting behaviors when there is a natural frequency spread and their corresponding characterization in Supplementary Figs. 16f-j. The global attraction between agents and their inherent circular motion enables the collective to form a single vortex when $K = 0, J = 1$ (Supplementary Fig. 16a); a small ring of phase-disordered agents on the boundary of the collective keeps the collective from having as high of a value for $S$ as when $F2$. The agents on the boundary each have a very small revolution radius and high natural frequency; this prohibits them from matching the phase of nearby agents and causes them to be repelled to the outer region of the collective. Supplementary Fig. 16f also shows that the degree of synchrony remains low even with positive $K$ because of the spread in natural frequencies across negative and positive values. Supplementary Fig. 16b shows two revolving clusters with a group of phase-disordered agents that remains around the centroid of the collective while the two clusters revolve about it at an average radius of 1. The plot in Supplementary Fig. 16g shows that the collective's and each natural frequency group's degree of synchrony significantly increases at $K \approx 1.5, J = 1$, but remains relatively low because of the natural frequency spread.

Supplementary Fig. 16c results from $F3, K = 0, J = 1$ and shows a very similar vortex to the one shown in Supplementary Fig. 16a. The trends for spatial-phase order and degree of synchrony shown in Supplementary Fig. 16h also closely resemble those in Supplementary Fig. 16f; this demonstrates that even though the whole collective has an inherent circular motion where agents travel in the same direction, the uniform distribution for the natural frequencies keeps the agents from reaching a very high degree of synchrony when $K < 1, J = 1$.

A revolving synchronized cluster with a ring of phase-disordered agents is shown in Supplementary Fig. 16d; the agents forming the outer ring of the collective each have a small revolution radius and a high natural frequency. As shown in Supplementary Fig. 16i, the spatial-phase order remains above ~0.5 when $-1 < K < 1$ because the synchronized cluster is traveling in a circular trajectory around the collective's centroid. Even though there is no instantaneous distribution of phases about the centroid, as was the case when the collective formed a vortex, most agents are synchronized and have a position about the centroid that is phase-correlated. The spatial-phase order cannot be higher because of the phase-disordered agents that remain even when $K$ is positive.

When $J < 0$, agents with similar phases repel each other; as shown in Supplementary Fig. 16e, agents with a larger revolution radius and thus a lower natural frequency tend to synchronize more easily and form a large revolving orb. Agents with a low revolution radius and a high natural frequency are unable to synchronize and thus attract to each other to form a cluster of phase-disordered agents closer to the collective's centroid. Supplementary Fig. 16j shows that order remains relatively constant at a low value across negative and positive values of $K$ since the collective either has high phase disorder ($K < 0$) or has partial synchronization ($K > 0$) at which point agents with the same phase are evenly distributed circumferentially about the collective's centroid. In both cases there is low opportunity for the collective to achieve circumferential phase organization. The degree of synchrony, however, begins to increase to a higher value when $K \approx 0.7$; this enables the collective to partially synchronize and encapsulate a tight cluster of phase-disordered agents.

An overview of the chiral swarmalators' emergent behaviors with natural frequency coupling is shown in Supplementary Fig. 17 and the characterization of some of the emergent behaviors is shown in Supplementary Fig. 18. Supplementary Fig. 18a shows the emergent formation when $F2, K = 0, J = 1$; two non-concentric phase waves with counterrotating agents form side by side. The non-concentric formation occurs because of the phase shift in $\dot{x}_i$ and $\dot{\theta}_i$ determined by the agents' natural frequency. The distance between the centroids of the two natural frequency groups is shown by the heat maps in Supplementary Fig. 18b, with a maximum occurring around $0.2 < K < 1$ and $J \approx 1$. The distance between the group centroids is non-zero for almost all positive values of $K$. The frequency coupling enables the current model to produce behaviors very similar to spermatozoa vortex arrays when there is local coupling.

When $F4$, the collective begins to mix more easily because the natural frequency distribution prevents each group from synchronizing with itself and producing the strong oscillatory attraction and repulsion with the opposing group, as was the case when $F2$. The double vortex formation is shown in Supplementary Fig. 18c, where two vortices form next to each other because of the phase shift in $\dot{x}_i$ and $\dot{\theta}_i$. The heat map in Supplementary Fig. 18d shows that the greater distance between the centroids of the two natural frequency groups occurs when $J > 0$ and $K > 0$. This is reasonable since these are the regions of the $K - J$ parameter space in which the clusters are most likely to enable coupling between agents with a similar sign for their natural frequency and thus enable some oscillatory attraction and repulsion with agents from the other natural frequency group.

Discussion 7. Order parameter.

This discussion defines the frequency group phase coherence order parameter which is used throughout some supplementary plots.

$$Z_{FG} = \frac{1}{2}\left(\frac{1}{n}\sum_j^n e^{i\theta_j} + \frac{1}{m}\sum_k^m e^{i\theta_k}\right) \quad (1)$$

Eq. 1 is used to calculate the average degree of synchrony within the collective when there are multiple natural frequency groups. In Equation (1) there are $n$ agents within a natural frequency group and $m$ agents in another group. The degree of synchrony for the collective is the average of the degree of synchrony within each group. In this study we look at the average degree of synchrony within groups when there are two natural frequency groups.

## Discussion 8. Analysis of Sync State

### Global Coupling Model with Noise
The global coupling model can be rewritten with white noise terms $\xi_i$ and $\eta_i$ to study its effect on the emergent collective behaviors.

$$\dot{x}_i = v_i + \frac{1}{N}\sum_{j \neq i}^{N}\left[\frac{x_j - x_i}{|x_j - x_i|}(A + J\cos(\theta_j - \theta_i - Q_{\dot{x}})) - B\frac{x_j - x_i}{|x_j - x_i|^2}\right] + \xi_i(t) \quad (2)$$

$$\dot{\theta}_i = \omega_i + \frac{K}{N}\sum_{j \neq i}^{N}\frac{\sin(\theta_j - \theta_i - Q_{\dot{\theta}})}{|x_j - x_i|} + \eta_i(t) \quad (3)$$

**Global coupling, identical, non-chiral:** We freeze $(J, K, Q_{\dot{x}}, Q_{\dot{\theta}}, \sigma) = (1, 1, 0, 0, 10)$ and vary $d$ (noise). Supplementary Fig. 40 and Supplementary Movie 8 show how the sync state changes with increasing $d$. At $d = 0$, the swarmalators are frozen in a disk. For $d > 0$, they melt and begin to move around, blurring the disk edge (the radius increases) and degrading the overall phase coherence. We quantify this loss of synchrony by plotting the Kuramoto phase coherence order parameter (Z) versus $d$ (Supplementary Fig. 41, a global measure) and by plotting the two point orientation correlation function (Supplementary Fig. 42, a local measure) defined as

$$C_{n,n}(r) := \langle\frac{1}{N(N-1)}\sum_{i,j}\hat{n}_i \cdot \hat{n}_j \delta(r_{ij} - r)\rangle \quad (4)$$

Where $\hat{n}_i = (\cos(\theta_i + \frac{\pi}{2}), \sin(\theta_i + \frac{\pi}{2}))$ is the 'orientation' of the i-th paticle, $r_{ij}$ is the distance between i, j, and $\langle \cdot \rangle$ denotes the ensemble average. $R$ bifurcates from zero at $d \approx 1$. $C_{n,n}(r) > \approx 0.6$ for $d > 0.6$. For larger d, it begins to decay, before finally becoming zero as sync is lost. We also computed the (spatial) velocity auto-correlation function,

$$C_{v,v}(t; \tau) := \langle\frac{1}{N}\sum_i \hat{v}_i(t)\hat{v}_i(t + \tau)\rangle \quad (5)$$

Where $\hat{v}_i$ is the unit vector of the spatial velocity. At $d = 0$, the swarmalators relax slowly into a fixed point and so $C_{v,v}(t)$ declines slowly from 1 (Supplementary Fig. 43). For any finite $d$, however, the brownian motion dominates and $C_{v,v}$ has small fluctuations about 0.

**Global coupling, non-identical, non-chiral:** Next we explore how distributed $\omega_i$ affect the above findings. We focus on the case where the natural frequencies are drawn uniformly at random:

$$U(-\Omega, \Omega) \quad (6)$$

$$\omega_i = (-\Omega, -1] \cup (1, \Omega) \text{ for } \Omega = 3 \quad (7)$$

Supplementary Fig. 44 and Supplementary Movie 9 shows how the sync state deform for increasing $\Omega$ and zero noise $d = 0$ ( we freeze $(J, K, \sigma) = (1,1,10)$ as before). For small $\Omega < \Omega_c$, the static sync state survives: swarmalators stay motionless with their phases locked. Note, however, that because they are non-identical, in the locked state the agents do not have identical phases, but rather exhibit a phase gradient. This phase gradient deforms the disk-like spatial density (Supplementary Fig. 44a and b). Beyond a critical $\Omega_c = 1$, however, the static sync bifurcates into an unsteady partial sync state. Here, the disk appears to 'boil': the fast swarmalators near the disk edge break free and vacillate noiselessly around the boundary. The slow swarmalators stay locked in the core, which executes small deformations itself (Supplementary Fig. 43c-f). For $\Omega \gg \Omega_c$, all phase coherence dies out and an incoherent gas like state appears. Supplementary Fig. 45 charts this trend: $Z(\Omega)$ decreases for increasing $\Omega$. Supplementary Fig. 46 shows a log-log plot the auto-correlation of the velocity $C_{v,v}(t)$. For $\omega < 1$, each $\Omega$ decreases to zero at a rate increasing with $\Omega$ (left panel). For $\omega > 1$, however, $C_{v,v}$ is wild and oscillatory, indicating the erratic motion of the 'boiling' in the partial sync state. Finally, Supplementary Fig. 47(a) shows the orientation correlation $C_{n,n}(r)$ (for $d = 0$; the right panel we discuss later). We see $C(r) \sim 1$ for small $\Omega$ consistent with Supplementary Fig. 44a. For larger $\Omega$, we see $C_{n,n}$ varies non-monotonically, picking up the full phase wave in Supplementary Fig. 44c-f. Turning on noise creates an interesting effect: non-monotonicity in $Z$ for $\Omega \sim 1$ (Supplementary Fig. 48a). The 'boiling' of the fast swarmalators in the partial sync state gets drowned out by the noise and the blurred sync state is born which has higher phase coherence. As $d$ is increased, this state blurs until becoming fully incoherent. Supplementary Fig. 48 shows this behavior by plotting $Z(d)$ for different $\Omega$ (the effect is clearer when plotted as $Z(d)$ for different $\Omega$; as opposed to $Z(\Omega)$ for different $d$). The $\Omega = 1$ curve, the regime in which the partial sync state exists, exhibits the non-monontonicty. The other $\Omega$ curves are shown for the sake of completion. The $C_{v,v}(t)$ for $d > 0$ are uninteresting; a rapid decline to brownian motion is observed for $d > 0.1$ (not plotted). Similarly the $C_{n,n}(r)$ becomes more blurry until the non-monotonicty disappears and the async state is born (Supplementary Fig. 47b).

**Local coupling, non-chiral:** Two things happen when coupling changes from global to local. For $\sigma$ the single cluster (Supplementary Fig. 49a) splits into multiple clusters surrounded by a gas of spatially free, but synchronized swarmalators (Supplementary Fig. 49b-e). Then for $\sigma \lessapprox 1$, all clusters disappear leaving a mass of sync'd oscillators (Supplementary Fig. 49f). To explore the length scales of the clusters, as well as the inter-cluster separation, we calculate

$$g(r) := \frac{1}{N(N-1)} \sum_{i,j} \delta(r - r_{ij}) \qquad (8)$$

for different $\sigma$. Supplementary Fig. 52 shows the results for $N = 500$ swarmalators placed uniformly at random in a box of length $L = 4$. For large $\sigma$, there is one cluster and $g$ has only one peak. For smaller $\sigma$ multiple clusters form and $g$ exhibits multiple peaks. Notice, however, that the location of the first peak $r^*$ is the same for each $\sigma$; the radius of each cluster does not depend on $\sigma$. For small amounts of disorder, both with quenched $\Omega > 0$ and active $d > 0$, multiple clusters persist, but eventually disappear for large $d, \Omega$. This is evident by the disappearance of the multiple peaks in Supplementary Fig. 51.

Supplementary Fig. 52 demonstrates the inter-cluster phase correlation by plotting $C_{n,n}(r)$ (bottom row) for $(d, \Omega) = (0.1, 0.2)$. Scatter plots of the corresponding states are shown in the top row for convenience. The first column shows the simple single cluster behavior: $C_{n,n}$ decreases slightly over the cluster and then drops to zero. There is a small fluctuation for large $r$, corresponding to stray swarmalators which have not been absorbed. The second column shows the multiple-cluster regime where $C_{n,n}$ varies non-monotonically. The first interval of decrease captures the intra-cluster correlation, the second the inter-cluster correlation, and the third the cluster-gas correlation. Finally the third column shows the gas regime where $C_{n,n}$ has small local correlations which disappear quickly as $r$ increases. Supplementary Fig. 53 shows the three $C$ for larger $\Omega$ in which the phase gradient within clusters is wider, which manifests as an oscillation in $C$ from positive to negative values.

# Discussion 9. Analysis of phase waves / vortices

**Global coupling:** Next we study the vortex like active phase wave. We set $(J, K) = (1, 0, 5)$ throughout this section. Supplementary Fig. 55 shows a scatter plot of the state for increasing amounts of disorder $d$, $\Omega > 0$ (see Supplementary Movie 12 for evolution of states). The state gradually blurs until becoming incoherent. Supplementary Fig. 55 quantifies this loss of coherence by plotting $S = \max(S_+, S_-)$, which is more than 0 in this state, against $d$ (left panel) and $\Omega$ (right panel).

**Local coupling:** When the coupling becomes local, we see a transition from a single vortex to multiple vortices to a incoherent gas (Supplementary Movie 12). Supplementary Fig. 56 plots the collective states (top row), the correlation functions (middle row), and the radial density (bottom row) for $\sigma = 10, 3, 1$ for small amounts of disorder $(d, \Omega) = (0.1, 0.2)$. For large coupling (left column): notice $C_{\theta,\theta}$ decreases from 1 to $\approx -0.6$ over the length scale of the single vortex, consistent with the phase gradient. The radial density g has a single peak, since there is only one cluster. For intermediary length scales $\sigma = 3$ (middle column), $C_{\theta,\theta}$ is non-zero over the length of a single vortex; there is no inter-vortex correlation. $g$ on the other hand has multiple peaks. Finally, for short range coupling $\sigma = 1$ phase-phase correlation is sparse and a single broad mode in $g$, consistent with the picture of an incoherent gas.

## Discussion 10: Analysis of chirality

We analyze chirality and draw the swarmalators' natural frequencies from the $F2$ distribution. This splits the population into two (half rotating clockwise ($\omega_i < 0$), the other half revolving counterclockwise ($\omega_i > 0$). As defined in the main text, this distribution is to ensure that the radii of revolution are bounded; recall we set the speed to be magnitude unity $v_i = c_i \boldsymbol{n}_i = 1\boldsymbol{n}_i = \boldsymbol{n}_i$. This means the radius of revolution (when uncoupled) of each swarmalators is $R_i = \frac{|v_i|}{\omega_i} = \frac{1}{\omega_i}$, so if we allow small $\omega_i$ then we get large $R_i$. The main effect of this distribution is that the phase wave / vortex state splits into two, so that the winding number of the overall state is $k = 2$ (Supplementary Fig. 57). The sync state bifurcates into the bouncing cluster state (see main text; not plotted here). To capture the phase order, we use the order parameter $S_{+,2} = |\frac{1}{N}\sum_j e^{2i(x \pm \theta)}|$. Supplementary Fig. 58 plots $S_2 := \max(S_{+,2}, S_{-,2})$ along with $S$ for increasing $d$ and $\Omega$ (left panel) and for increasing $\Omega$ and fixed $d$ (right panel) which shows a steady decline to incoherence. The velocity-velocity correlation functions were uninformative (similar to Supplementary Fig. 43) so we do not display them here. Supplementary Fig. 59 shows local coupling produces no new new effects. The same the single vortex to multi-vortex to gas-like bifurcations are observed for increasing local coupling; i.e. Supplementary Fig. 59 for chiral swarmalators is qualitatively the same as Supplementary Fig. 56 for non-chiral swarmalators. The only difference is the phase-phase correlation function $C_{\theta,\theta}$ has two peaks, oscillating from positive to negative back to positive, reflecting the phase wave with winding number $k = 2$.

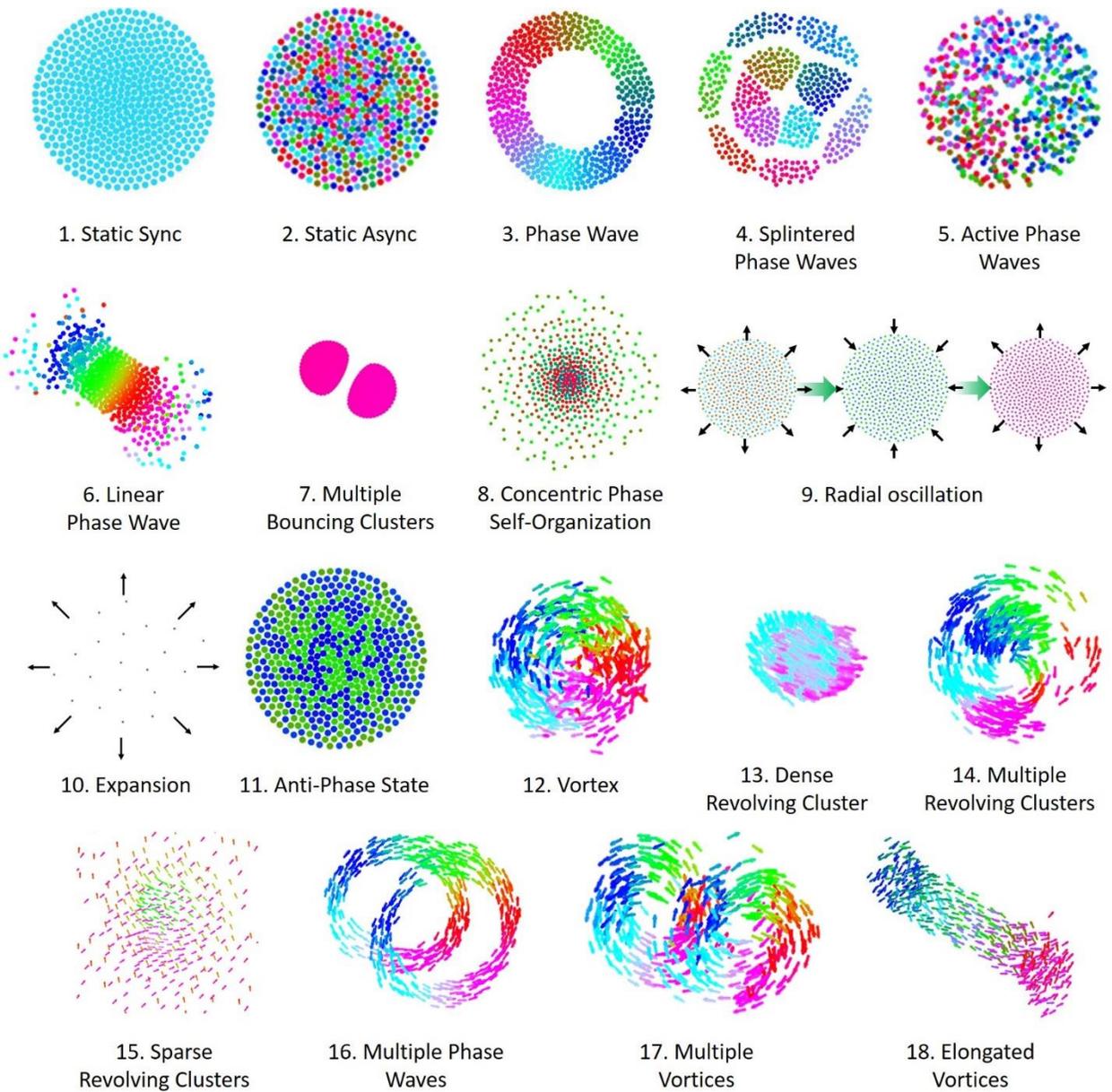

**Supplementary Fig. 1. Summary of emergent states.** The behaviors shown above summarize the major behaviors found through our model and their numbers correspond to their number in Table 1.

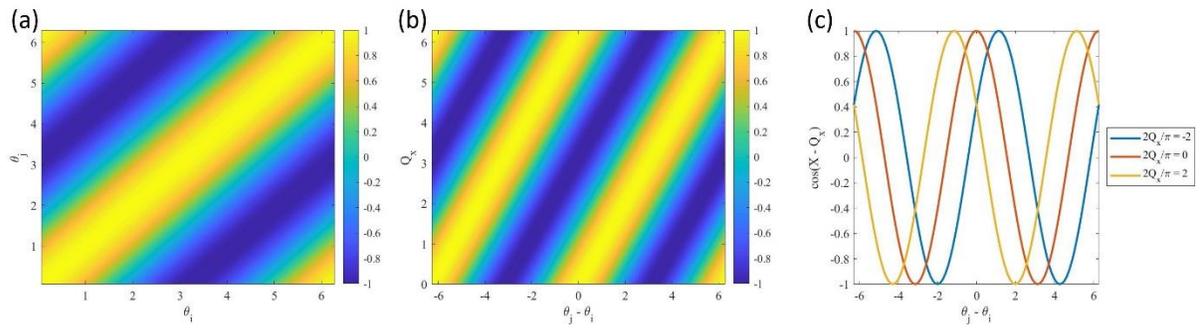

**Supplementary Fig. 2. Phase-interaction parameter space.** Value given by the spatial phase interaction term in Eq. 1: **(a)** Cases explored for $Q_{\dot{x}} = 0$ and different phase values. The color bar is defined by $\cos(\theta_j - \theta_i)$. **(b)** Cases explored for a constant phase shift value $Q_{\dot{x}}$ and different phase differences $(\theta_j - \theta_i)$. The color bar is defined by $\cos(\theta_j - \theta_i - Q_{\dot{x}})$. **(c)** Oscillation behavior of the phase coupling-induced motion term in $\dot{x}_i$.

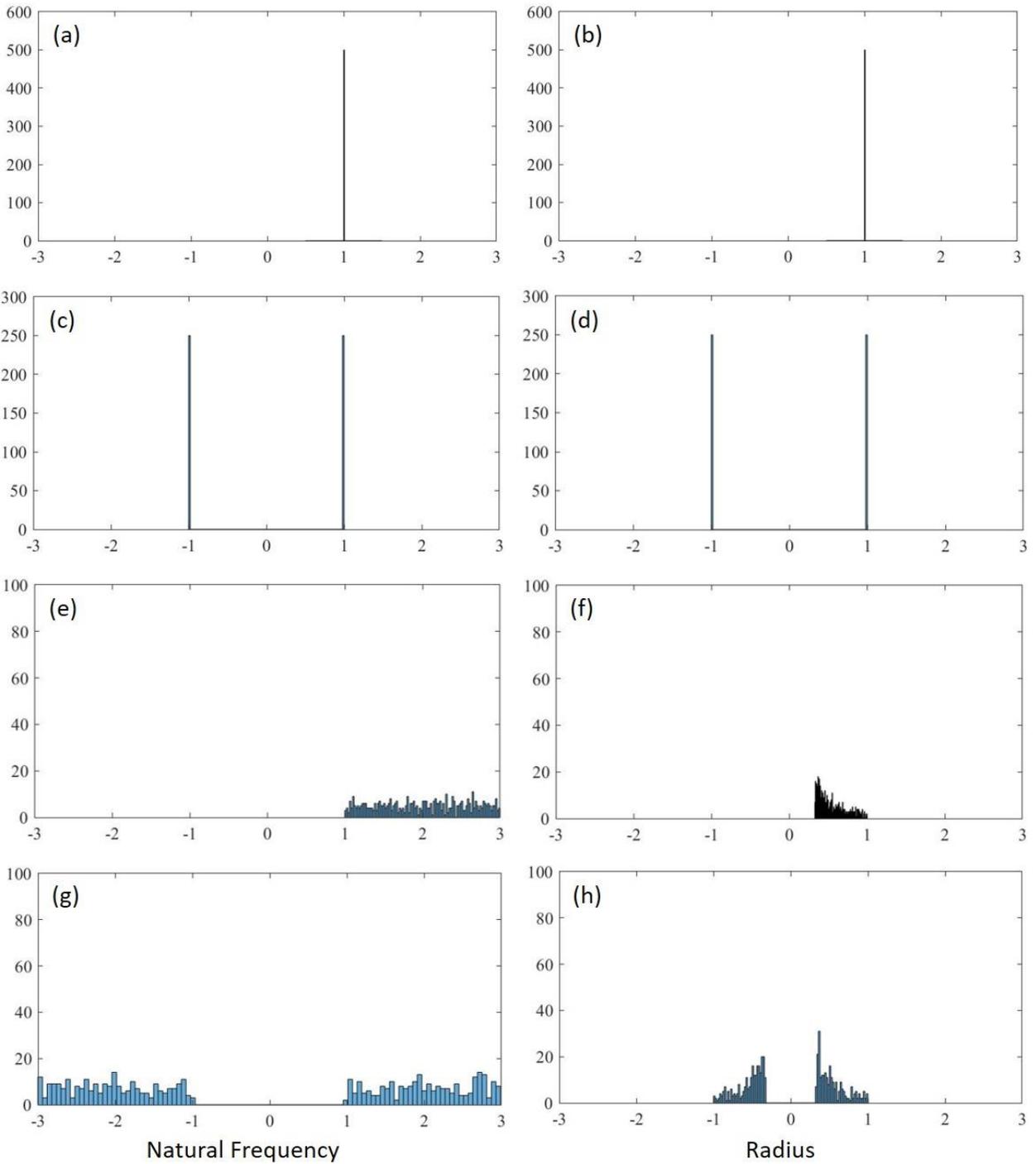

**Supplementary Fig. 3. Natural frequency and revolution radius distributions.** Histogram for each of the natural frequency distributions used throughout the numerical studies. **(a-b)** $F1$; **(c-d)** $F2$; **(e-f)** $F3$; **(g-h)** $F4$. The distributions shown in (a,c,e,g) correspond to the natural frequencies and the distributions shown in (b,d,f,h) correspond to revolution radius.

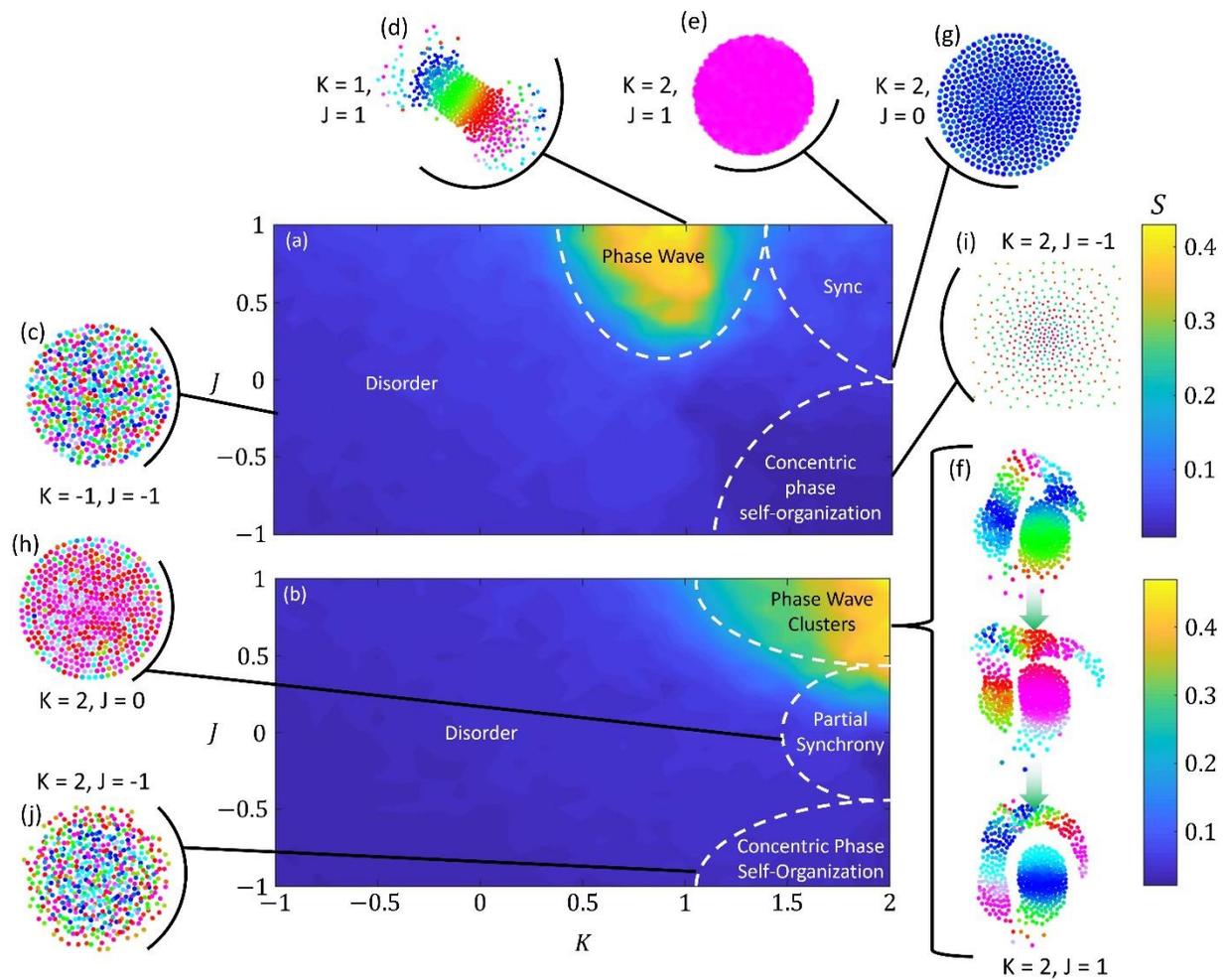

**Supplementary Fig. 4. Non-chiral swarmalators with a natural frequency spread.** Heat maps of $S$ across $K - J$ parameter space is shown for test cases with **(a)** $F3$ and **(b)** $F4$. **(c)** Static async. **(d)** Partial sync with phase wave along an axis. **(e)** Partial sync with synchronized cluster forms at the center. **(f)** Periodic bouncing of partial sync clusters. **(g-h)** Partial sync within a circular formation. **(i-j)** Concentric phase self-organization.

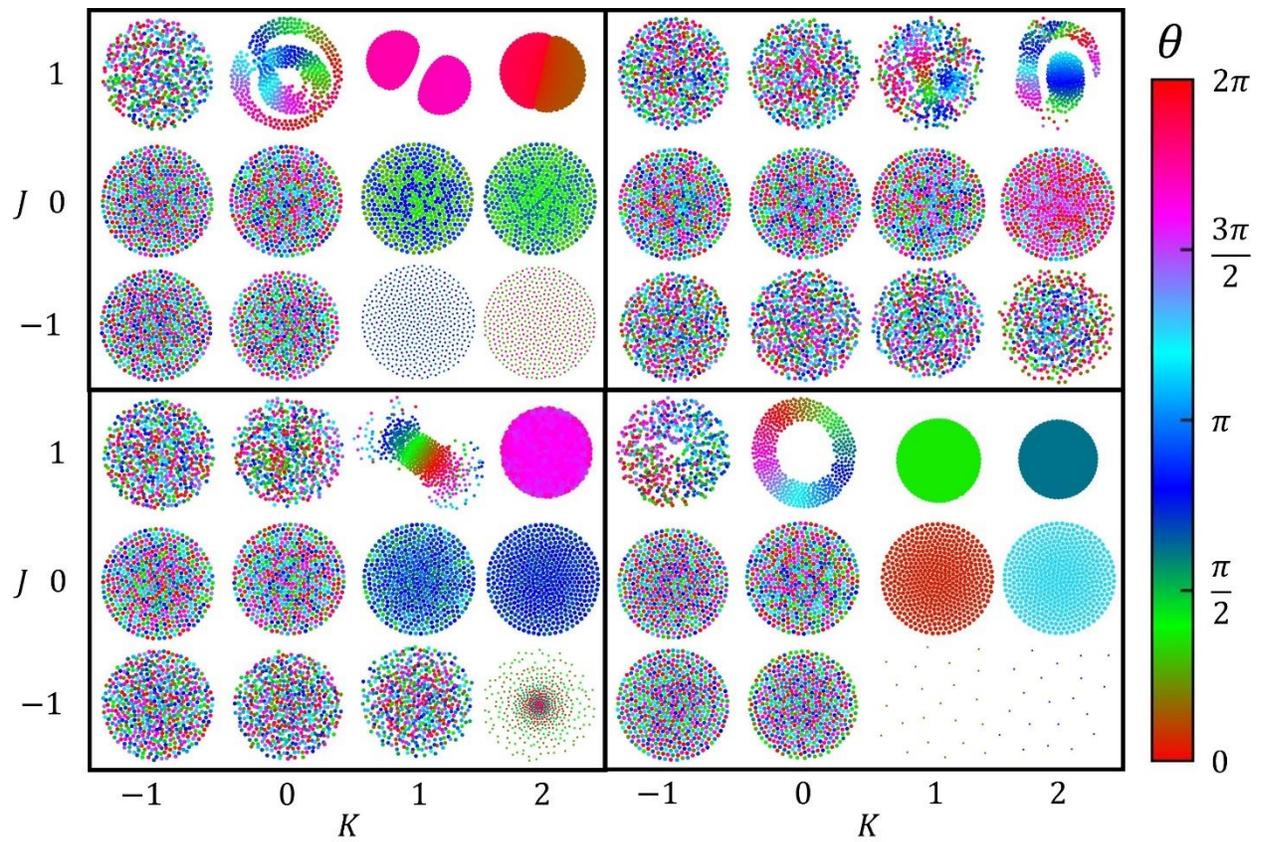

**Supplementary Fig. 5. Collective behaviors of non-chiral swarmalators with no frequency coupling.** $F2$ (Upper left). $F4$ (Upper right). $F3$ (Lower left). $F1$ (Lower right).

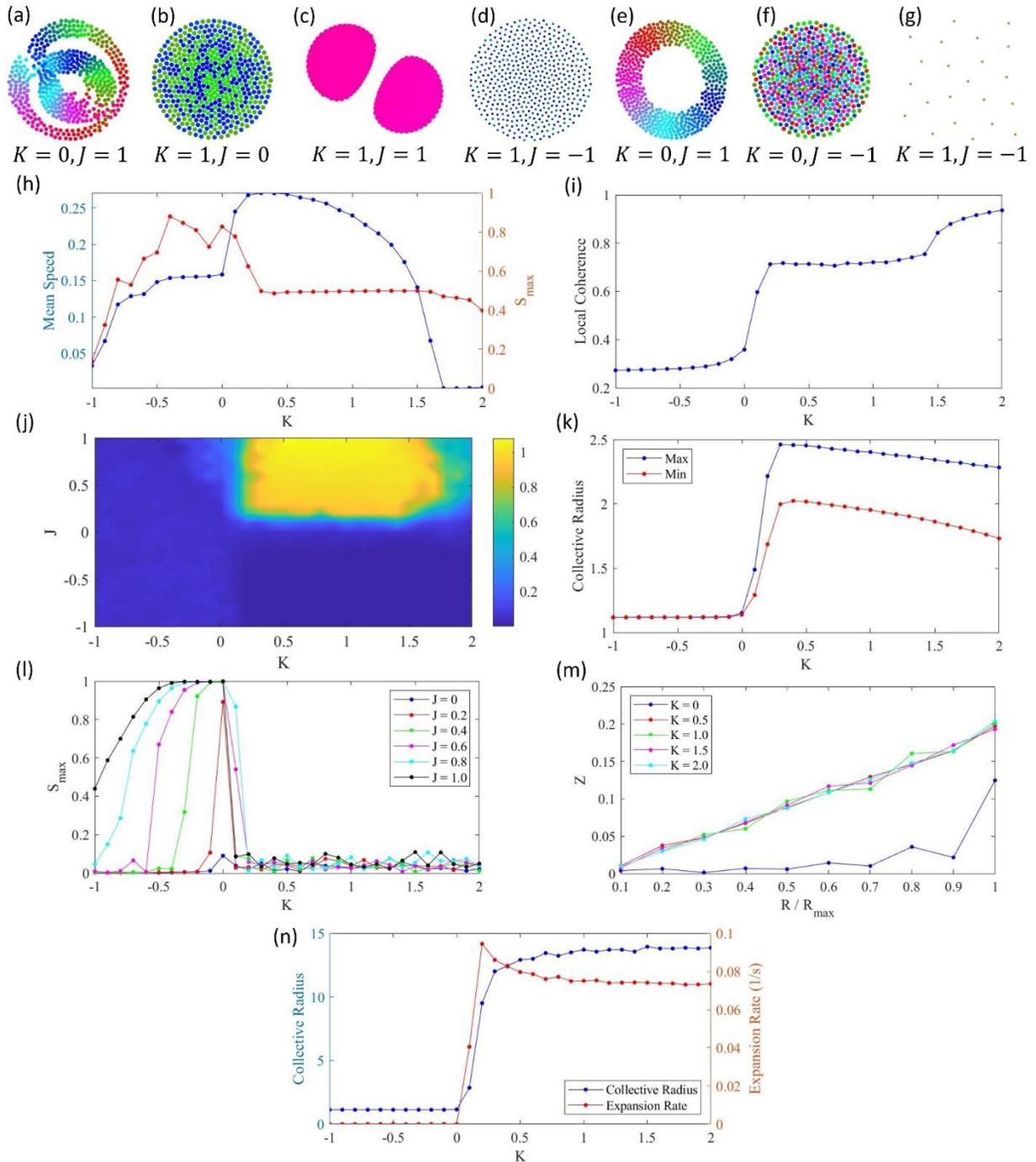

**Supplementary Fig. 6. Characterization of collective behaviors in non-chiral swarmalators with discrete natural frequencies and no frequency coupling. (a-g)** Various emergent configurations of non-chiral swarmalators. **(h-k)** Collective behavior characterizations for the same $\omega$ distributions listed in **(a-k)**, respectively. **(a)** $F2$; **(b)** $F2$; **(c)** $F2$; **(d)** $F2$; **(e)** $F1$; **(f)** $F1$; **(g)** $F1$. **(h)** Mean speed and $S$ order when $J = 1$; **(i)** Max distance between the two natural frequency group centroids; **(j)** Local coherence with neighboring agents when $J = 0$; **(k)** collective radius at peak expansion and contraction when $J = -1$; **(l)** $S$ order for various values of $J$ and $K$; **(m)** phase coherence as a function of distance from the collective's centroid when $J = -1$; **(n)** collective radius and expansion rate when $J = -1$.

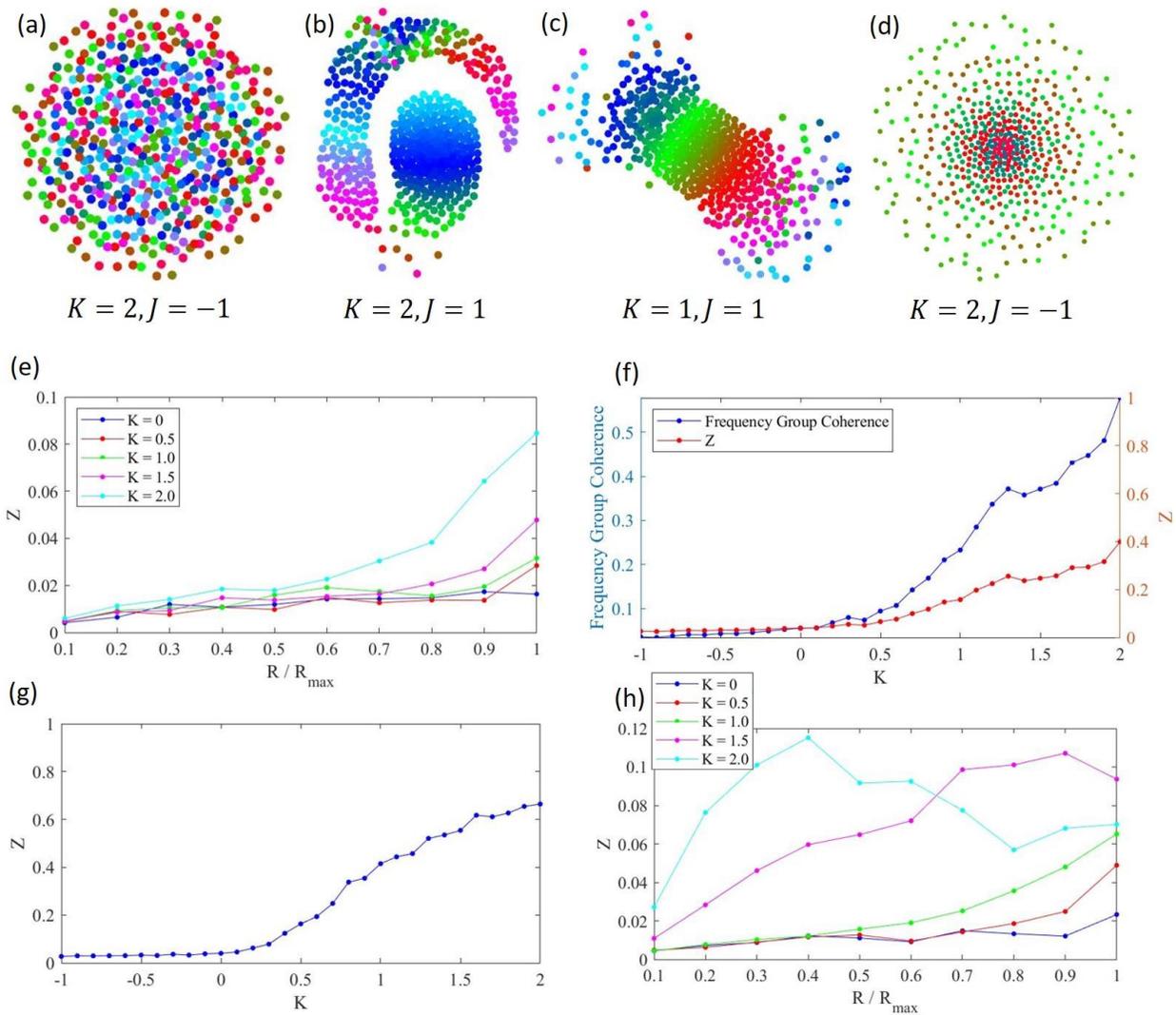

**Supplementary Fig. 7. Characterization of collective behaviors in non-chiral swarmalators with a natural frequency spread and no frequency coupling.** (**a-d**) Various emergent configurations of non-chiral swarmalators. (**e-h**) Collective behavior characterizations for the same natural frequency distributions listed in (**a-d**), respectively. (**a**) $F4$; (**b**) $F4$; (**c**) $F3$; (**d**) $F3$. (**e**) phase coherence as function of distance from the collective centroid when $J = -1$; (**f**) Natural frequency group and global phase coherence when $J = 1$; (**g**) phase coherence when $J = 1$; (**h**) phase coherence when $J = -1$.

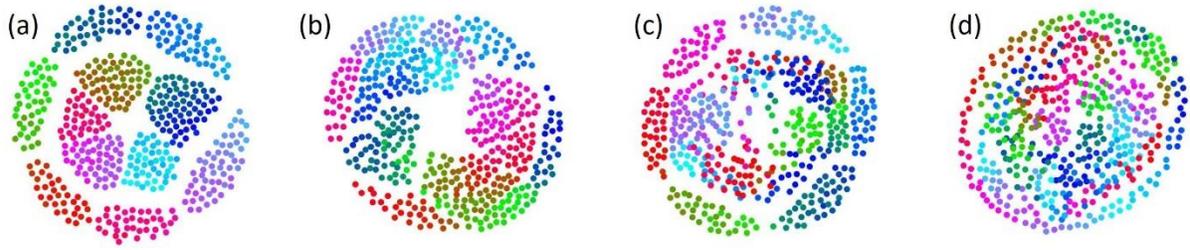

**Supplementary Fig. 8. Splintered phase waves.** Images of collectives with different numbers natural frequency groups when $c, Q_{\dot{x}}, Q_{\dot{\theta}} = 0, K = -0.1, J = 1$. **(a)** $\omega_i \in [-1], [1]$. **(b)** $\omega_i \in [-1], [0], [1]$. **(c)** $\omega_i \in [-1], [-0.5], [0.5], [1]$. **(d)** $\omega_i \in [-1], [-0.5], [0], [0.5], [1]$.

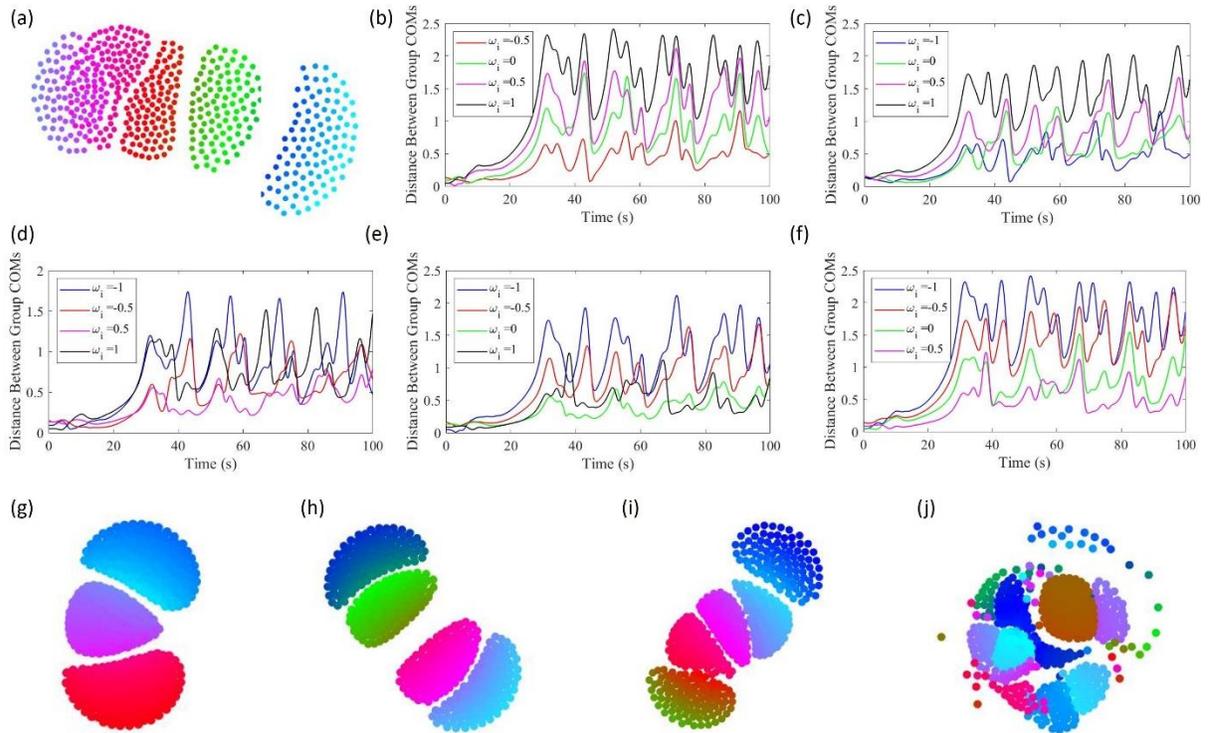

**Supplementary Fig. 9. Frequency group separation. (a)** A collective with five natural frequency groups ($\omega_i \in [-1], [-0.5], [0], [0.5], [1]$) when $c = 0$, and $K, J = 1$. **(b-f)** The distance between each natural frequency group's center of mass and all other frequency groups' center of masses over time. Distance from the center of mass of the group of agents with a natural frequency of **(b)** $\omega_i = -1$, **(c)** $\omega_i = -0.5$, **(d)** $\omega_i = 0$, **(e)** $\omega_i = 0.5$, **(f)** $\omega_i = 1$. **(g-j)** Collectives with differing numbers of natural frequency groups shown after 775 time steps. **(g)** $\omega_i \in [-1], [0], [1]$. **(h)** $\omega_i \in [-1], [-0.5], [0.5], [1]$. **(i)** $\omega_i \in [-1], [-0.5], [0], [0.5], [1]$. **(j)** $\omega_i \in [-4], [-3], [-2], [-1], [0], [1], [2], [3], [4]$.

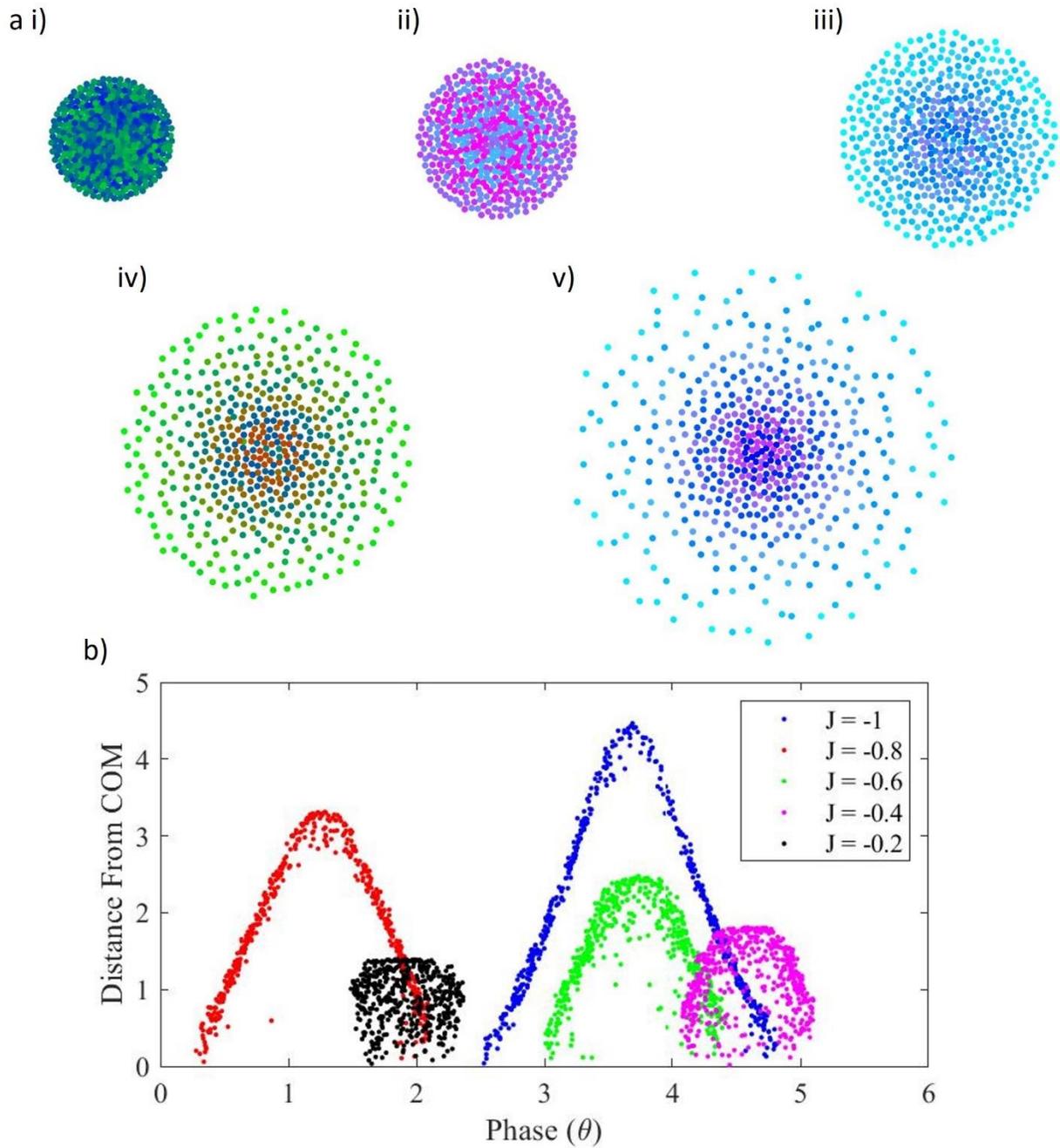

**Supplementary Fig. 10. Concentric phase self-organization. (a-e)** Collectives with $F4$ and different negative values of $J$ that self-organize concentric rings of agents with similar phases. **(a)** $J = -0.2$, **(b)** $J = -0.4$, **(c)** $J = -0.6$, **(d)** $J = -0.8$, **(e)** $J = -1.0$. **(f)** Plot of agents' distance from collective centroid as a function of phase demonstrates that after a certain distance there is distance-based phase self-organization.

(a)

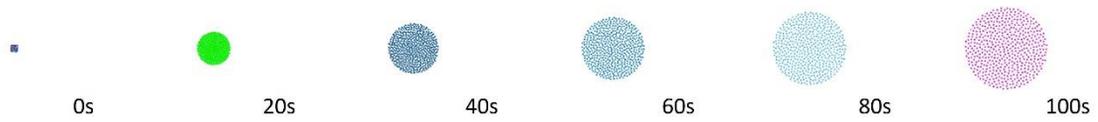

(b)

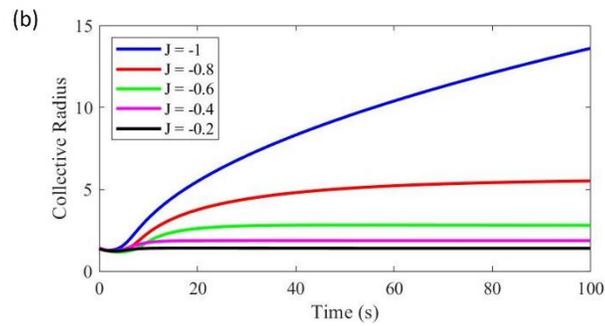

**Supplementary Fig. 11. Collective expansion.** (**a**) Collective expanding indefinitely when $c = 0, F1, K = 1,$ and $J = -1$. (**b**) Collective radius plotted as a function of time for various values of $J$.

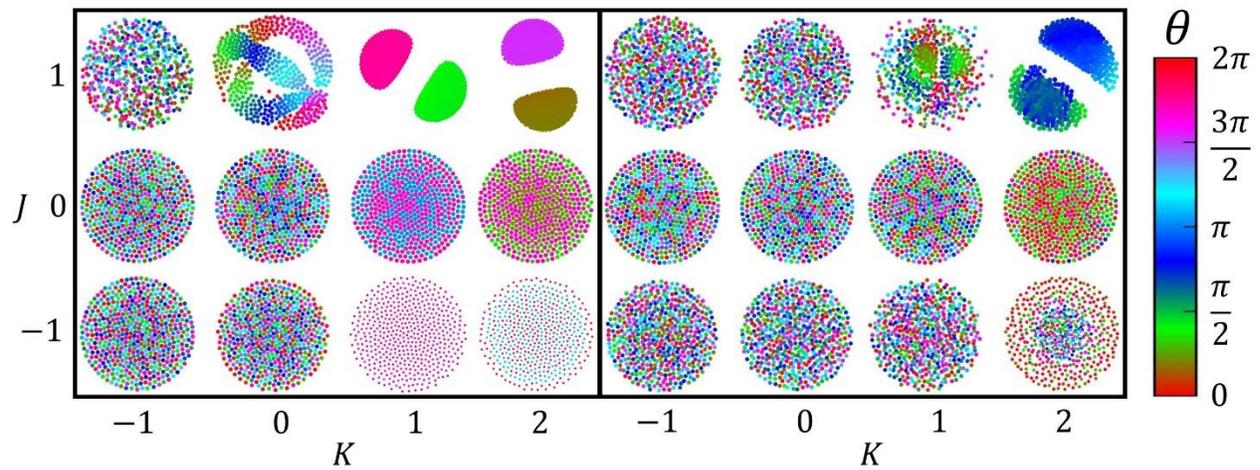

**Supplementary Fig. 12. Collective behaviors of non-chiral swarmalators with frequency coupling.** $Q_{\dot{x}} = \frac{\pi}{2}\left|\frac{\omega_j}{|\omega_j|} - \frac{\omega_i}{|\omega_i|}\right|, Q_{\dot{\theta}} = \frac{\pi}{4}\left|\frac{\omega_j}{|\omega_j|} - \frac{\omega_i}{|\omega_i|}\right|. F2$ (Left). $F4$ (Right).

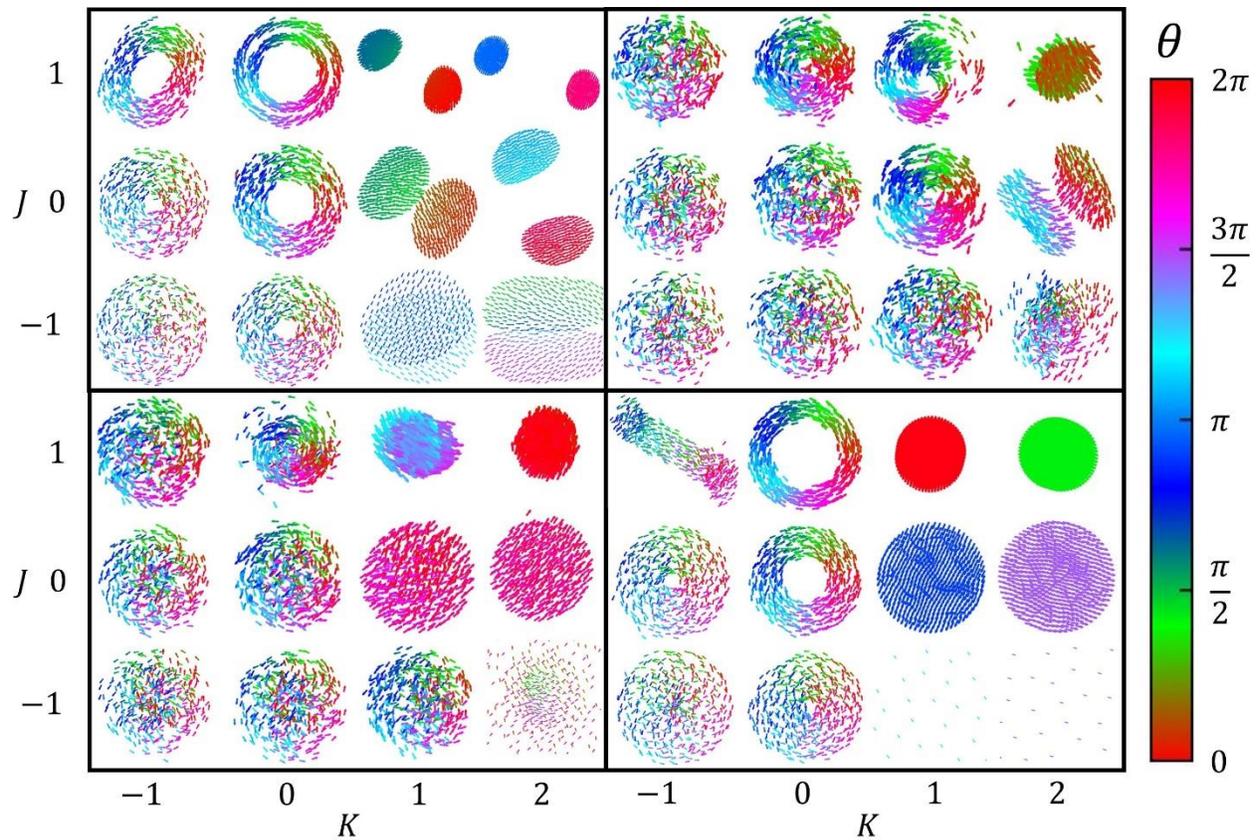

**Supplementary Fig. 13. Collective behaviors of chiral swarmalators with no frequency coupling.** $F2$ (Upper left). $F4$ (Upper right). $F3$ (Lower left). $F1$ (Lower right). Quiver plots are shown here instead of instantaneous positions because chiral swarmalators have an inherent motion driving them.

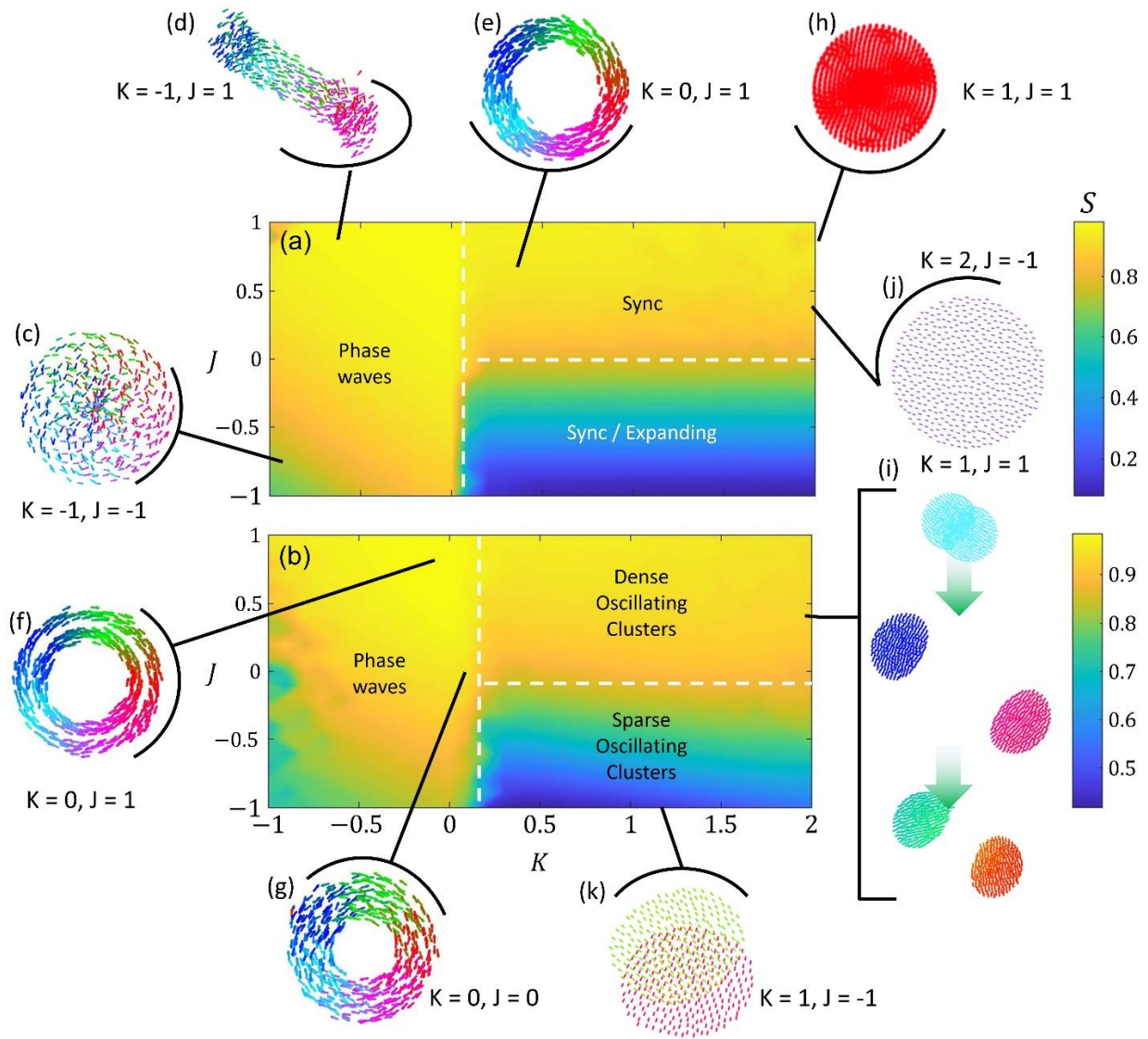

**Supplementary Fig. 14. Revolving swarmalators with no natural frequency spread.** Heat maps of $S$ across $K - J$ parameter space is shown for test cases with (**a**) $F1$ and (**b**) $F2$. (**c**) Phase wave. (**d**) Phase wave along a slim ellipse. (**e**) Phase wave. (**f**) Concentric double phase waves. (**g**) Phase wave. (**h**) Synchronized cluster. (**i**) Dense revolving clusters. (**j**) Expanding synchronized revolving cluster. (**k**) Sparse oscillating clusters.

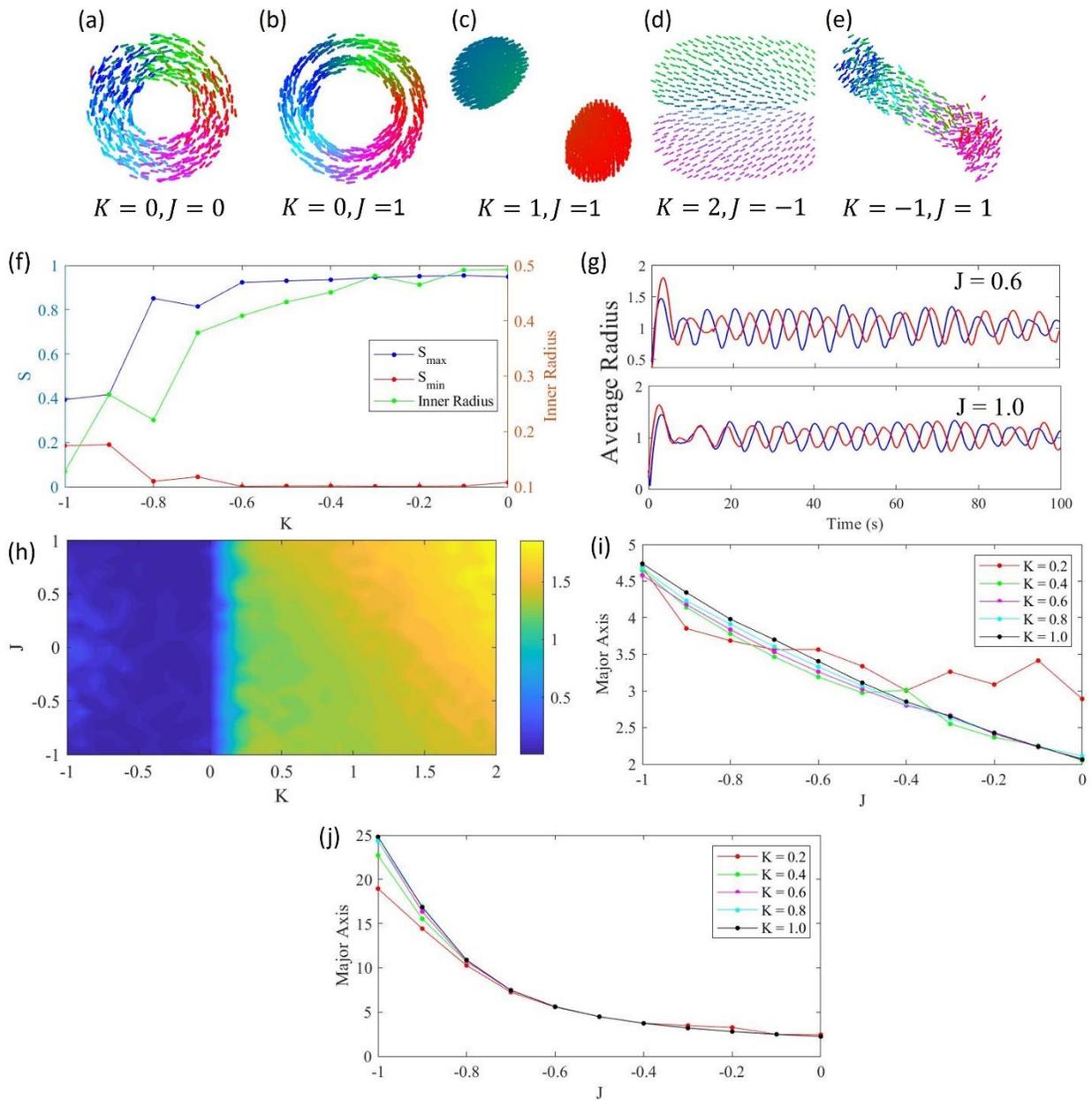

**Supplementary Fig. 15. Characterization of collective behaviors in chiral swarmalators with discrete natural frequencies and no frequency coupling.** (a-e) Various emergent configurations of chiral swarmalators. (f-j) Collective behavior characterizations for the same ω distributions listed in (a-e), respectively. (a) $F2$; (b) $F2$; (c) $F2$; (d) $F2$; (e) $F1$; (f) $S_{max}$, $S_{min}$, and inner ring radius when $J = 0$; (g) average radius of each natural frequency group over time when $J = 0.6, 1$; (h) max distance between the two clusters; (i) major axis length for different $K$; (j) major axis length when $K = -1$.

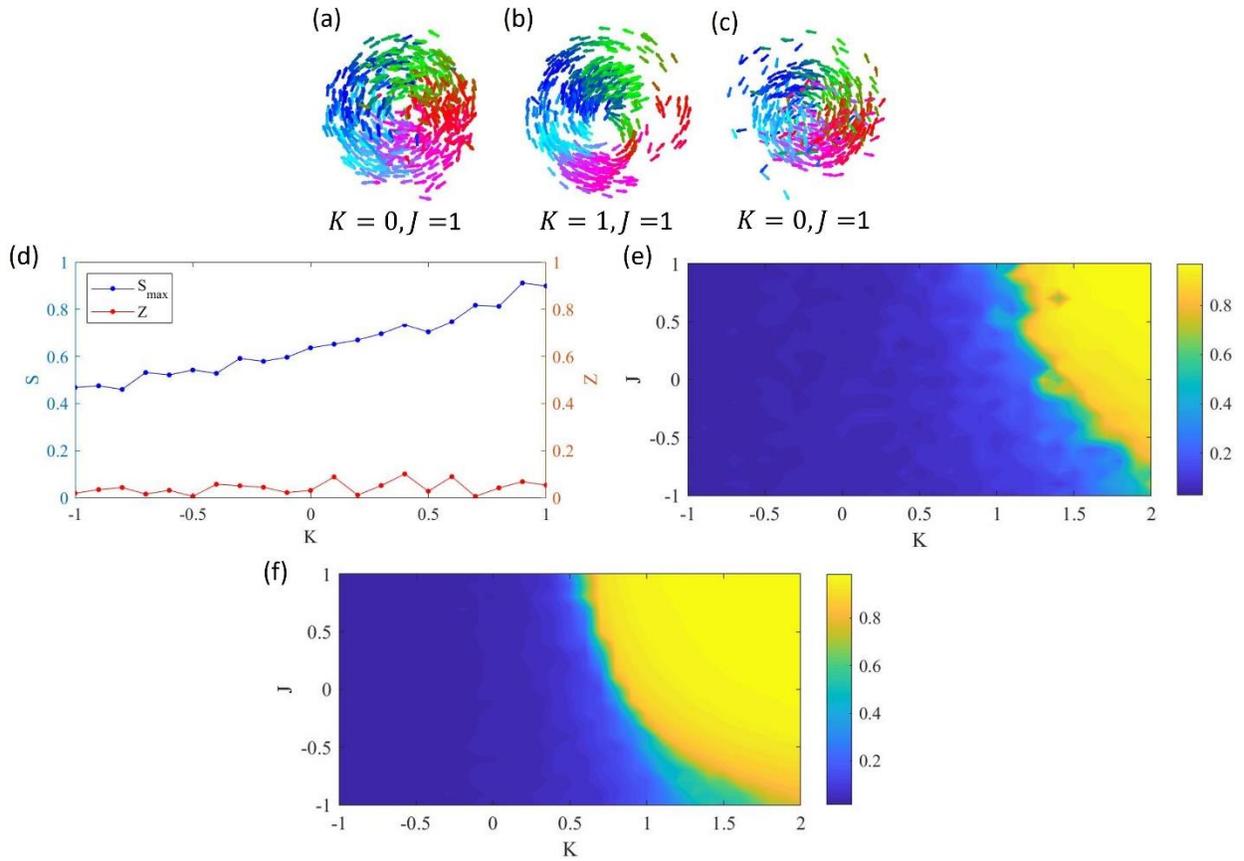

**Supplementary Fig. 16. Characterization of collective behaviors in chiral swarmalators with a natural frequency spread and no frequency coupling. (a-e)** Emergent configurations of chiral swarmalators. **(f-j)** Collective behavior characterizations for the same ω distributions listed in **(a-e)**, respectively. **(a)** $F4$; **(b)** $F4$; **(c)** $F3$; **(d)** $S$ order and phase coherence when $J = 0$; **(e)** Heat map of the average natural frequency group S order. **(f)** Heat map of phase coherence.

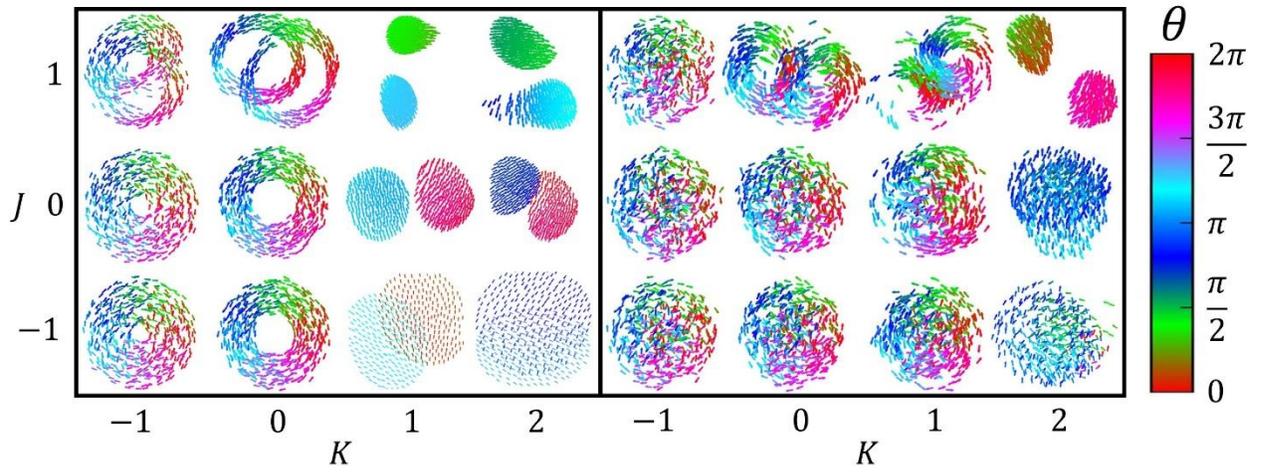

**Supplementary Fig. 17. Collective behaviors of frequency-coupled chiral swarmalators.** $Q_{\dot{x}} = \frac{\pi}{2}\left|\frac{\omega_j}{|\omega_j|} - \frac{\omega_i}{|\omega_i|}\right|, Q_{\dot{\theta}} = \frac{\pi}{4}\left|\frac{\omega_j}{|\omega_j|} - \frac{\omega_i}{|\omega_i|}\right|, F2$ (Left). $F4$ (Right).

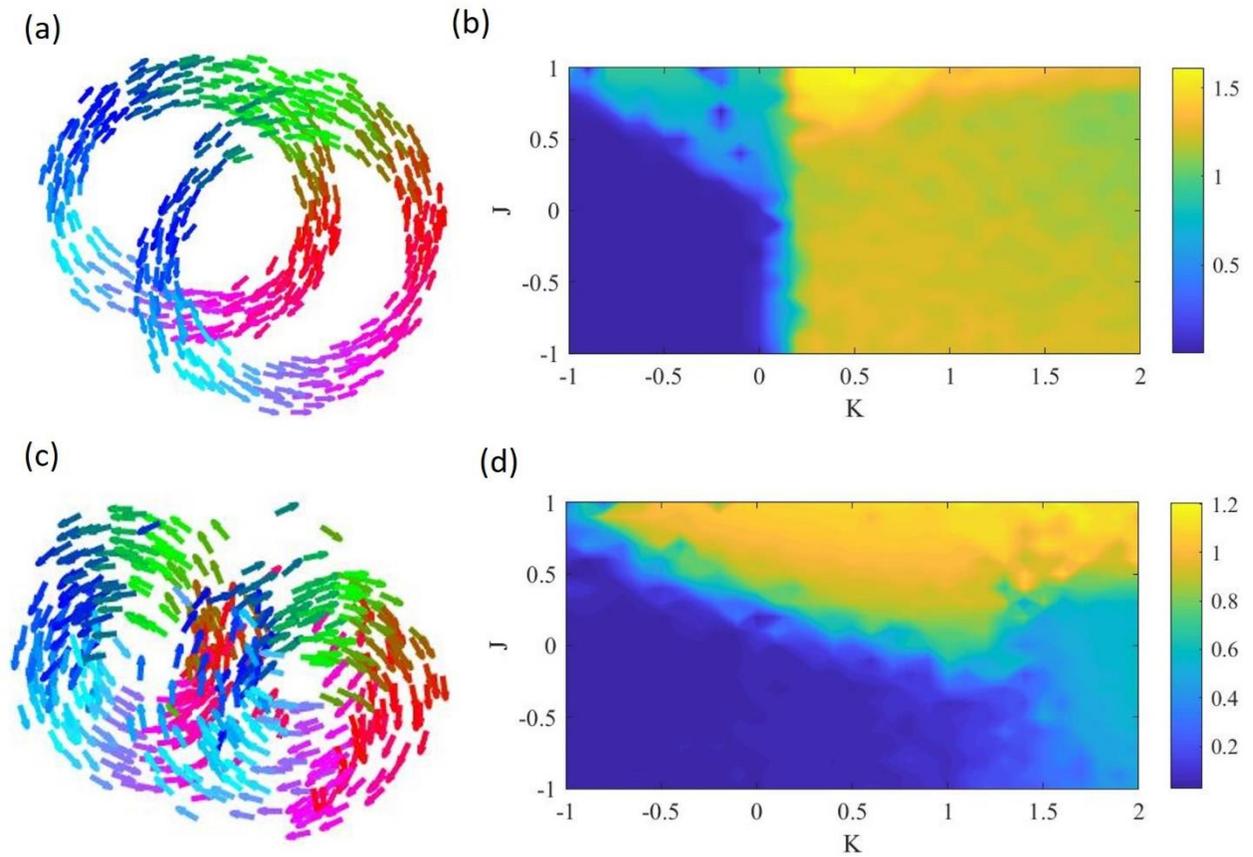

**Supplementary Fig. 18. Characterization of collective behaviors in frequency-coupled chiral swarmalators.** (a) $K = 0, J = 1, \ g(\omega) = \{-\Omega, \Omega\}$; (c) $K = 0, J = 1, \ F4$. (b,d) Average distance between the natural frequency groups' centroids for (a) and (c), respectively.

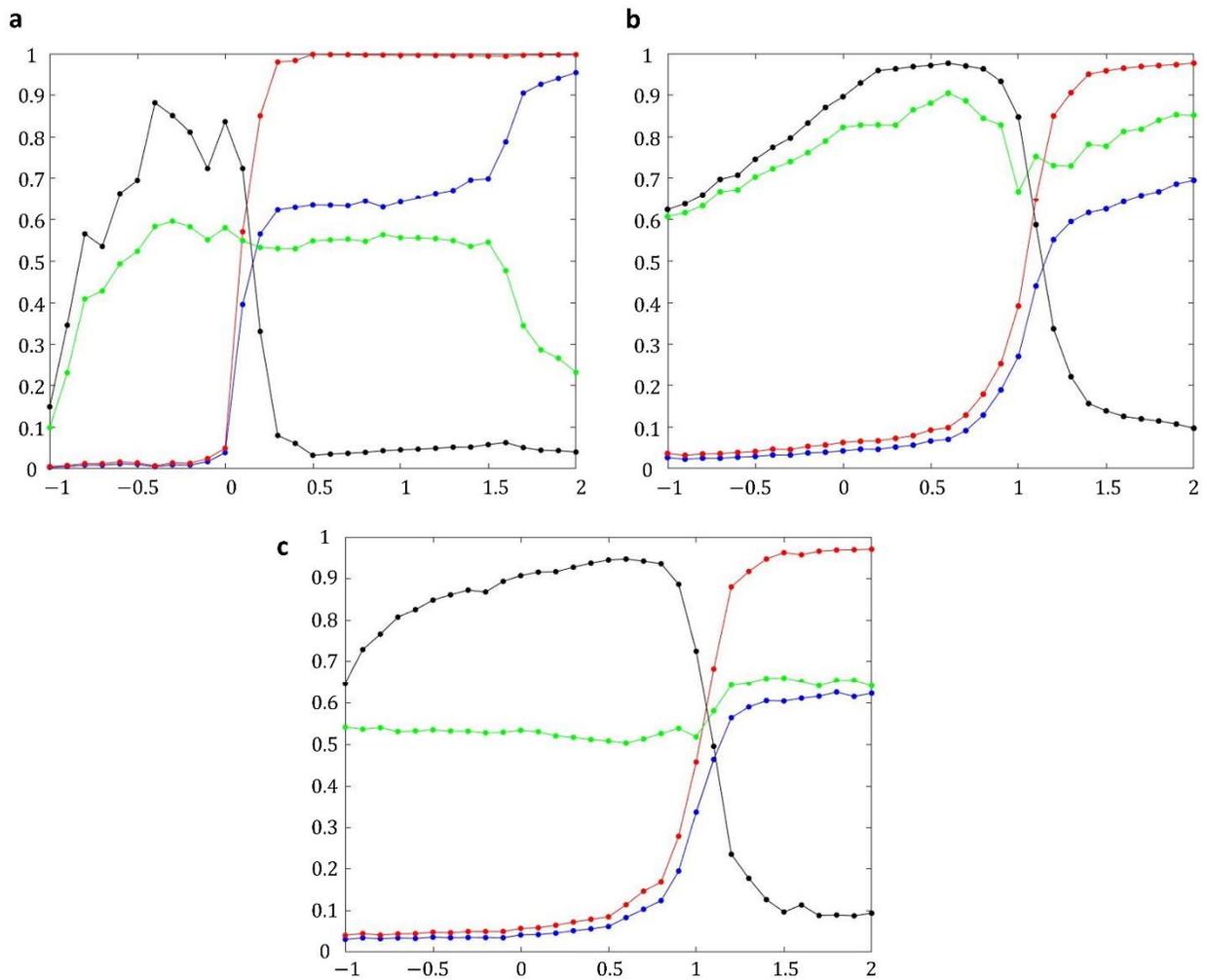

**Supplementary Fig. 19. Order parameters.** Four order parameters are plotted for **(a)** non-chiral swarmalators with the natural frequency distribution $F2$, **(b)** revolving swarmalators with the natural frequency distribution $F3$, and **(c)** frequency-coupled revolving swarmalators with the distribution $F4$. The four order parameters are global phase coherence (blue), frequency group phase coherence (red), global spatial phase order (green), frequency group spatial phase order (black).

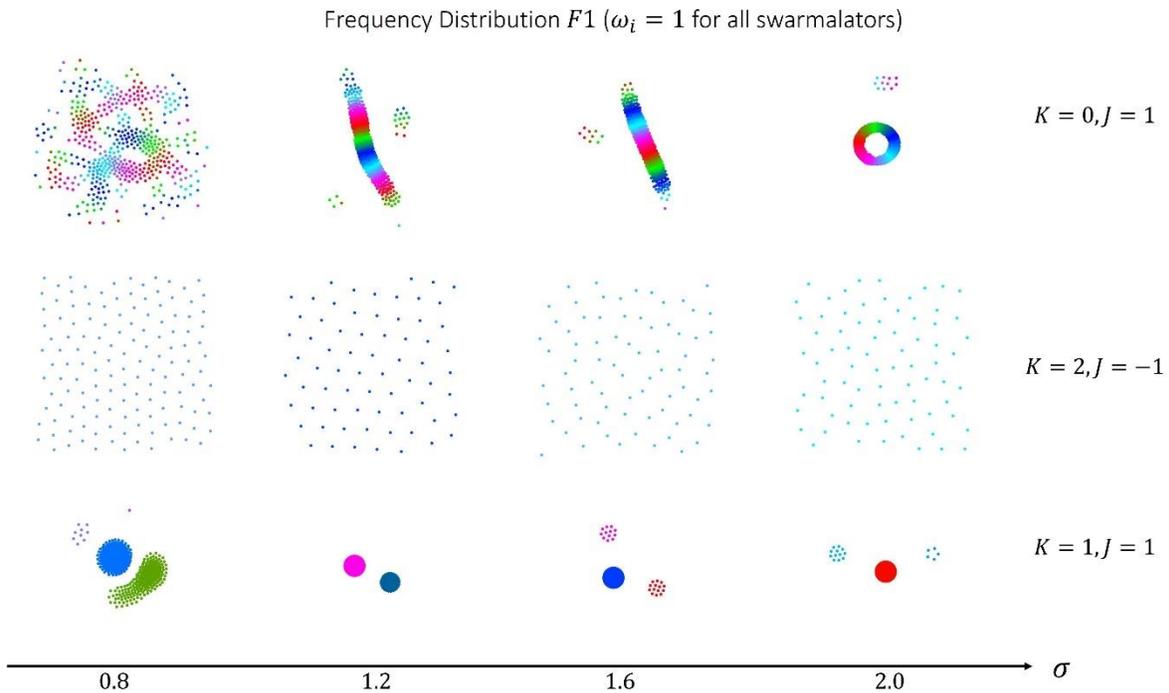

**Supplementary Fig. 20. Collective behaviors of locally coupled non-chiral swarmalators with one natural frequency and no frequency coupling.** The $x$-axis increases the coupling distance ($\sigma$). (Top) $K = 0, J = 1$. (Middle) $K = 2, J = -1$. (Bottom) $K = 1, J = 1$.

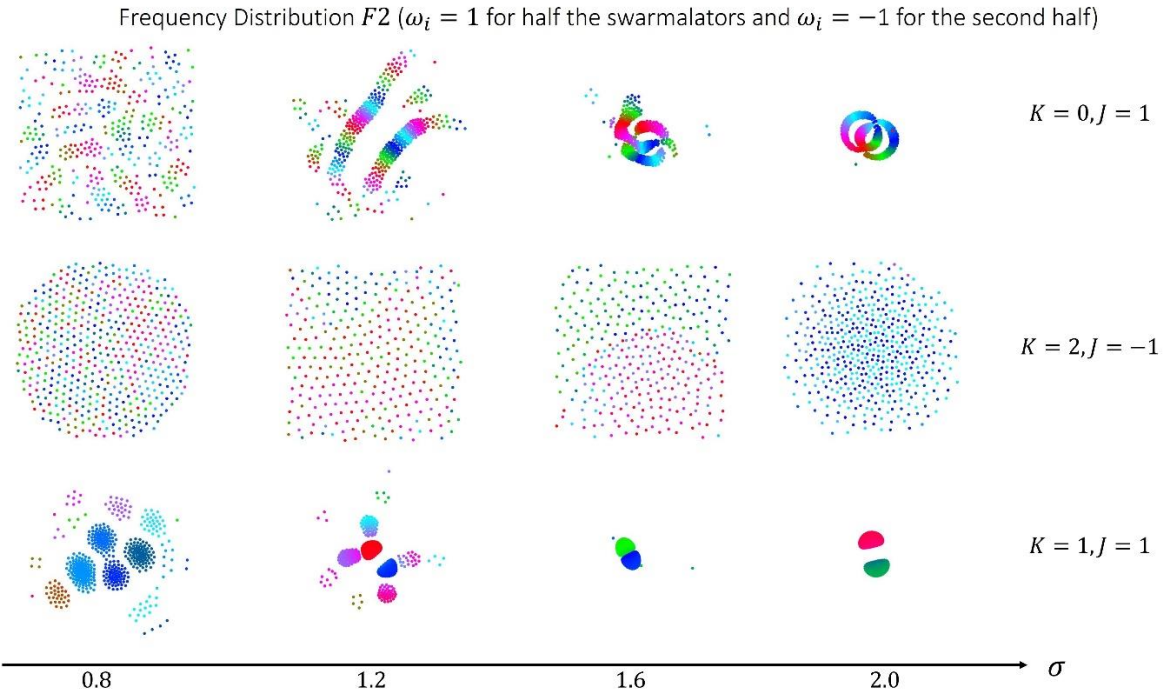

**Supplementary Fig. 21. Collective behaviors of locally coupled non-chiral swarmalators with two natural frequencies and no frequency coupling.** The $x$-axis increases the coupling distance ($\sigma$). (Top) $K=0, J=1$. (Middle) $K=2, J=-1$. (Bottom) $K=1, J=1$.

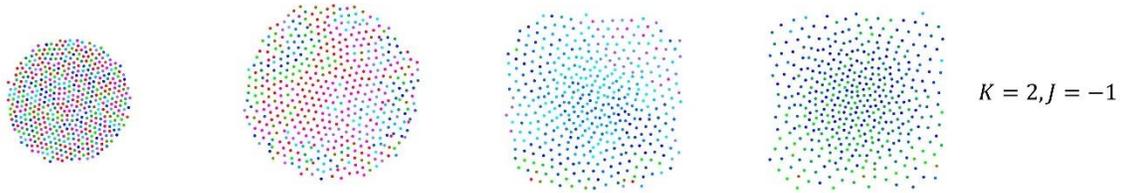

Frequency Distribution $F3$: $\omega_i \sim U(1,3)$ for all of the swarmalators

$K = 2, J = -1$

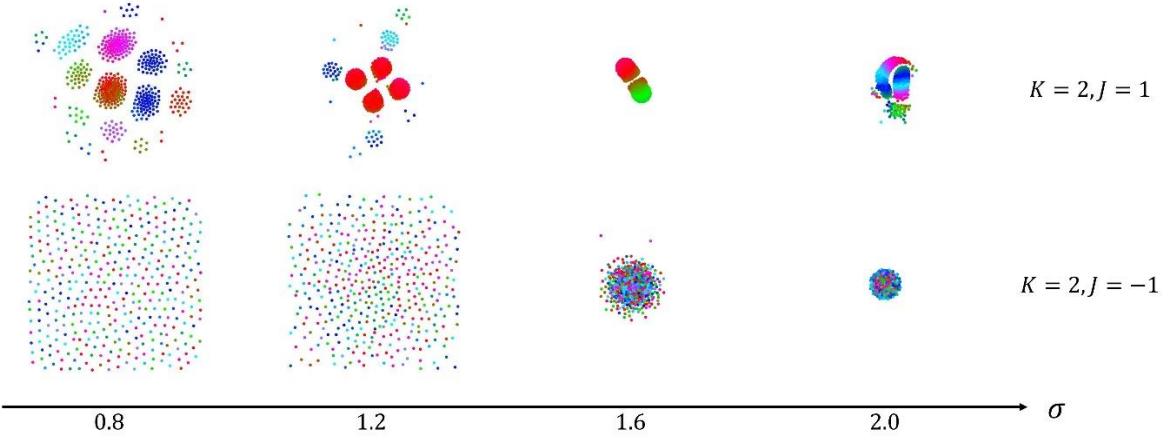

Frequency Distribution $F4$: $\omega_i \sim U(1,3)$ for half the collective and $\omega_i \sim U(-3,-1)$ for the second half

$K = 2, J = 1$

$K = 2, J = -1$

0.8     1.2     1.6     2.0     $\sigma$

**Supplementary Fig. 22. Collective behaviors of locally coupled non-chiral swarmalators with a natural frequency spread and no frequency coupling.** The $x$-axis increases the coupling distance ($\sigma$). (Top) $K = 2, J = -1$. (Middle) $K = 2, J = 1$. (Bottom) $K = 2, J = -1$.

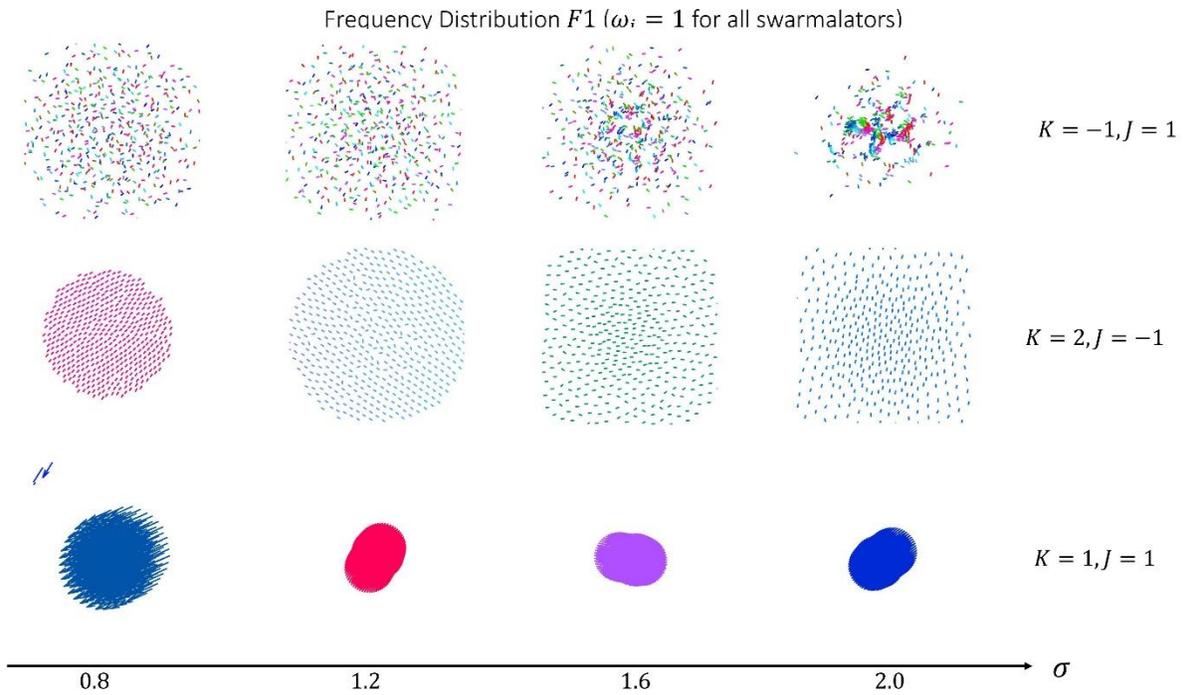

**Supplementary Fig. 23. Collective behaviors of locally coupled chiral swarmalators with one natural frequency and no frequency coupling.** The $x$-axis increases the coupling distance ($\sigma$) and is normalized by the maximum radius of revolution ($R = 1$). (Top) $K = 0, J = 1$. (Middle) $K = 2, J = -1$. (Bottom) $K = 1, J = 1$.

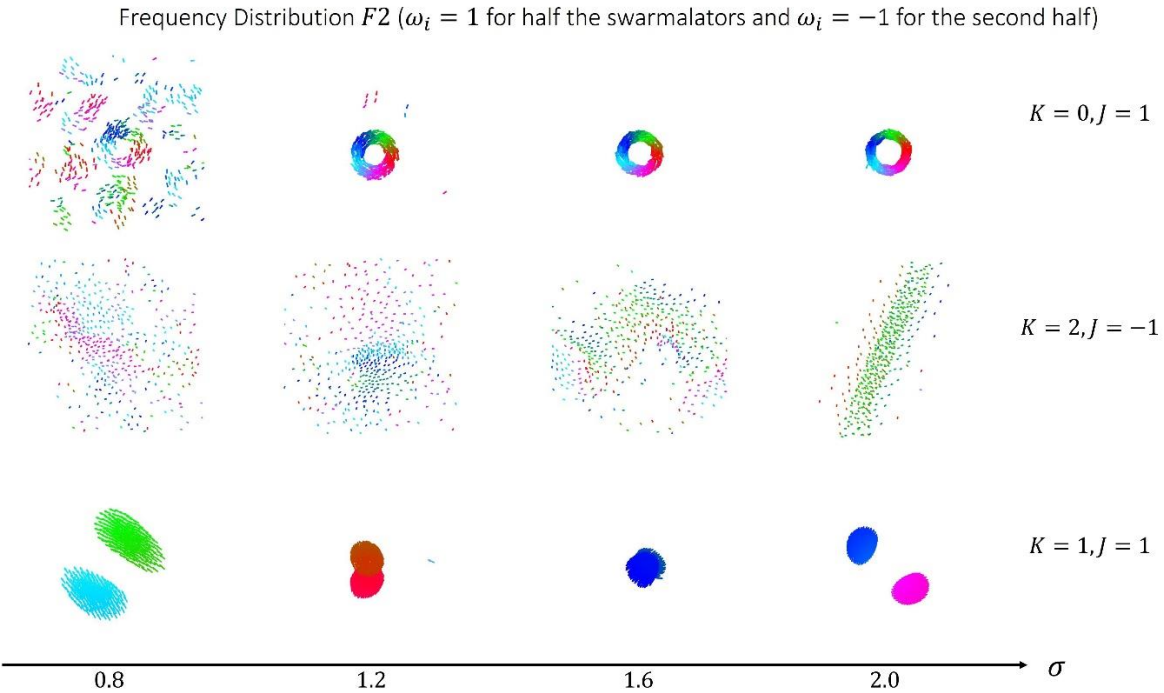

**Supplementary Fig. 24. Collective behaviors of locally coupled chiral swarmalators with one natural frequency and no frequency coupling.** The $x$-axis increases the coupling distance ($\sigma$) and is normalized by the maximum radius of revolution ($R = 1$). (Top) $K = 0, J = 1$. (Middle) $K = 2, J = -1$. (Bottom) $K = 1, J = 1$.

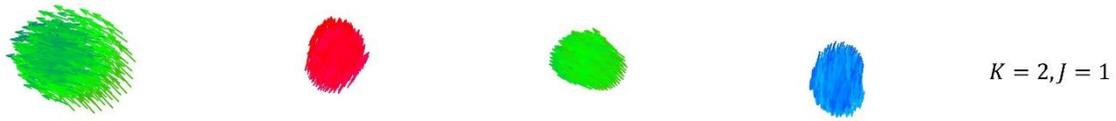

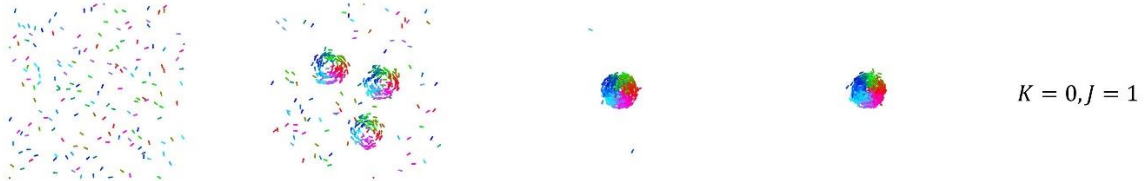

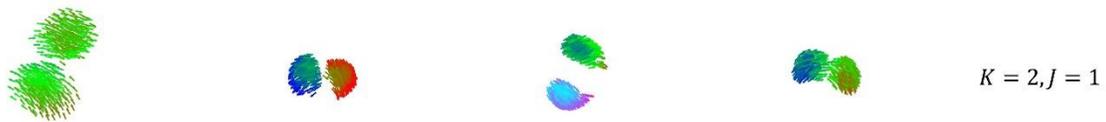

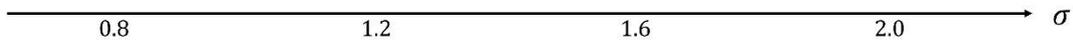

**Supplementary Fig. 25. Collective behaviors of locally coupled chiral swarmalators with a natural frequency spread and no frequency coupling.** The $x$-axis increases the coupling distance ($\sigma$) and is normalized by the maximum radius of revolution ($R = 1$). (Top) $K = 2, J = 1$. (Middle) $K = 0, J = 1$. (Bottom) $K = 2, J = 1$.

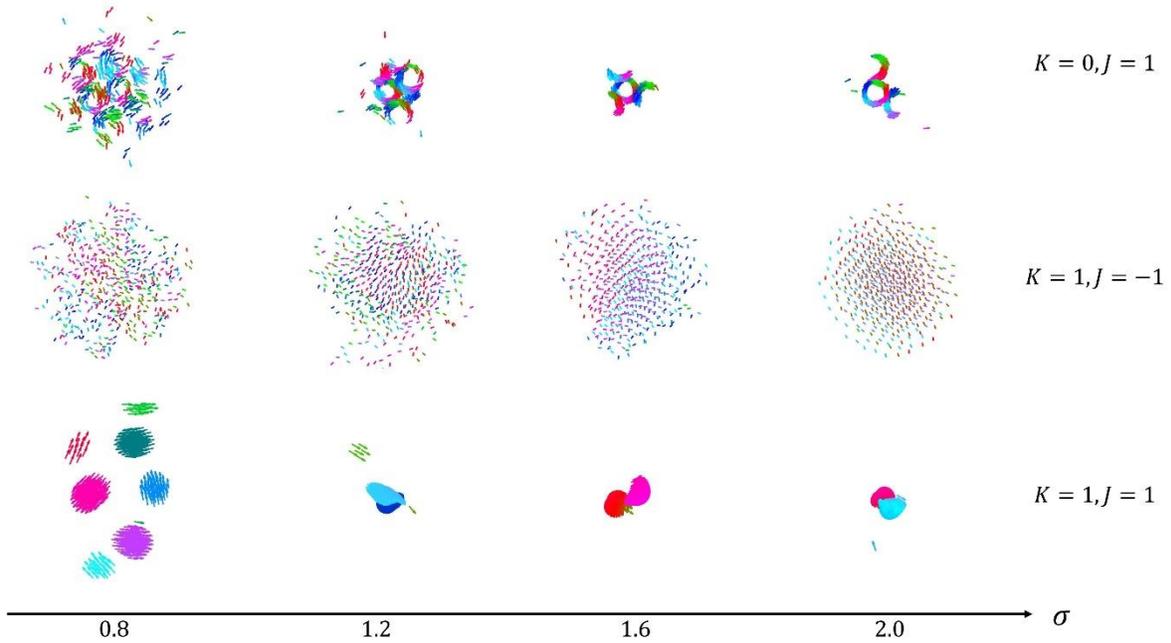

**Supplementary Fig. 26. Collective behaviors of locally coupled chiral swarmalators with no natural frequency spread and frequency coupled.** The $x$-axis increases the coupling distance ($\sigma$). (Top) $K = 0, J = 1$. (Middle) $K = 1, J = -1$. (Bottom) $K = 1, J = 1$.

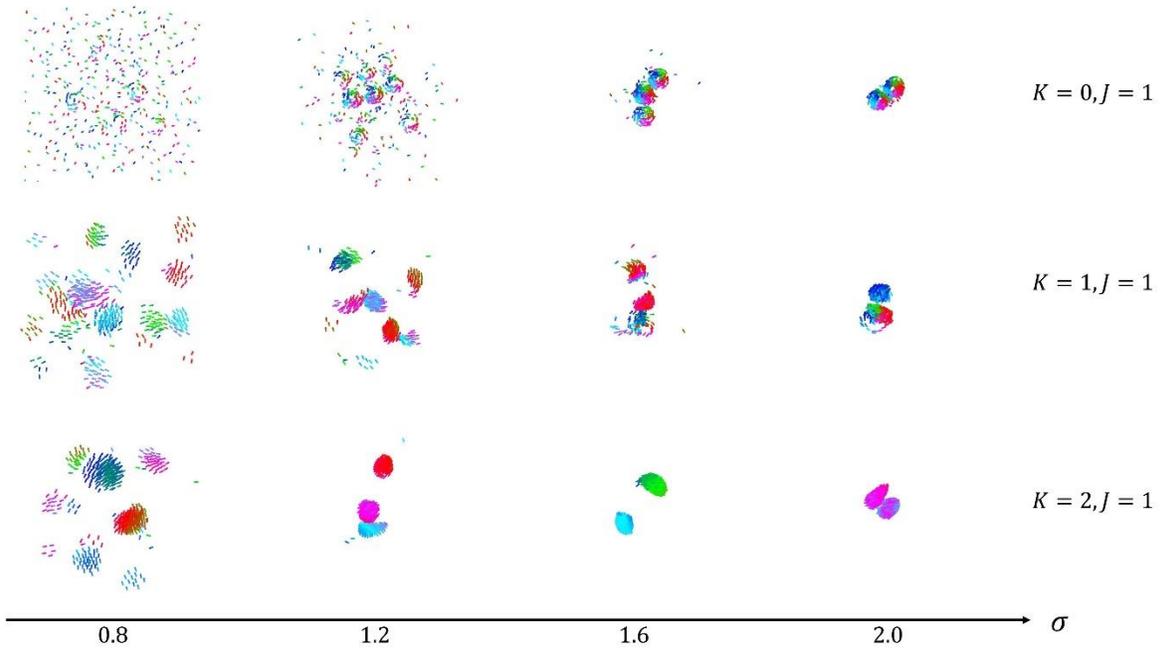

Frequency Distribution $F4$: $\omega_i \sim U(1,3)$ for half the collective and $\omega_i \sim U(-3,-1)$ for the second half

**Supplementary Fig. 27. Collective behaviors of locally coupled chiral swarmalators with a natural frequency spread and frequency coupling.** The $x$-axis increases the coupling distance ($\sigma$). (Top) $K = 0, J = 1$. (Middle) $K = 1, J = 1$. (Bottom) $F4$. $K = 2, J = 1$.

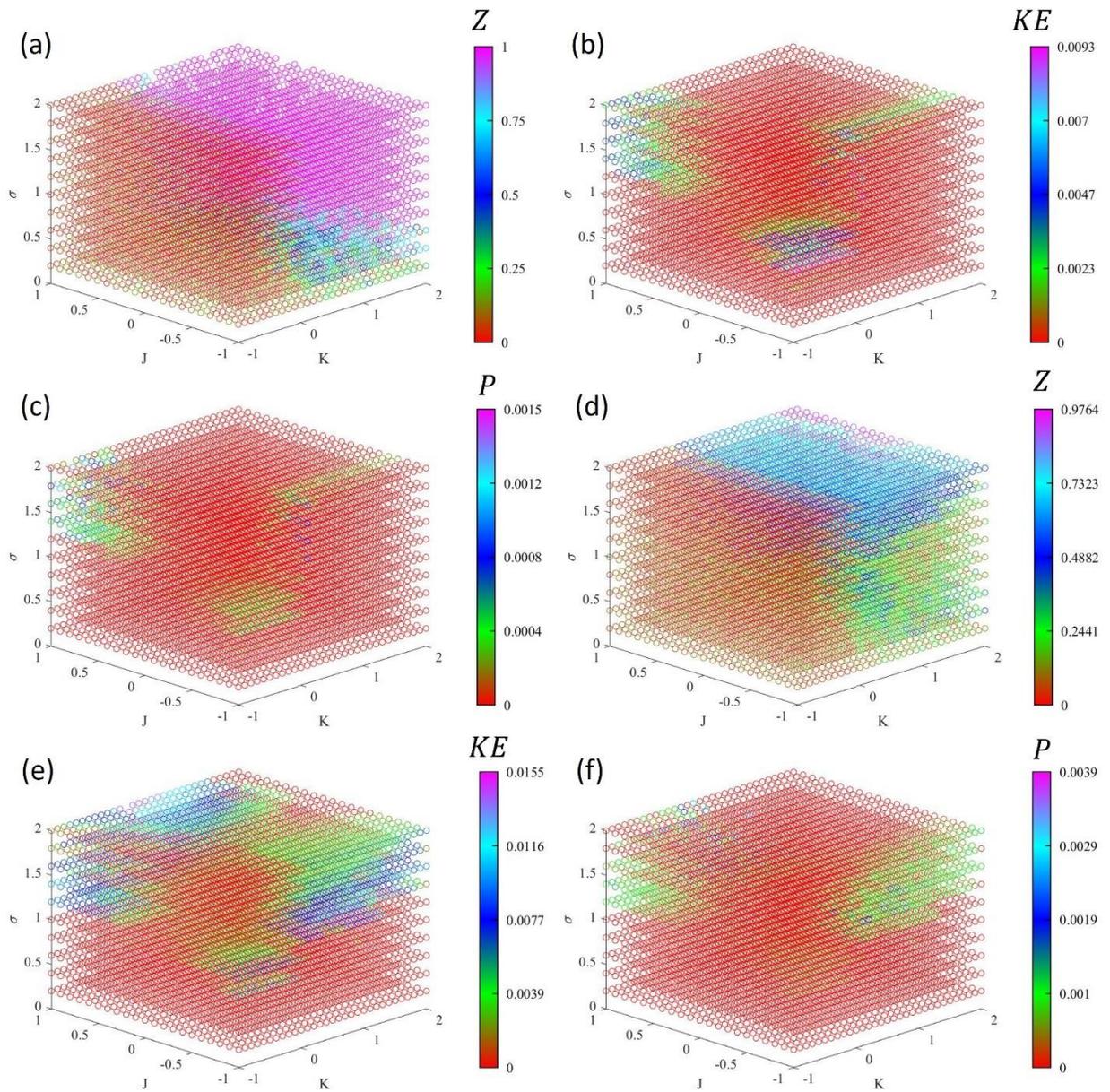

**Supplementary Fig. 28. 3D Maps for locally coupled non-chiral swarmalators with no frequency spread and no frequency coupling.** (a-c) There is a single natural frequency, $\omega_i = 1$, throughout the collective. (d-f) Half the collective has $\omega_i = -1$ and the other half has $\omega_i = 1$. **(a,d)** Degree of synchrony ($Z$). **(b,e)** Kinetic energy ($KE$). **(c,f)** Linear momentum ($P$).

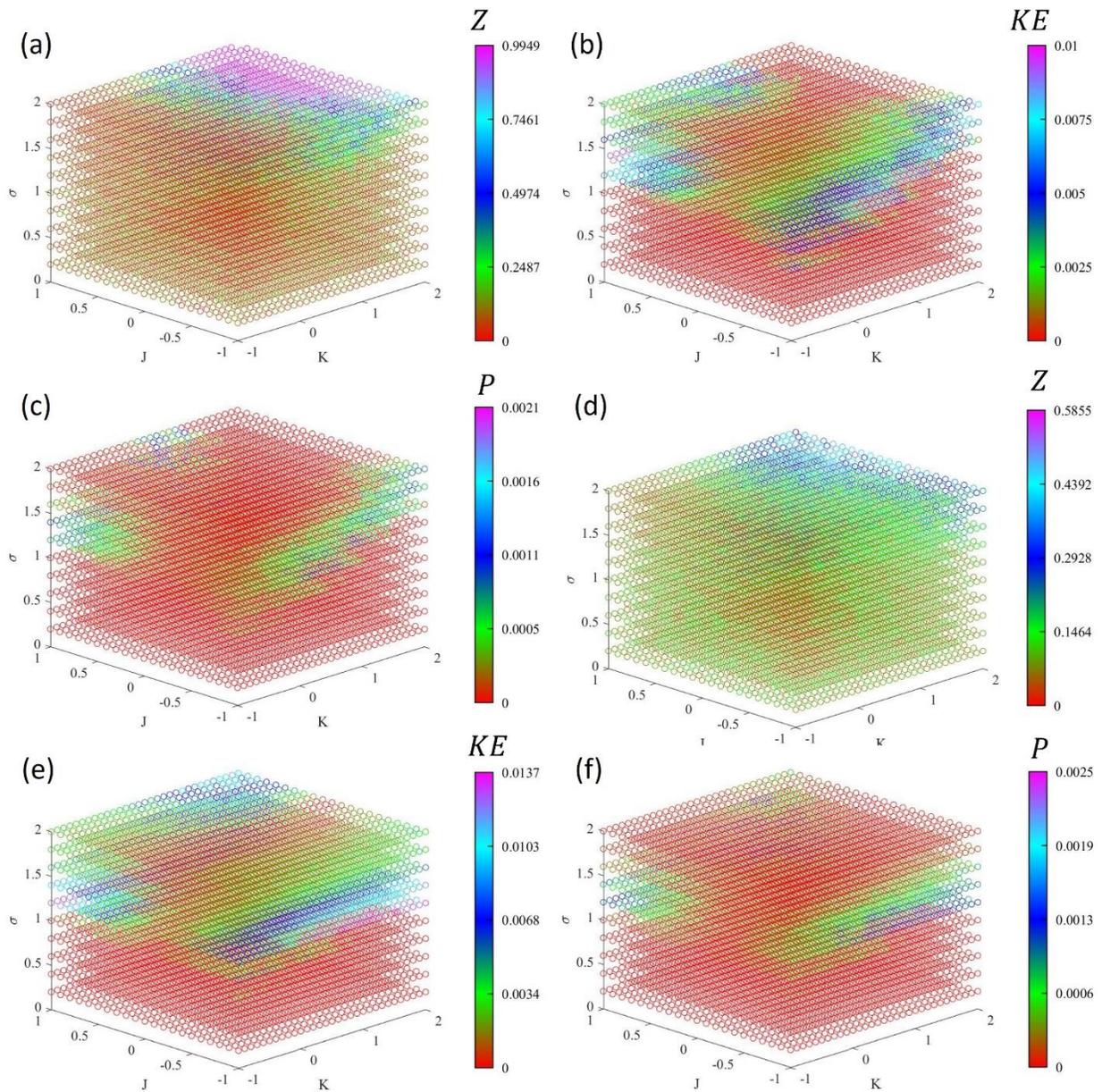

**Supplementary Fig. 29. 3D Maps for locally coupled non-chiral swarmalators with a frequency spread and no frequency coupling.** (a-c) All swarmalators have their natural frequency randomly selected from single uniform distribution, such that $\omega_i \sim U(\Omega, 3\Omega)$. (d-f) Half the collective has their natural frequencies randomly selected from the uniform distribution $U(1,3)$ and the second half have their natural frequency selected from the uniform distribution $U(-3,-1)$. **(a,d)** Degree of synchrony ($Z$). **(b,e)** Kinetic energy ($KE$). **(c,f)** Linear momentum ($P$).

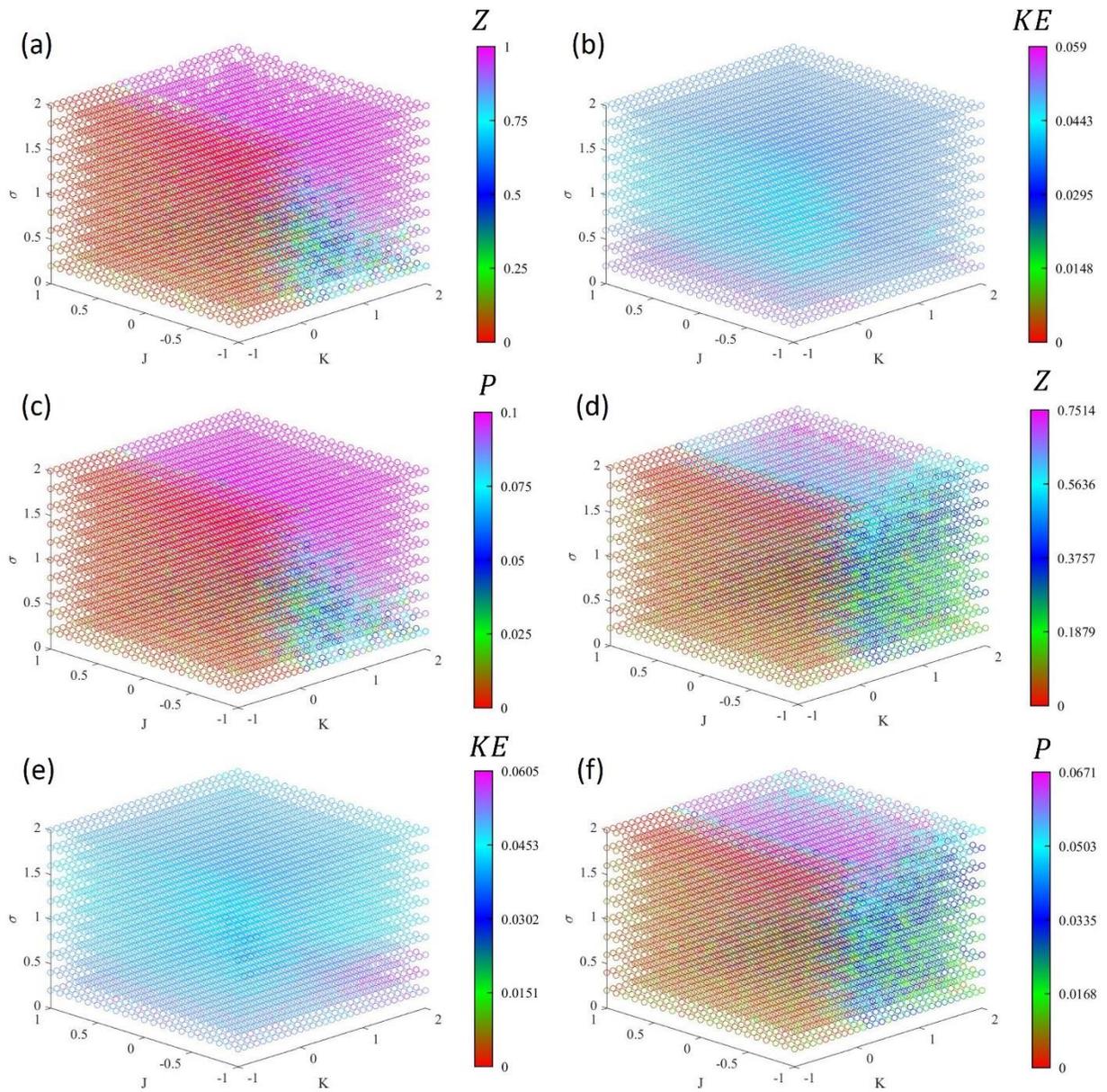

**Supplementary Fig. 30. 3D Maps for locally coupled chiral swarmalators with no frequency spread and no frequency coupling.** (a-c) There is a single natural frequency, $\omega_i = 1$, throughout the collective. (d-f) Half the collective has $\omega_i = -1$ and the other half has $\omega_i = 1$. **(a,d)** Degree of synchrony ($Z$). **(b,e)** Kinetic energy ($KE$). **(c,f)** Linear momentum ($P$).

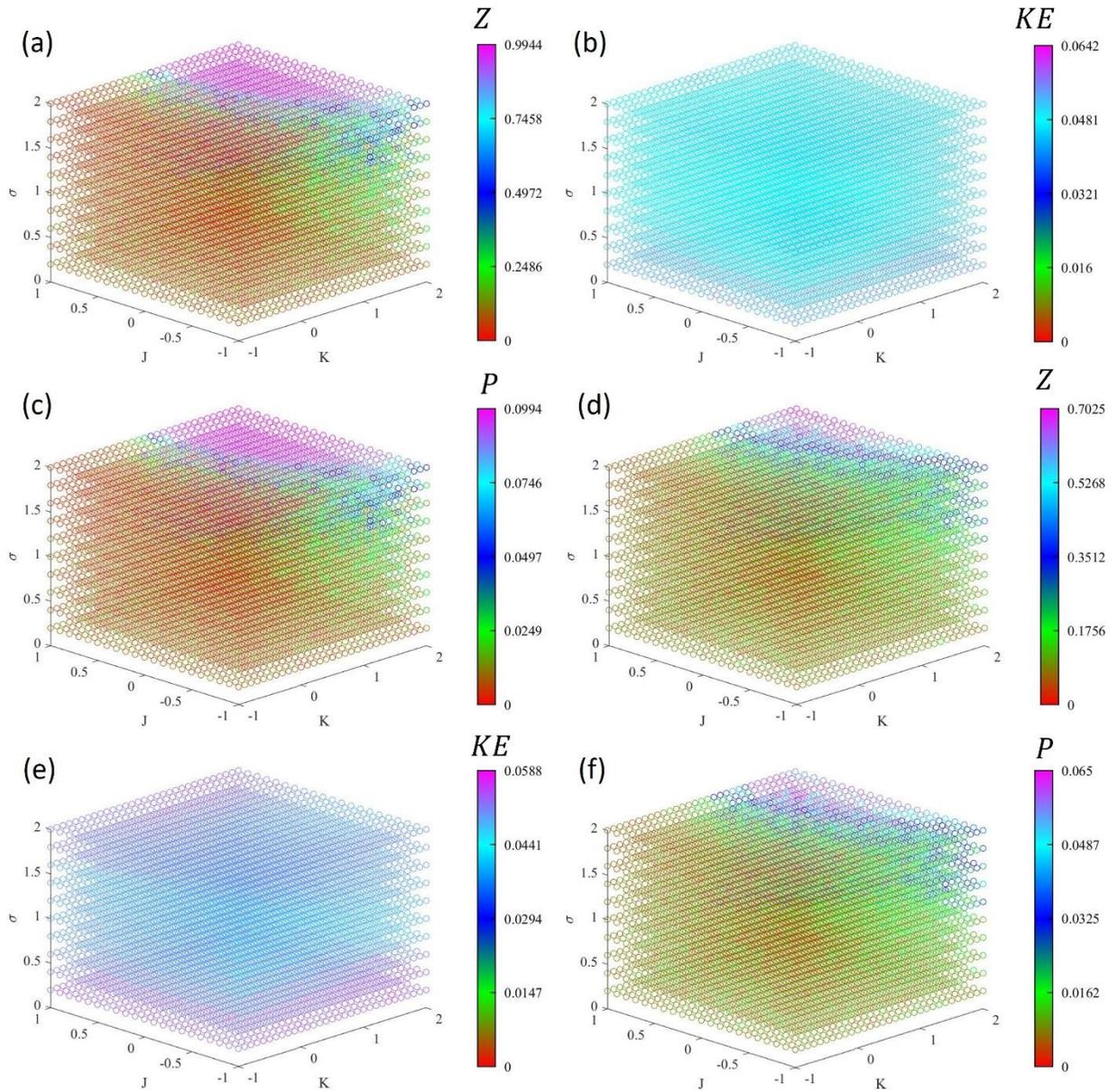

**Supplementary Fig. 31. 3D Maps for locally coupled chiral swarmalators with a frequency spread and no frequency coupling.** (a-c) All swarmalators have their natural frequency randomly selected from single uniform distribution, such that $\omega_i \sim U(\Omega, 3\Omega)$. (d-f) Half the collective has their natural frequencies randomly selected from the uniform distribution $U(1,3)$ and the second half have their natural frequency selected from the uniform distribution $U(-3,-1)$. **(a,d)** Degree of synchrony ($Z$). **(b,e)** Kinetic energy ($KE$). **(c,f)** Linear momentum ($P$).

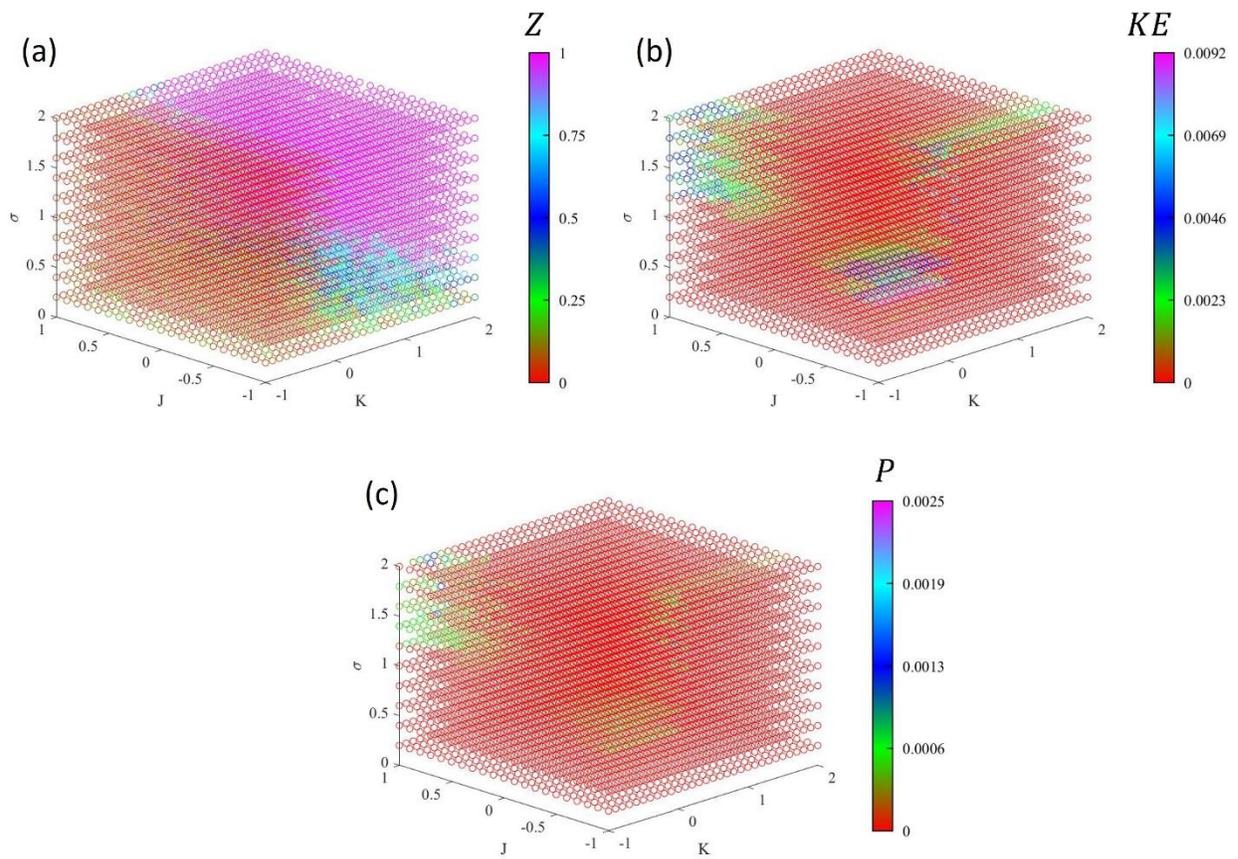

**Supplementary Fig. 32. 3D Maps for locally coupled non-chiral swarmalators with no frequency spread and frequency-coupled.** Half the collective has $\omega_i = -1$ and the other half has $\omega_i = 1$. **(a)** Degree of synchrony ($Z$). **(b)** Kinetic energy ($KE$). **(c)** Linear momentum ($P$).

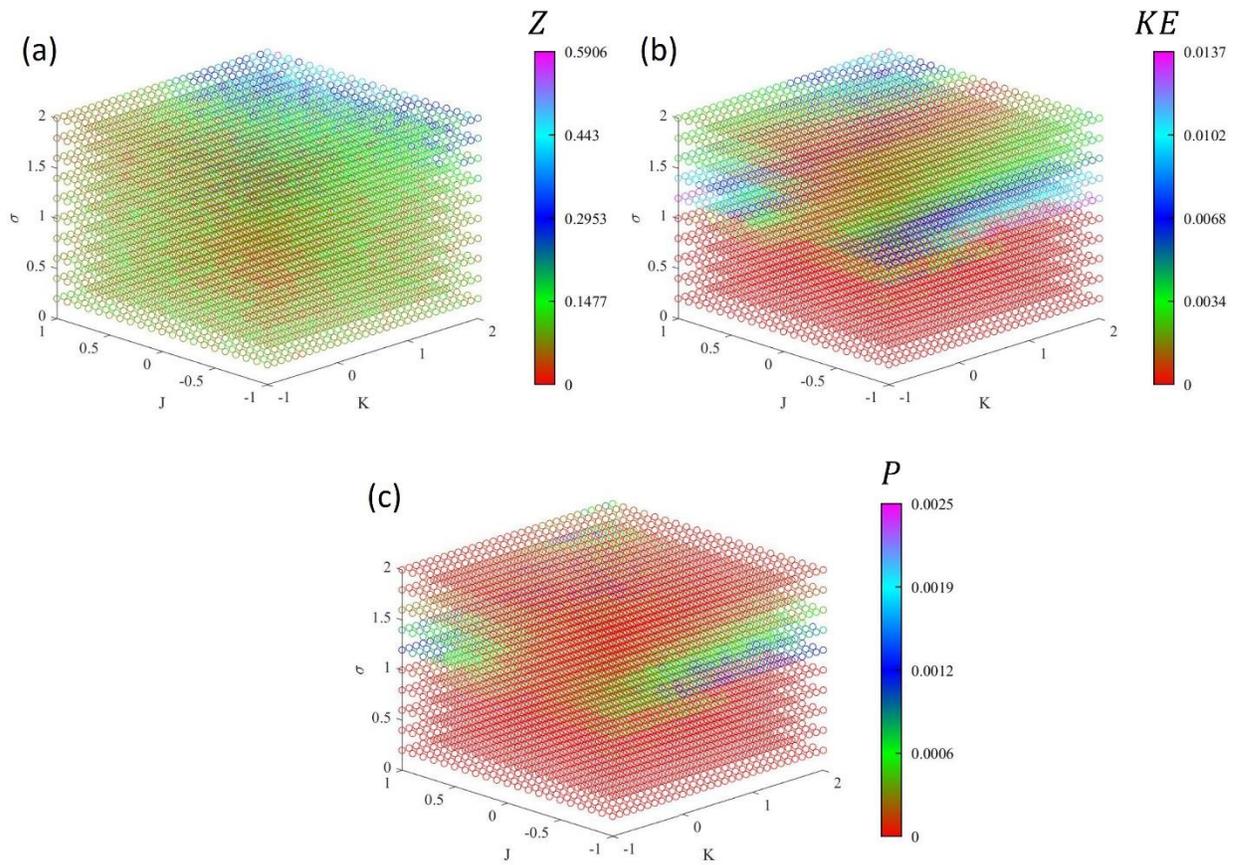

**Supplementary Fig. 33. 3D Maps for locally coupled non-chiral swarmalators with a frequency spread and frequency-coupled.** Half the collective has their natural frequencies randomly selected from the uniform distribution $U(1,3)$ and the second half have their natural frequency selected from the uniform distribution $U(-3,-1)$. **(a)** Degree of synchrony ($Z$). **(b)** Kinetic energy ($KE$). **(c)** Linear momentum ($P$).

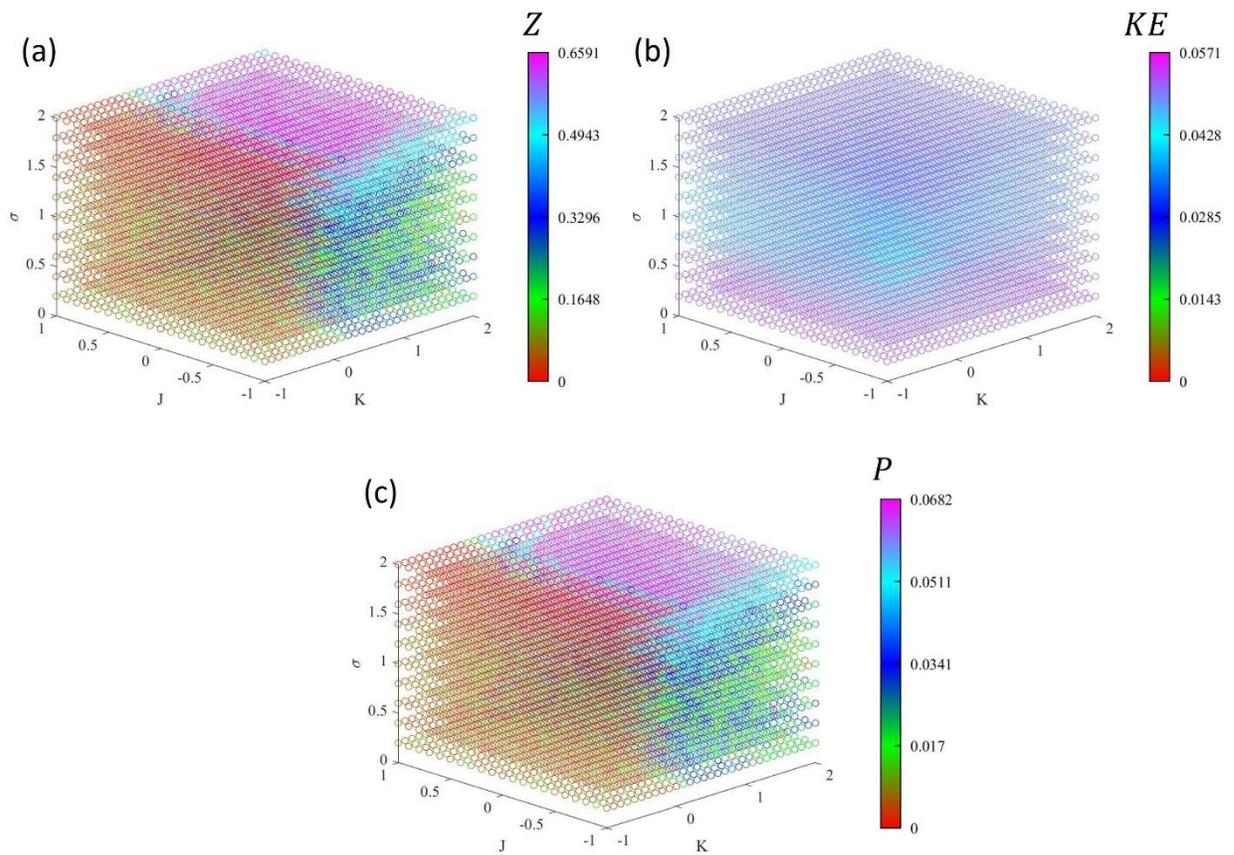

**Supplementary Fig. 34. 3D Maps for locally coupled chiral swarmalators with no frequency spread and frequency-coupled.** Half the collective has $\omega_i = -1$ and the other half has $\omega_i = 1$. **(a)** Degree of synchrony ($Z$). **(b)** Kinetic energy ($KE$). **(c)** Linear momentum ($P$).

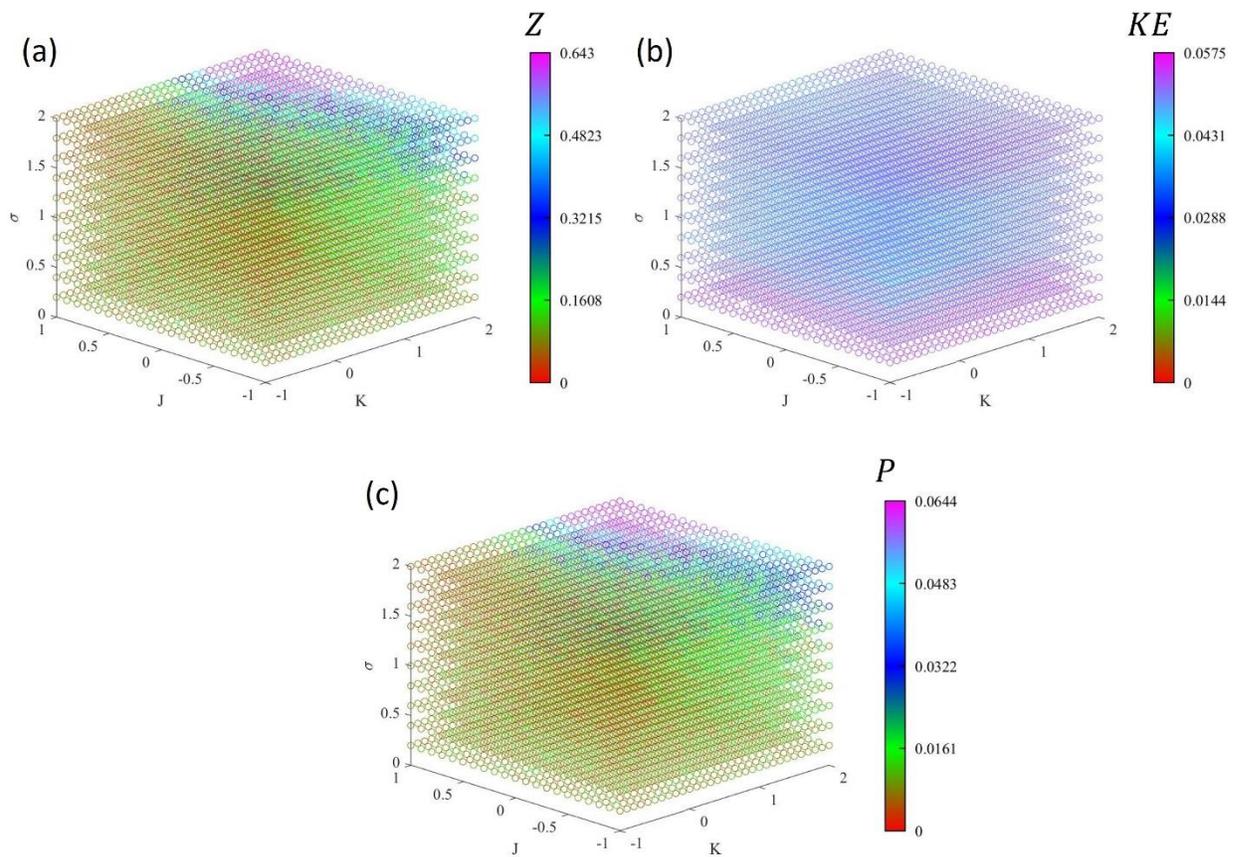

**Supplementary Fig. 35. 3D Maps for locally coupled chiral swarmalators with a frequency spread and frequency-coupled.** Half the collective has their natural frequencies randomly selected from the uniform distribution $U(1, 3)$ and the second half have their natural frequency selected from the uniform distribution $U(-3, -1)$. **(a)** Degree of synchrony ($Z$). **(b)** Kinetic energy ($KE$). **(c)** Linear momentum ($P$).

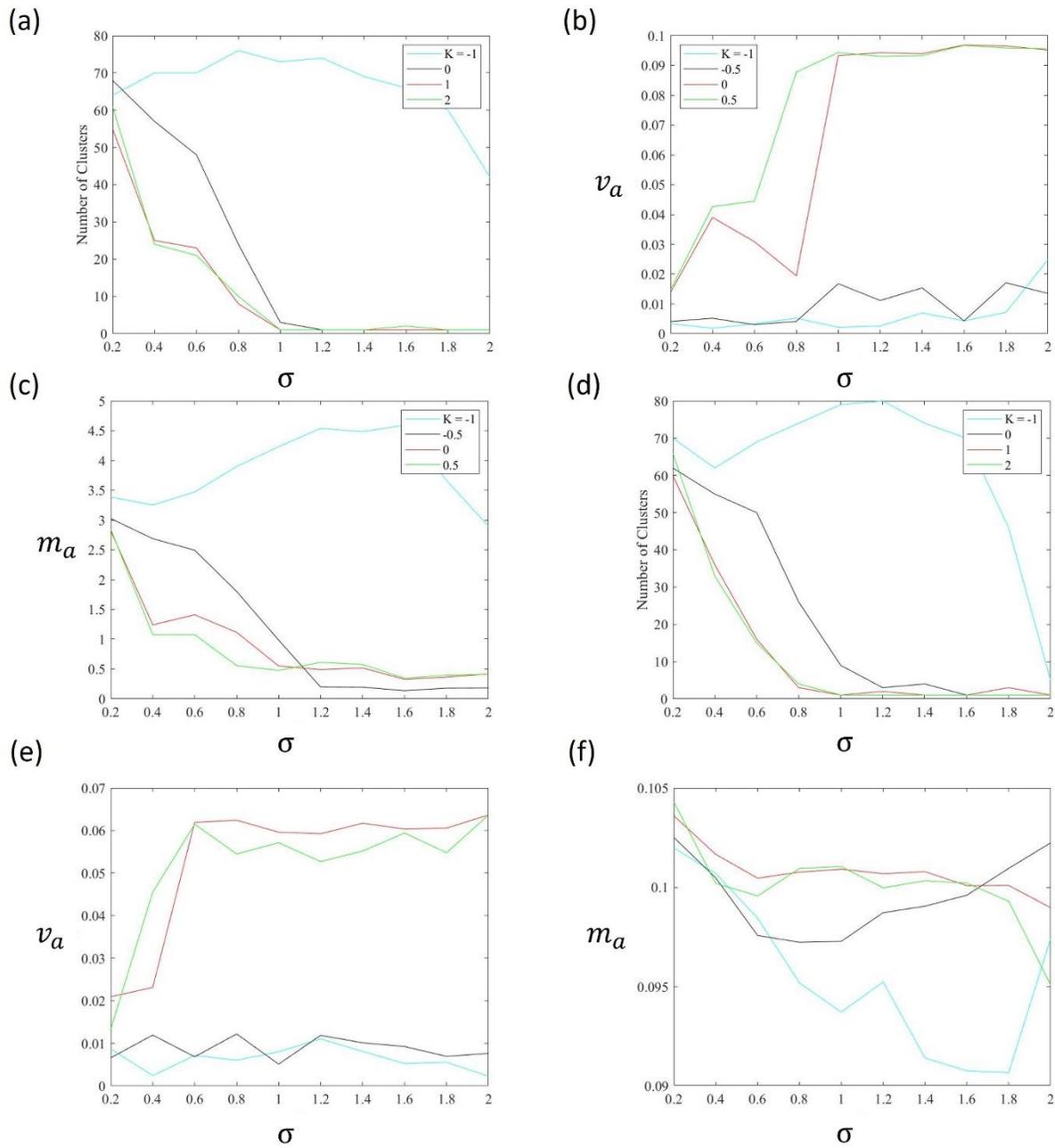

**Supplementary Fig. 36. Cluster count and motion characterization of locally coupled chiral swarmalators with no frequency spread and no frequency coupling.** (a-c) There is a single natural frequency, $\omega_i = 1$, throughout the collective. (d-f) Half the collective has $\omega_i = -1$ and the other half has $\omega_i = 1$. **(a,d)** Number of clusters. **(b,e)** Normalized velocity ($v_a$). **(c,f)** Normalized angular momentuum ($m_a$).

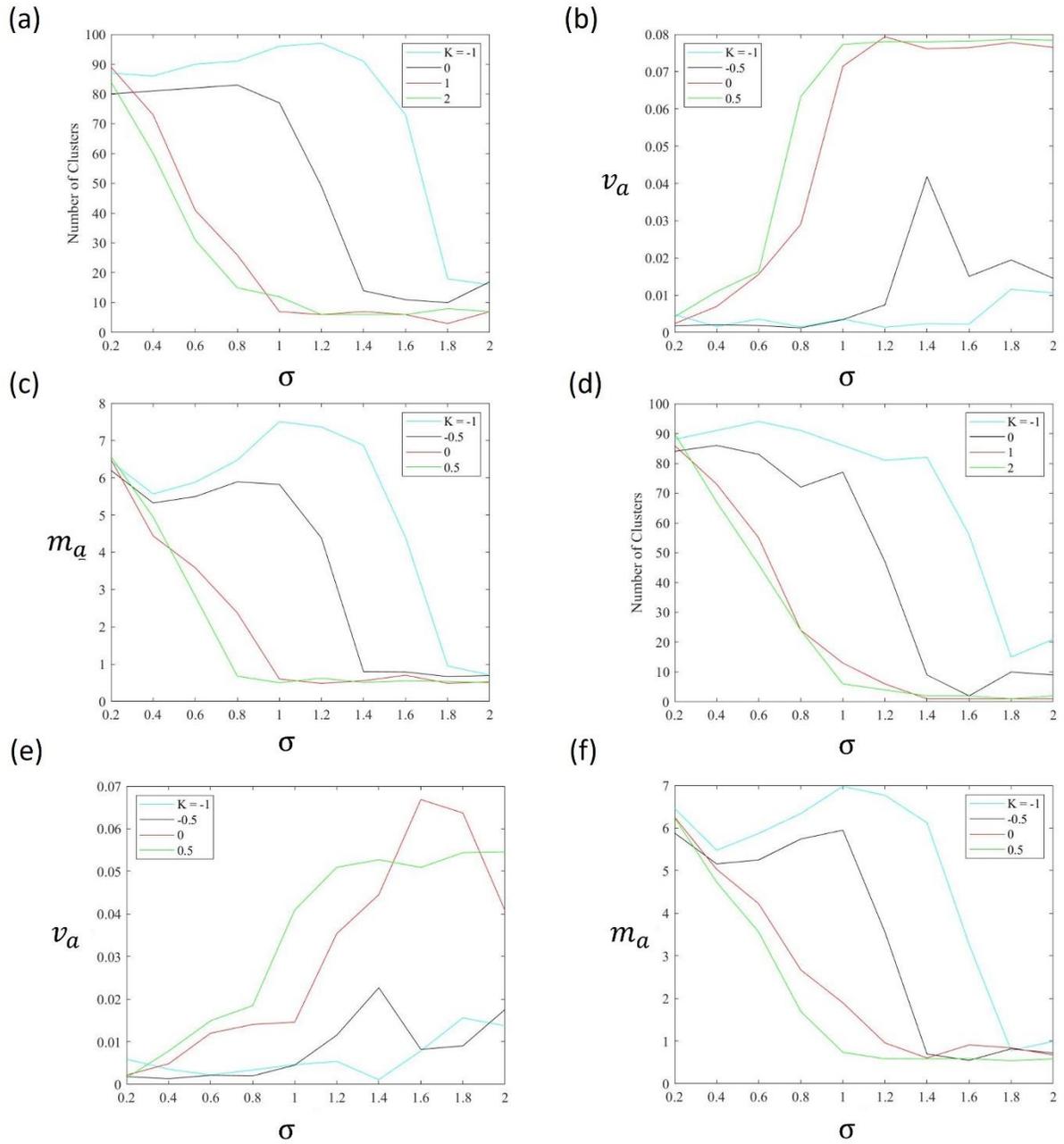

**Supplementary Fig. 37. Cluster count and motion characterization of locally coupled chiral swarmalators with a frequency spread and no frequency coupling.** (a-c) All swarmalators have their natural frequency randomly selected from single uniform distribution, such that $\omega_i \sim U(\Omega, 3\Omega)$. (d-f) Half the collective has their natural frequencies randomly selected from the uniform distribution $U(1,3)$ and the second half have their natural frequency selected from the uniform distribution $U(-3,-1)$. **(a,d)** Number of clusters. **(b,e)** Normalized velocity ($v_a$). **(c,f)** Normalized angular momentuum ($m_a$).

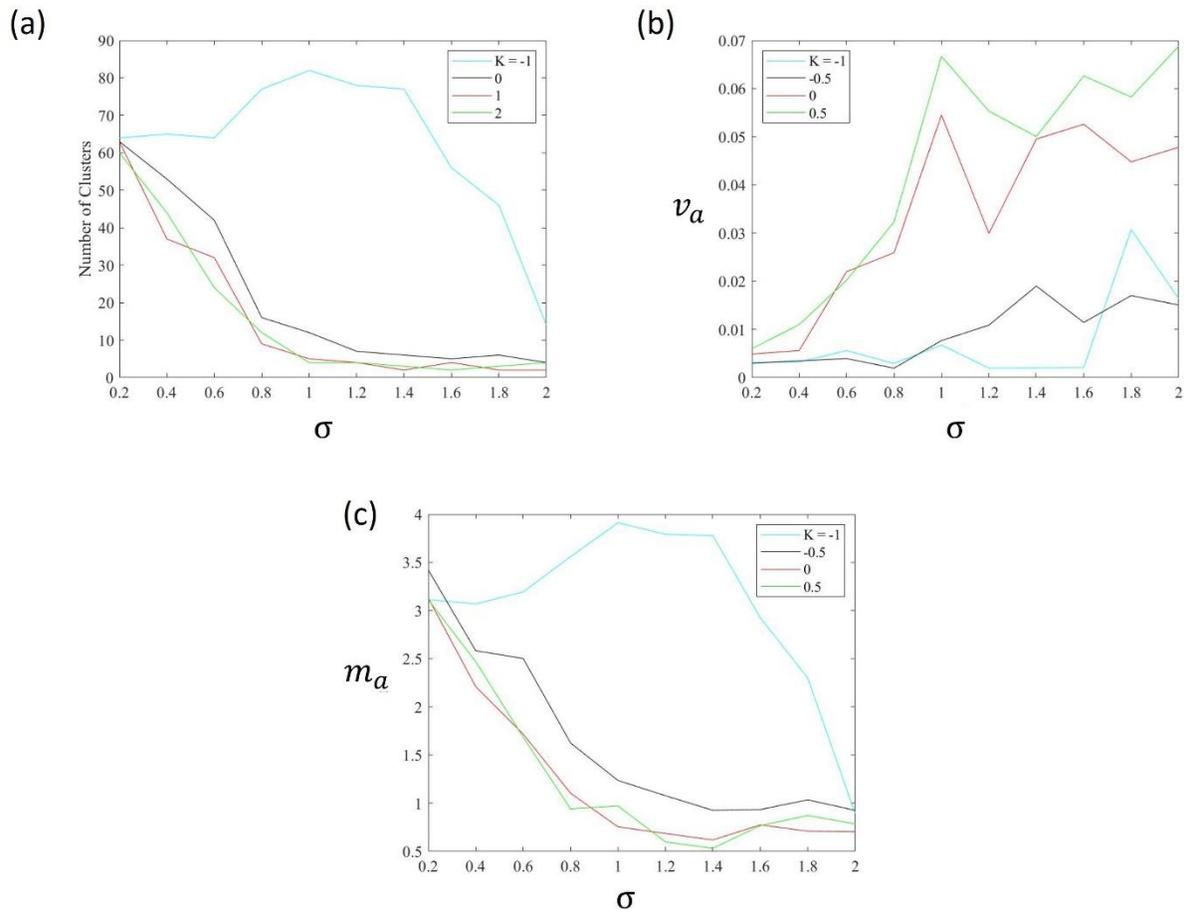

**Supplementary Fig. 38. Cluster count and motion characterization of locally coupled chiral swarmalators with no frequency spread and frequency-coupled.** Half the collective has $\omega_i = -1$ and the other half has $\omega_i = 1$. **(a)** Number of clusters. **(b)** Normalized velocity ($v_a$). **(c)** Normalized angular momentuum ($m_a$).

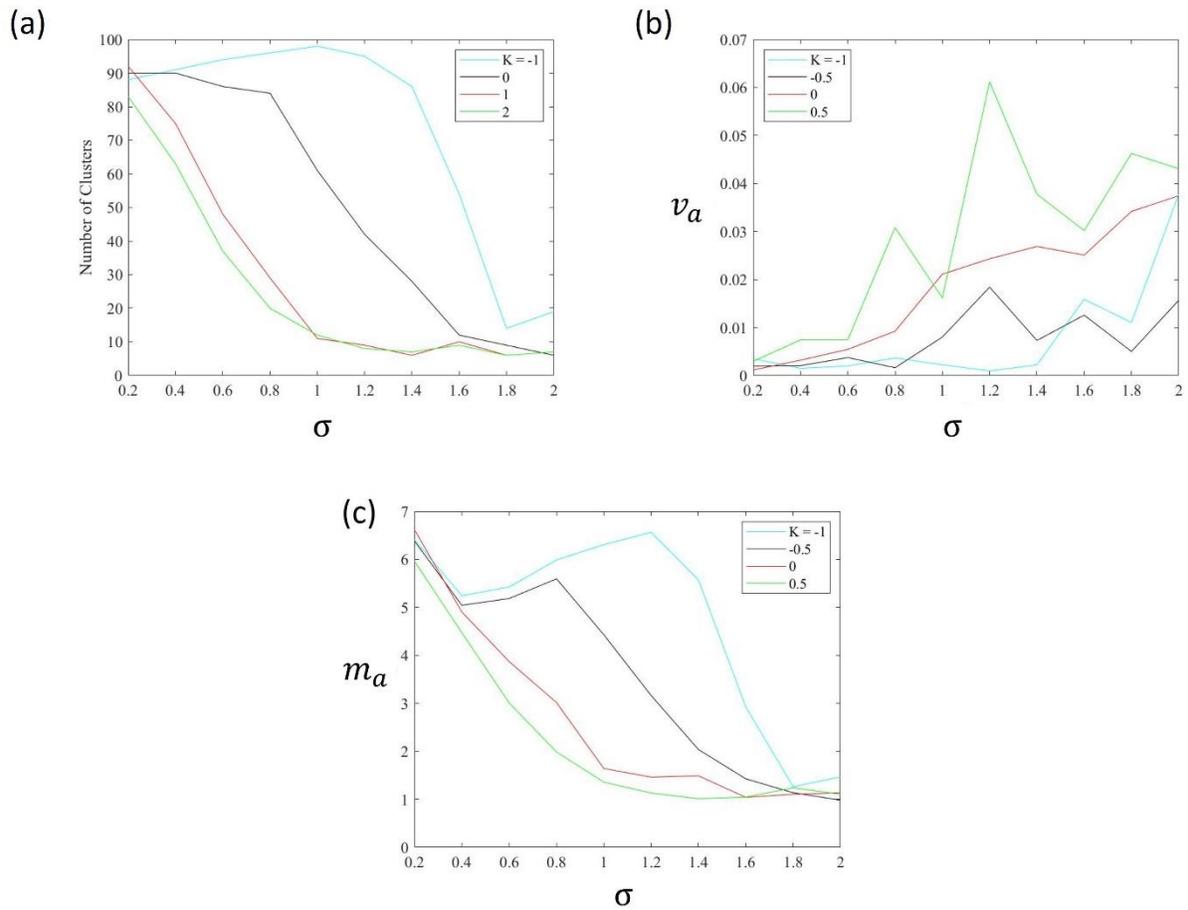

**Supplementary Fig. 39. Cluster count and motion characterization of locally coupled chiral swarmalators with a frequency spread and frequency-coupled.** Half the collective has their natural frequencies randomly selected from the uniform distribution $U(1, 3)$ and the second half has their natural frequency selected from the uniform distribution $U(-3, -1)$. **(a)** Number of clusters. **(b)** Normalized velocity ($v_a$). **(c)** Normalized angular momentuum ($m_a$).

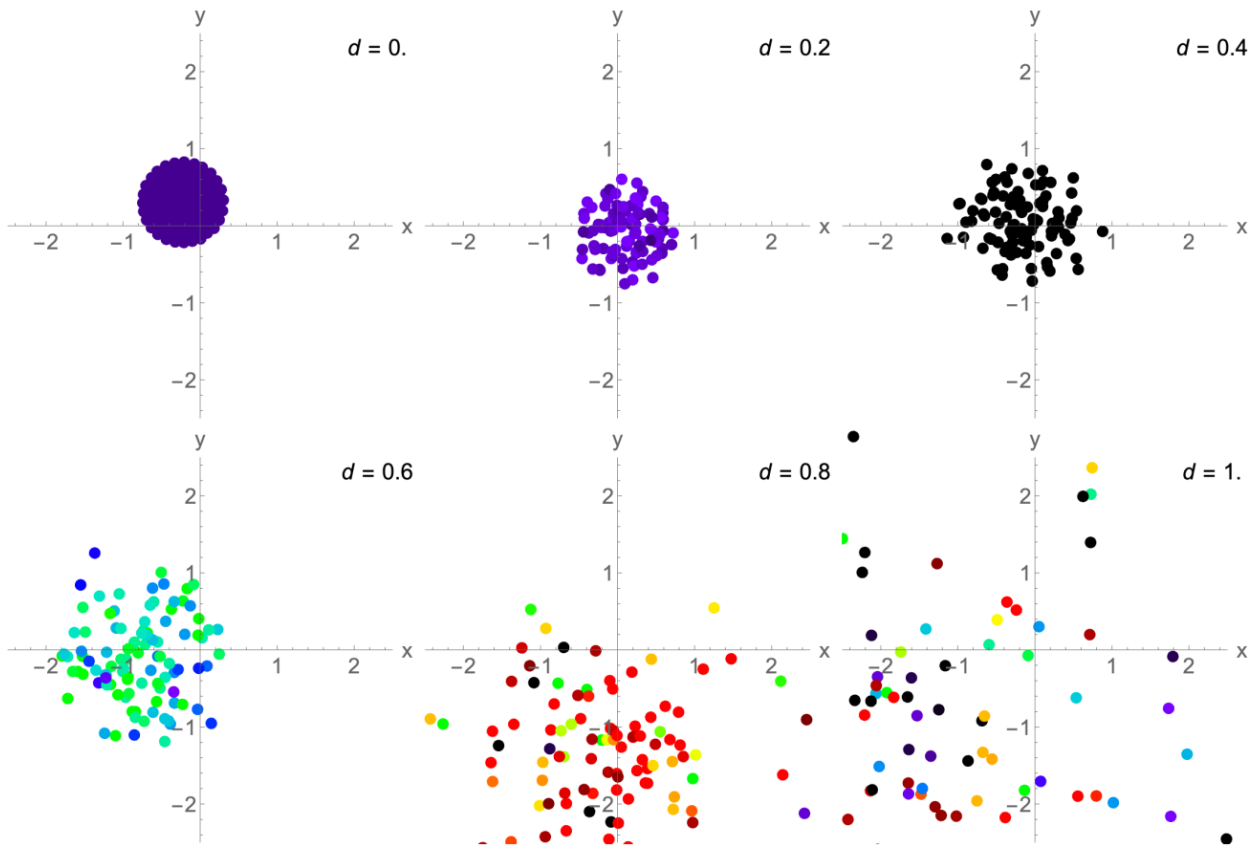

**Supplementary Fig. 40. Sync state for increasing $d$.** System becomes noticeably disordered as $d$ inreases.

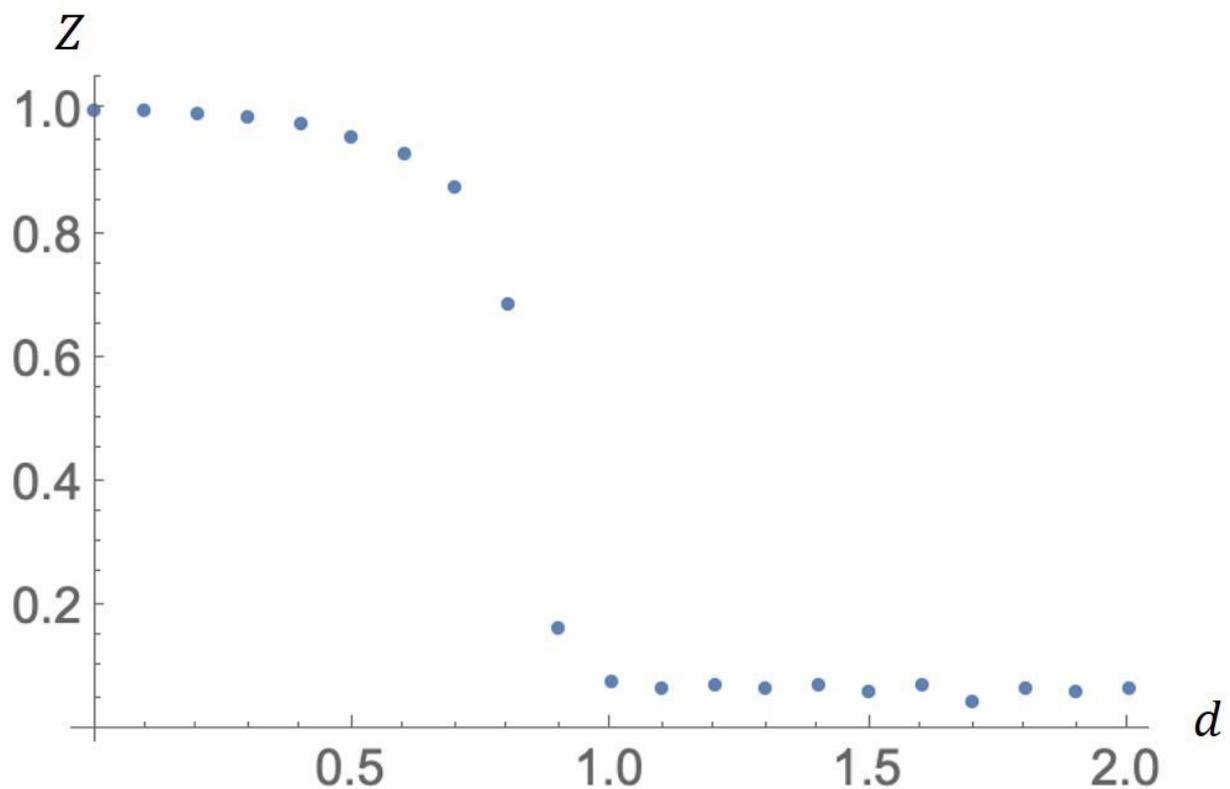

**Supplementary Fig. 41. Phase coherence for increasing $d$ in collectives with global coupling, identical, and non-chiral swarmalators.** On the y-axis, $Z$ decreases to almost zero with increasing $d$.

Page **83** of **108**

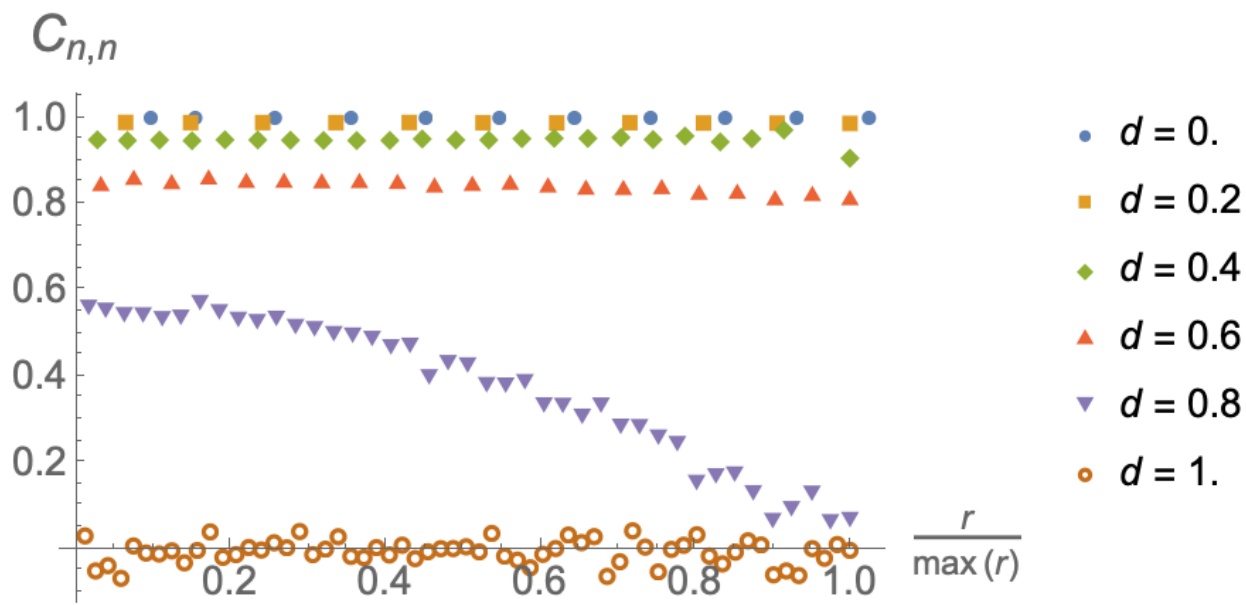

**Supplementary Fig. 42. Phase-phase correlation function.** Phase-phase correlation function $C_{\theta,\theta}(r)$ Pars: $(J, K, \sigma) = (1, 1, 10)$. $(dt, T, N) = (0.1, 200, 300)$. The first half of the data is dropped as transient. Each data point is the average of five simulations.

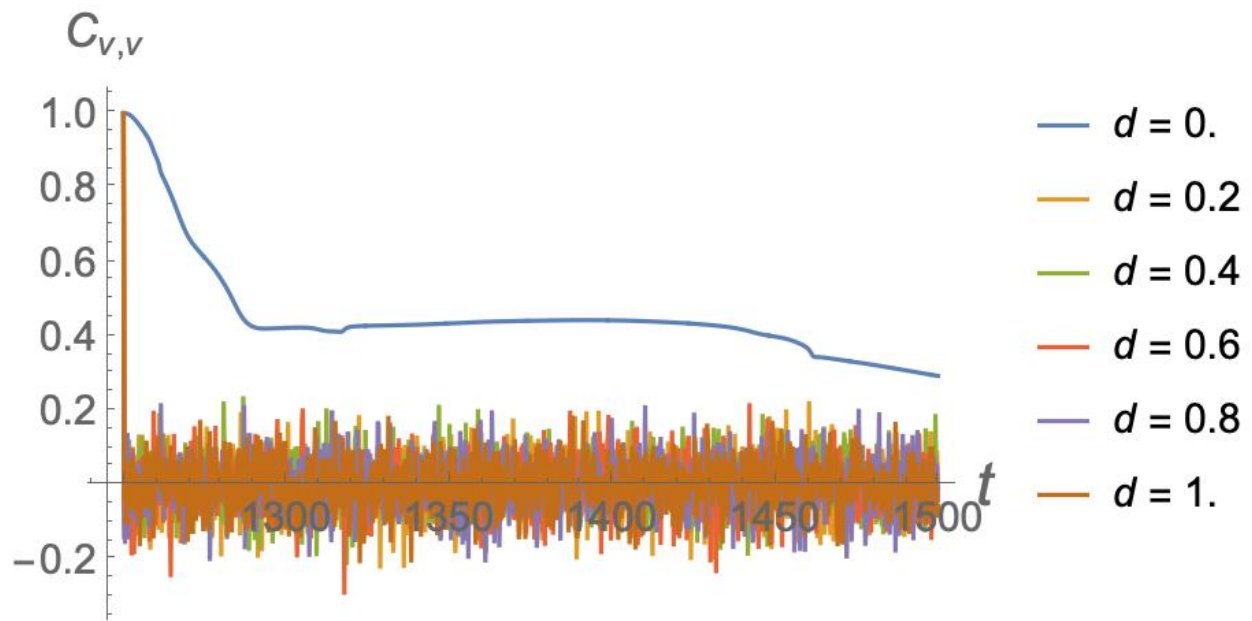

**Supplementary Fig. 43. Velocity auto-correlation function.** Simulation parameters: $(dt, T, N) = (0.1, 500, 200)$. The first half of data was dropped as transients (i.e. $\tau = NT/2$ in Eq. (2)).

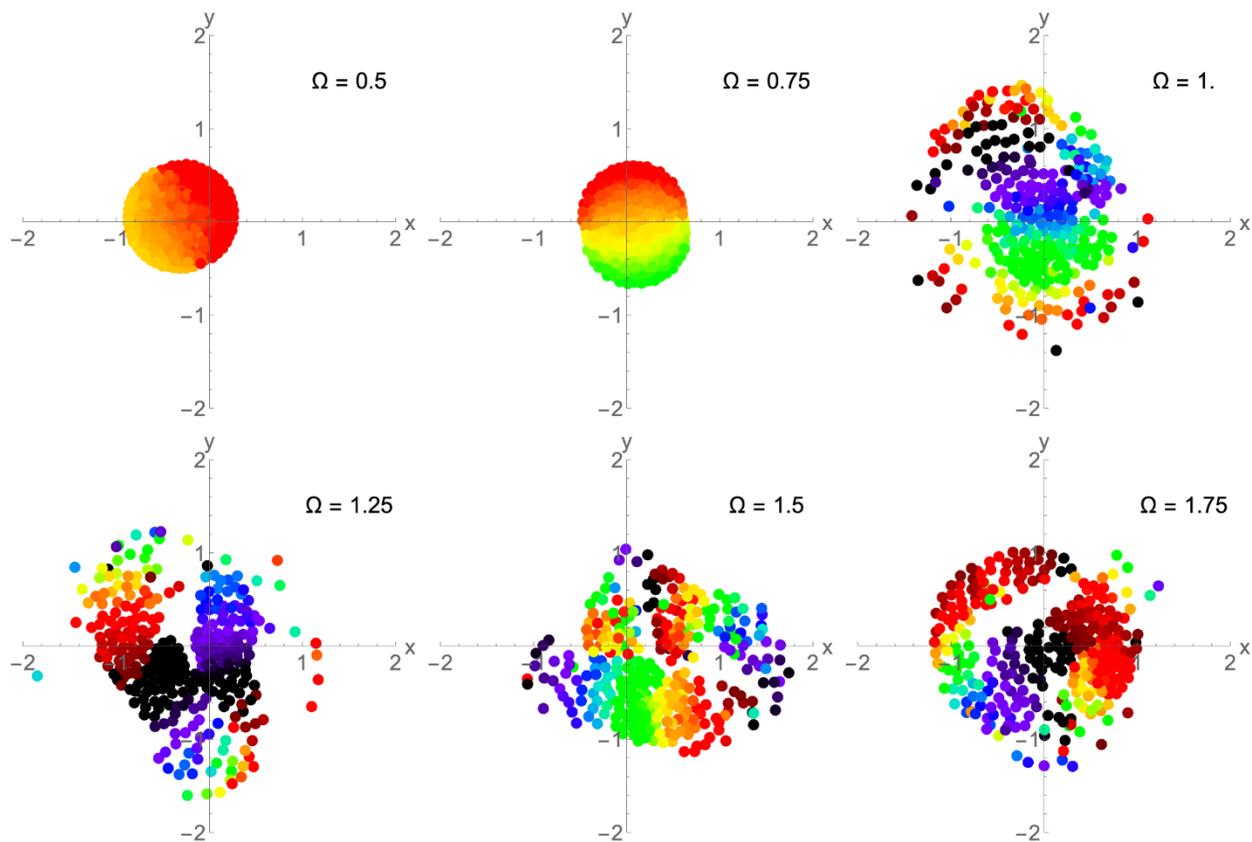

**Supplementary Fig. 44. Sync state for increasing Ω.** As Ω increases, the collective moves away from being synchronized in a cluster to being spatially organized by phase.

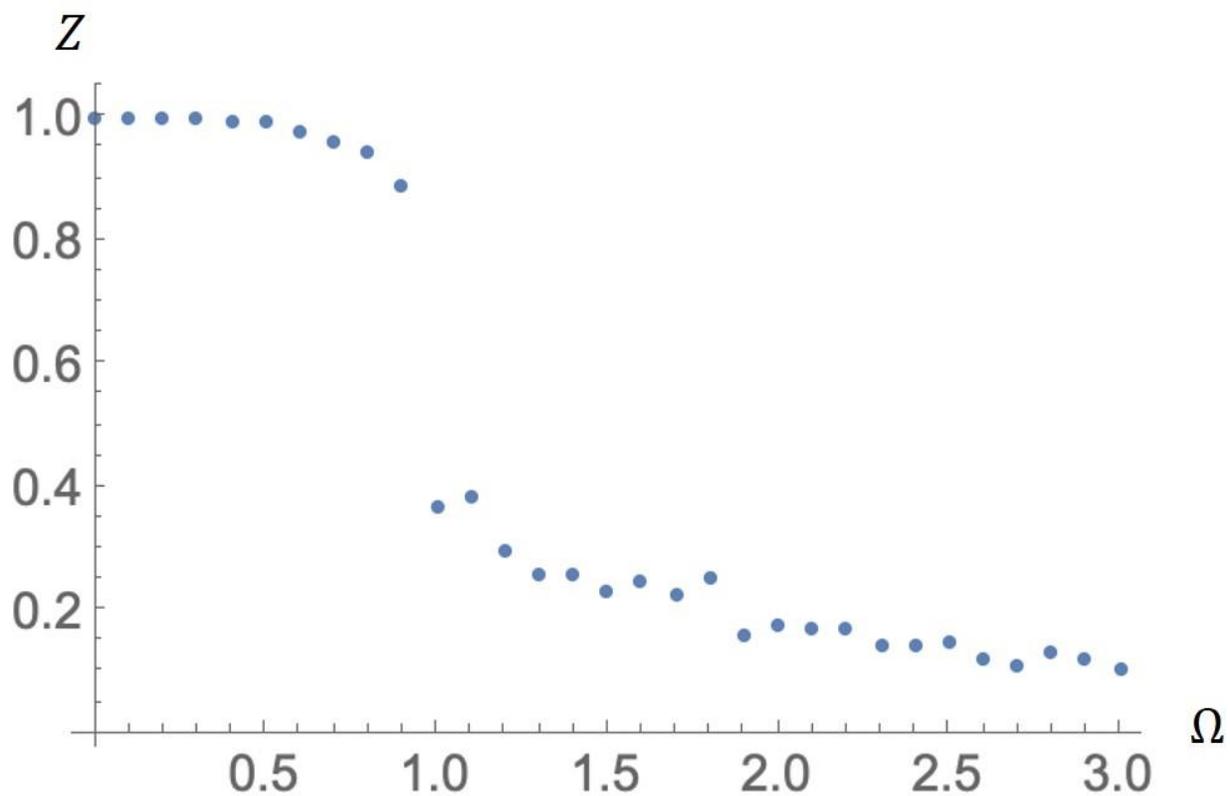

**Supplementary Fig. 45. Phase coherence for increasing $d$ in collectives with global coupling, non-identical, and non-chiral swarmalators.** On the y-axis, $Z$ decreases to almost zero with increasing $d$.

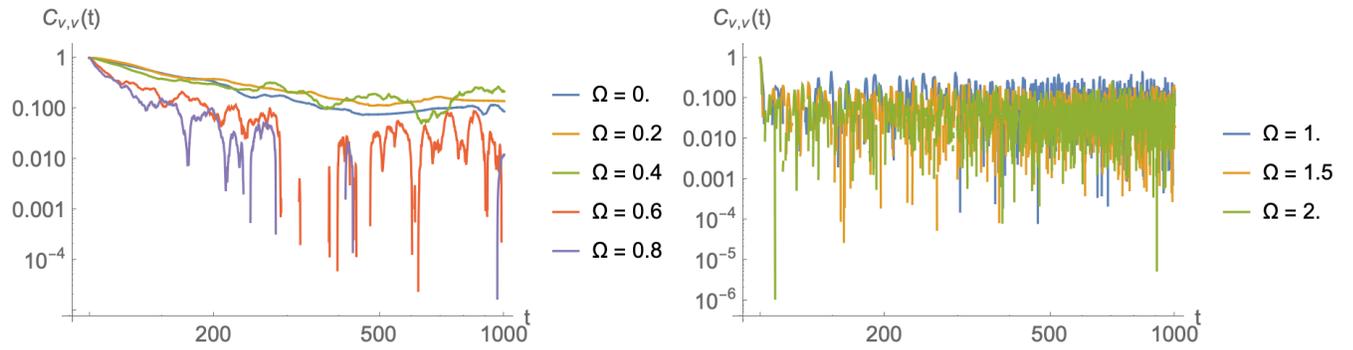

**Supplementary Fig. 46. Non-monotonic behavior of phase coherence for collectives with global coupling, non-identical, and non-chiral swarmalators.** Non-monotonic behavior of $R(d)$ for $\Omega = 1$. Simulation parameters: $(J, K, \sigma) = (1, 1, 10)$, $(dt, T, N) = (0.25, 1000, 200)$. Each data point is the average of five simulations.

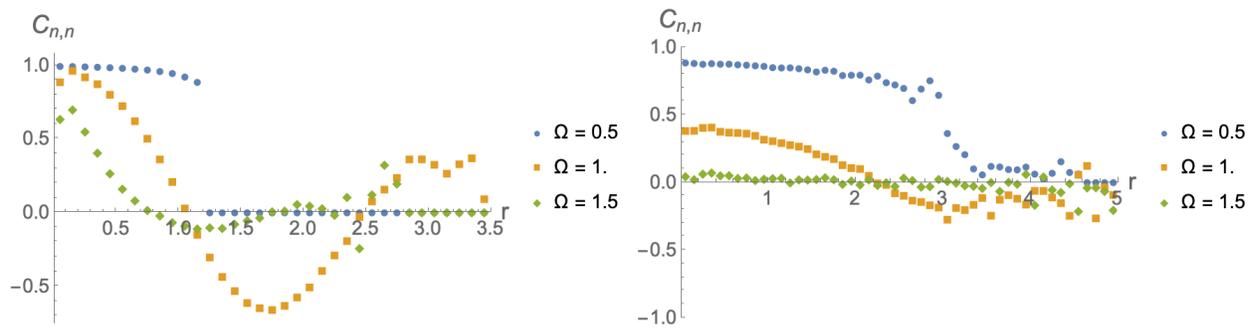

**Supplementary Fig. 47. Phase-phase correlation function for collectives with global coupling, non-chiral, and non-identical swarmalators.** Left: $d = 0$. Right: $d = 0.5$. Simulation parameters: $(J, K, \sigma) = (1, 1, 10)$, $(dt, T, N) = (0.25, 1000, 200)$. Each data point is the average of five simulations.

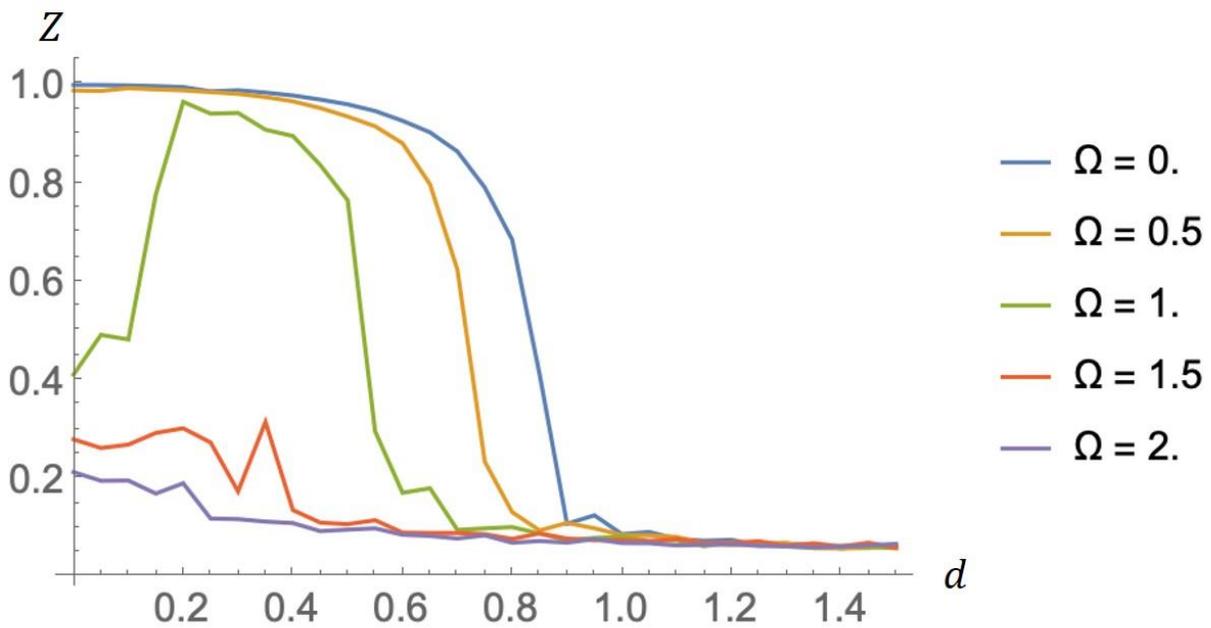

**Supplementary Fig. 48. Non-monotonic behavior of phase coherence for collectives with local coupling and non-chiral swarmalators.** Non-monotonic behavior of $Z(d)$ on the y-axis for $\Omega = 1$. Simulation parameters: $(J, K, \sigma) = (1, 1, 10)$, $(dt, T, N) = (0.1, 500, 300)$. Each data point is the average of five simulations.

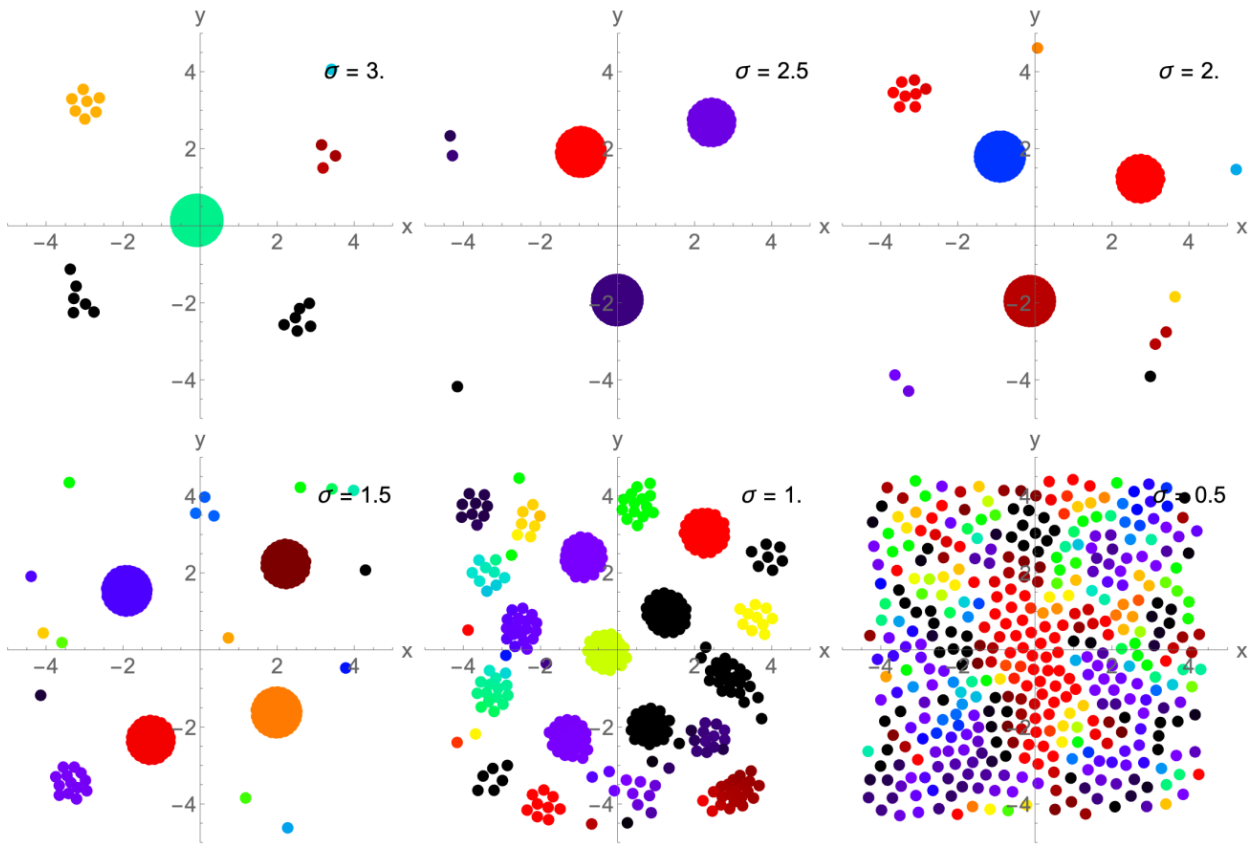

**Supplementary Fig. 49. Sync state for different $\sigma$ when there is local coupling.** Simulation parameters: $(J, K, d, \Omega) = (1, 1, 0.1, 0)$, $(dt, T, N) = (0.25, 200, 400)$. Each swarmalator is initially placed within a box of side length $L = 4$ and its phase is randomly selectred from a uniform distribution between 0 and $2\pi$.

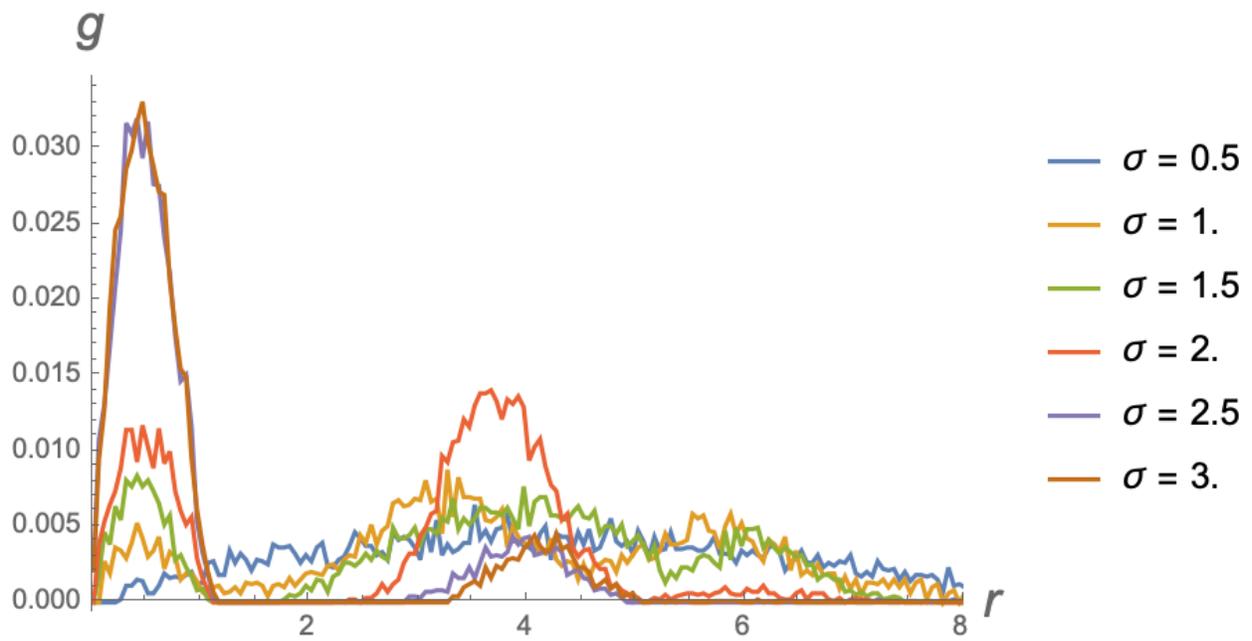

**Supplementary Fig. 50.** $g(r)$ **for different coupling ranges** $\sigma$. All swarmalators were initially placed randomly within a box of side length $L = 4$.

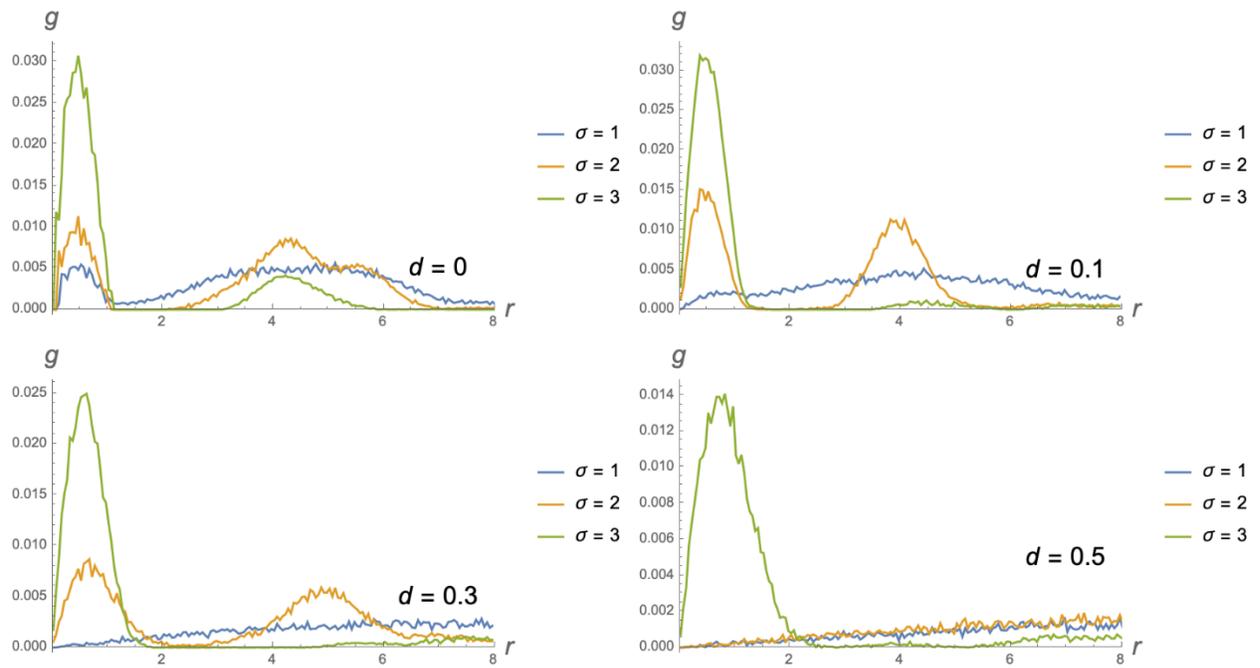

**Supplementary Fig. 51.** $g(r)$ **for different coupling ranges $\sigma$ when noise increases.** All swarmalators were initially placed randomly within a box of side length $L = 4$.

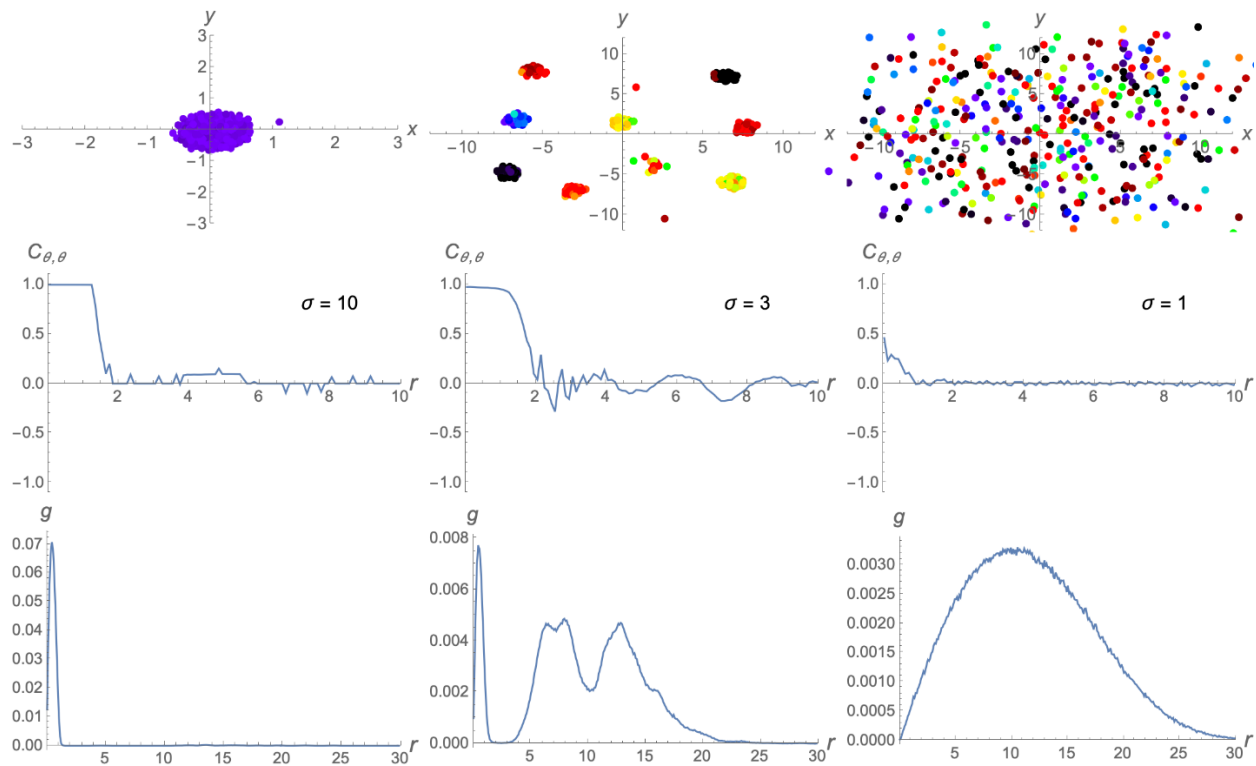

**Supplementary Fig. 52. Phase-phase correlation of collectives with local coupling ($\sigma = 10, 3, 1$), non-identical, non-chiral, and with noise.** Top row: scatter plots of states. Bottom row: phase-phase correlation function for different coupling ranges of $\sigma$. Simulation parameters: $(J, K, d, \Omega) = (1, 1, 0.1, 0.2)$, $(dt, T, N) = (0.25, 500, 300)$. Each data point is the average of 500 simulations.

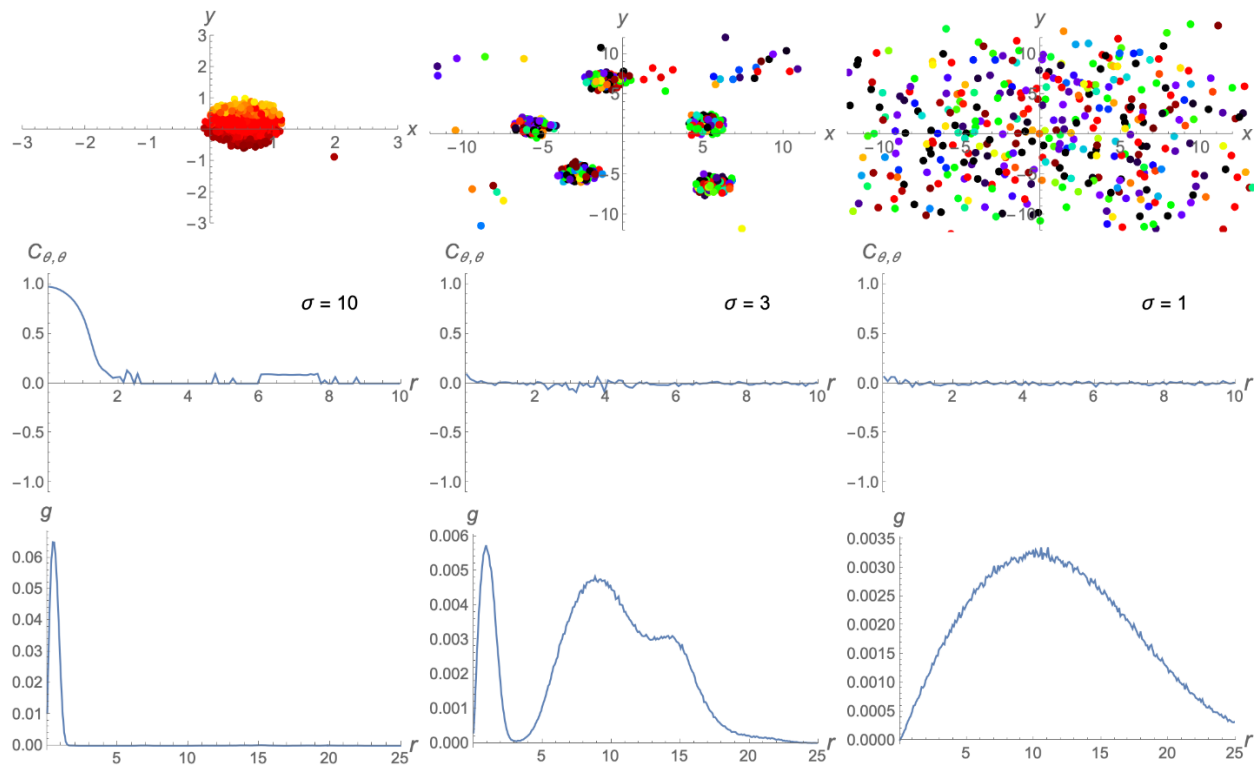

**Supplementary Fig. 53. Phase-phase correlation of collectives with local coupling ($\sigma = 10, 3, 1$), non-identical ($\Omega = 0.9$), non-chiral, and with noise.** Top row: scatter plots of states. Bottom row: phase-phase correlation function for different coupling ranges of $\sigma$. Simulation parameters: $(J, K, d, \Omega) = (1, 1, 0.1, 0.9)$, $(dt, T, N) = (0.25, 500, 300)$. Each data point is the average of 500 simulations.

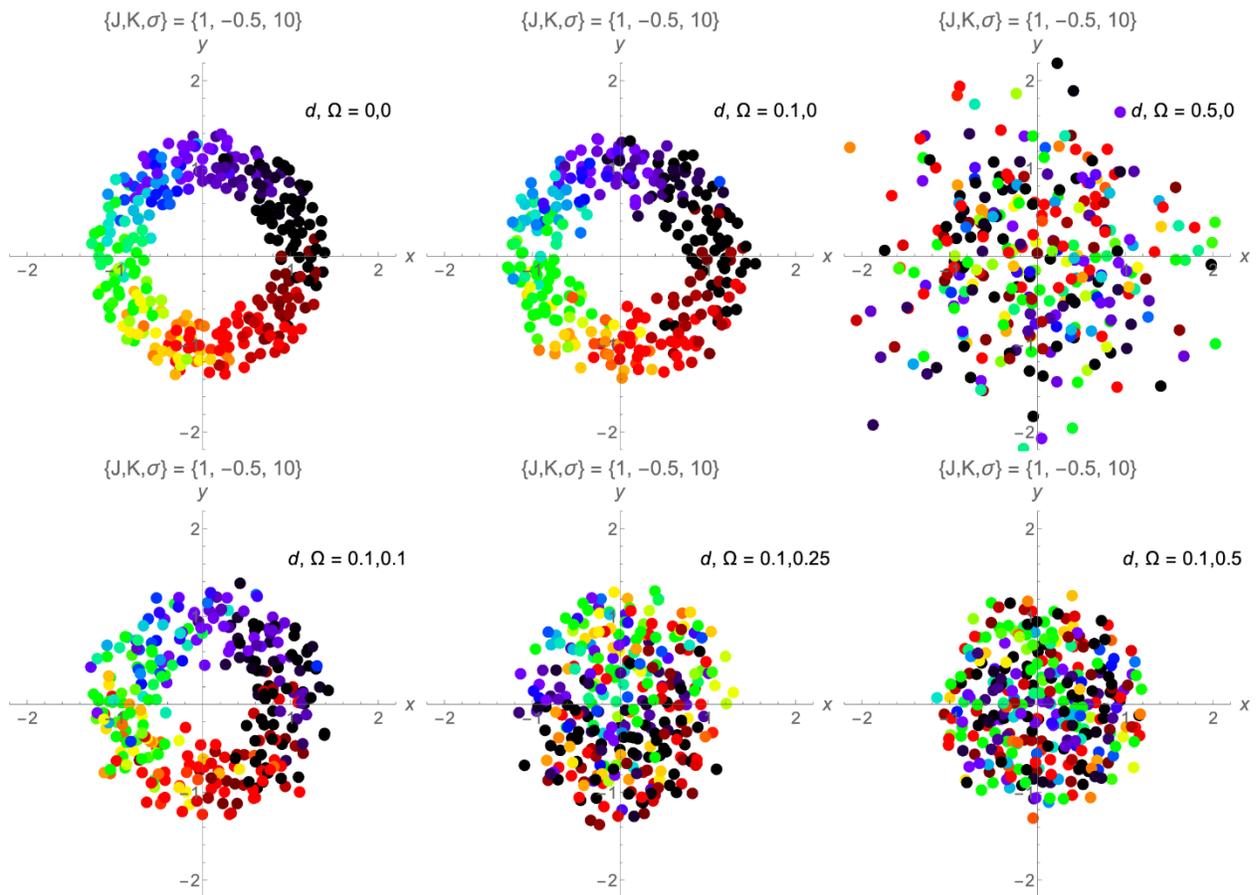

**Supplementary Fig. 54. Vortices with disorder.** Scatter plots of vortices for different amounts of disorder, both active $d > 0$ and quenched $\Omega > 0$.

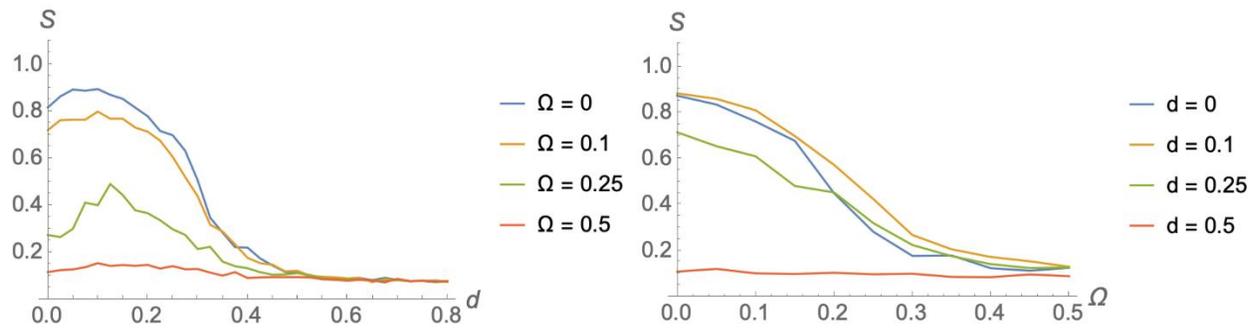

**Supplementary Fig. 55. Order parameter s for different $d$ and $\Omega$.** $S = \max(S_+, S_-)$. Left: $S(d)$ for different $\Omega$. Right: $S(\Omega)$ for different $d$. Simulation parameters: $(J, K) = (0, 1)$.

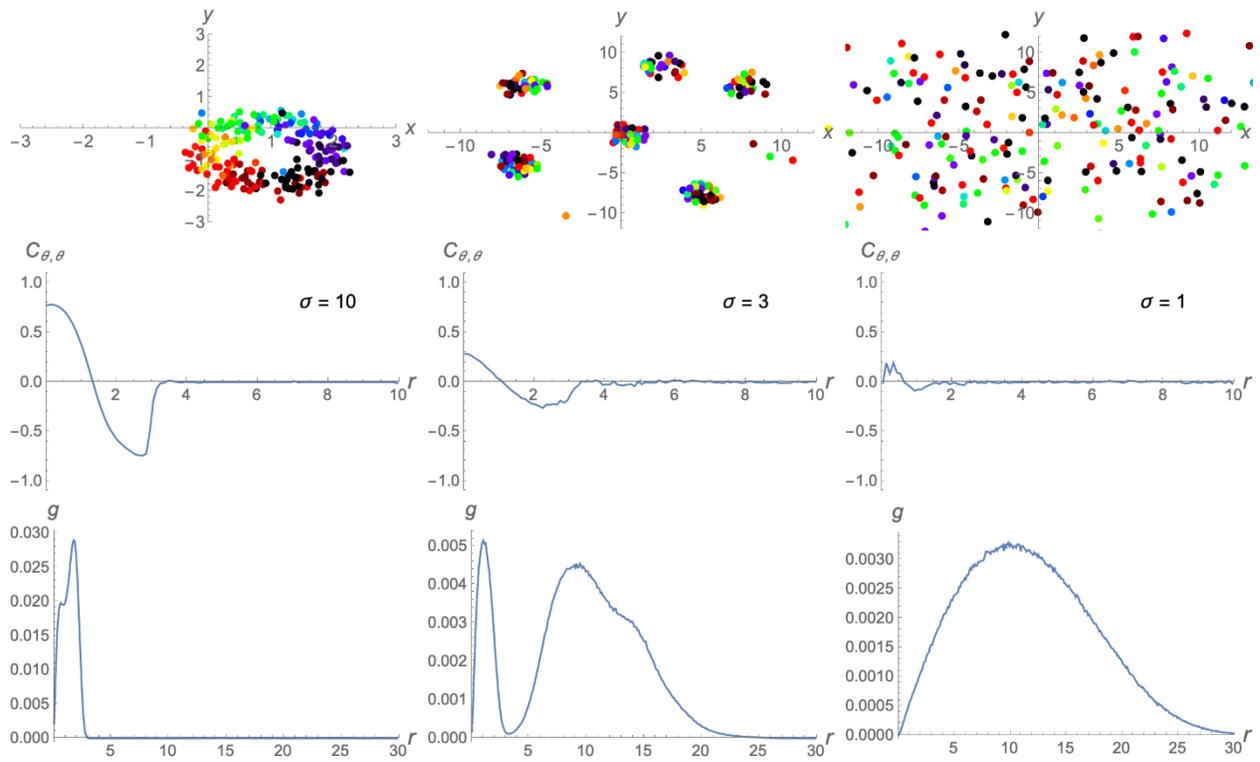

**Supplementary Fig. 56. Phase-phase correlation and $g(r)$ for vortices.** First column: long range coupling regime, with one cluster. Middle row: intermediary coupling regime with multiple vortices. Right row: short range regime, incoherence gas. Simulation parameters: $(dt, T, N) = (0.25, 800, 200)$, $(J, K, d, \Omega) = (1, -0.5, 0.1, 0.2)$. Each data point is the average of 100 simulations.

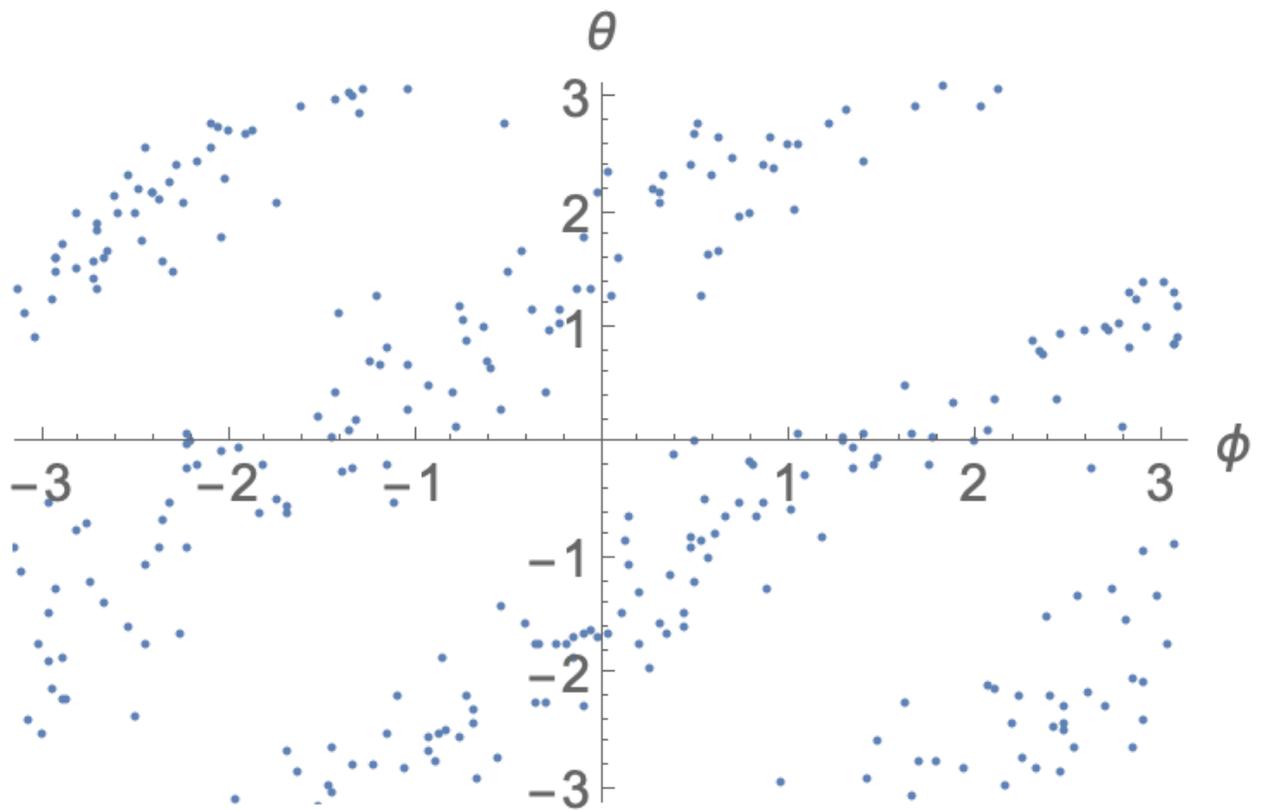

**Supplementary Fig. 57. Phase-phi space for phase wave state.** Scatter plot of phase wave state showing the winding number $k = 2$. Simulation parameters: $(J, K, \Omega, \sigma, d) = (1, -0.3, 1.5, 10, 0.1)$, $(dt, T, N) = (0.25, 100, 300)$.

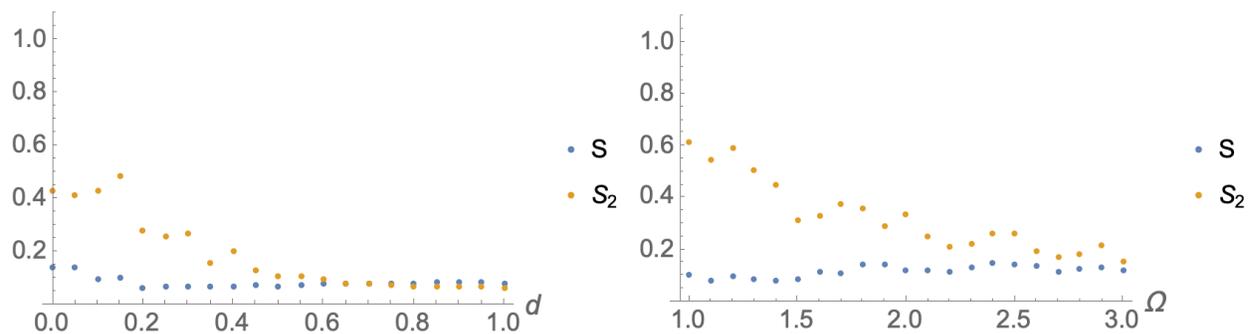

**Supplementary Fig. 58. Order parameters for active and quenched disorder.** Order parameters vs. active disorder (left panel) and quenched disorder (right panel). Simulation parameters: $(J, K, \sigma) = (1, -0.3, 10)$, $(dt, T, N) = (0.25, 100, 300)$. (Left) $\Omega = 1.5$. (Right) $d = 0.1$.

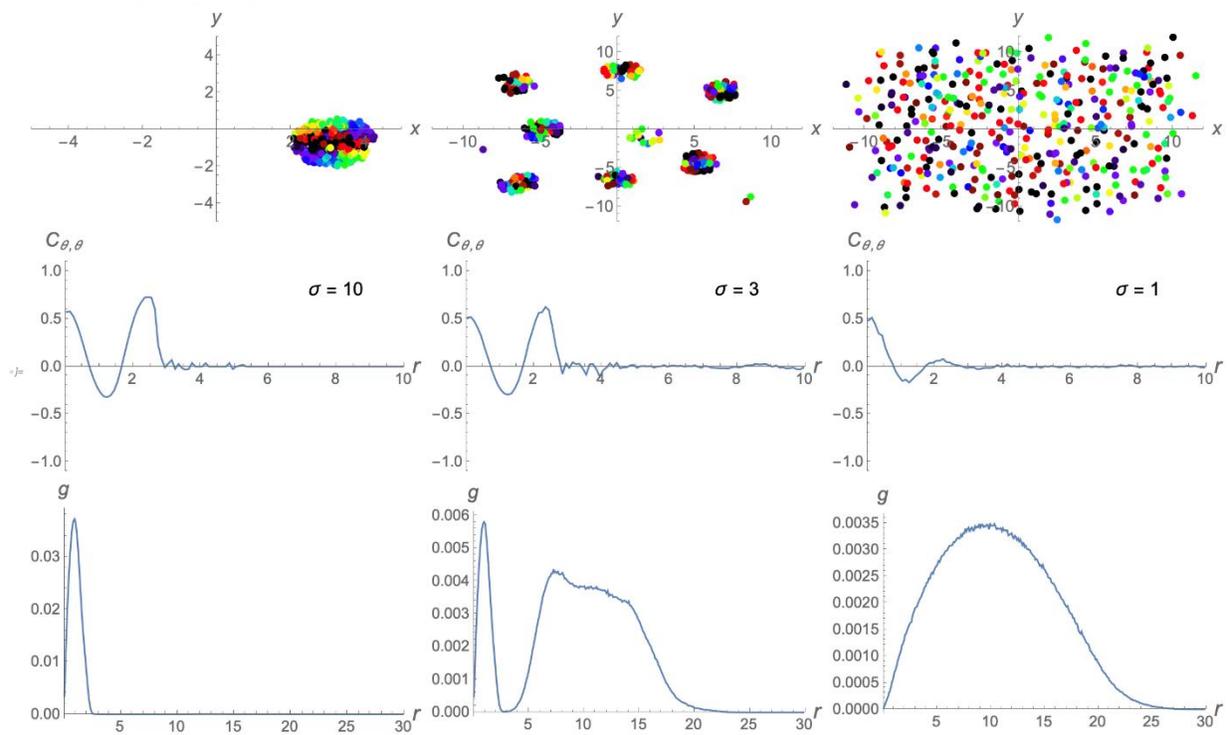

**Supplementary Fig. 59. Phase-phase correlation and $g(r)$ for revolving collectives in the vortex state.** First column: long range coupling regime, with one cluster. Middle row: intermediary coupling regime with multiple vortices. Right row: Short range regime, incoherence gas. Sim pars: (dt, T, N) = (0.25, 800, 200). Pars: (J, K, d, $\Omega$) = (1, −0.5, 0.1, 3). Each data point is the average of 100 simulations.

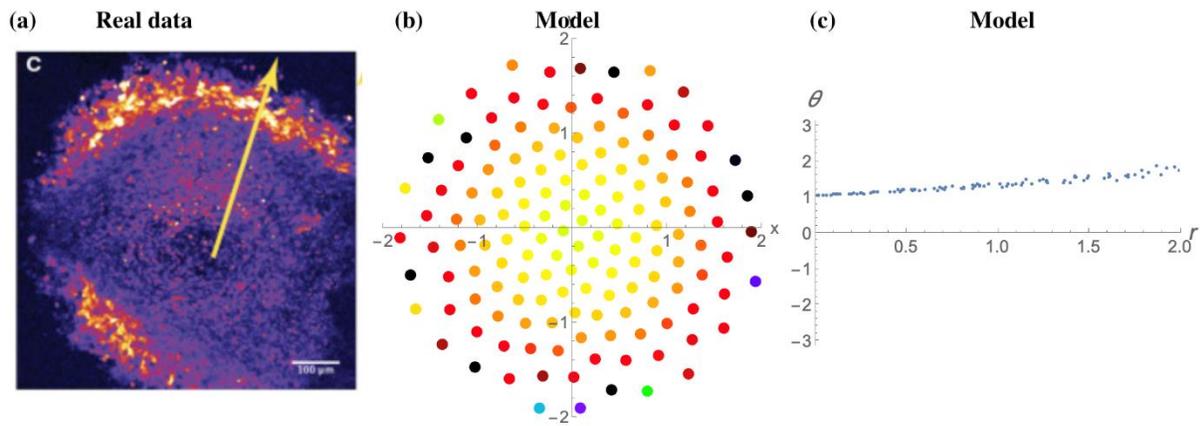

**Supplementary Fig. 60. Comparison to embryonic genetic oscillators.** **(a)** shows a sketch of the real data. The embryonic cells arrange themselves in disk-like pattern and a wave of genetic activity (highlighted by a bright yellow dye; see paper for details) travels radially outward as indicated by the yellow arrow. **(b)** and **(c)** demonstrate the model reproduces this pattern. See also Supplementary Movie 17.

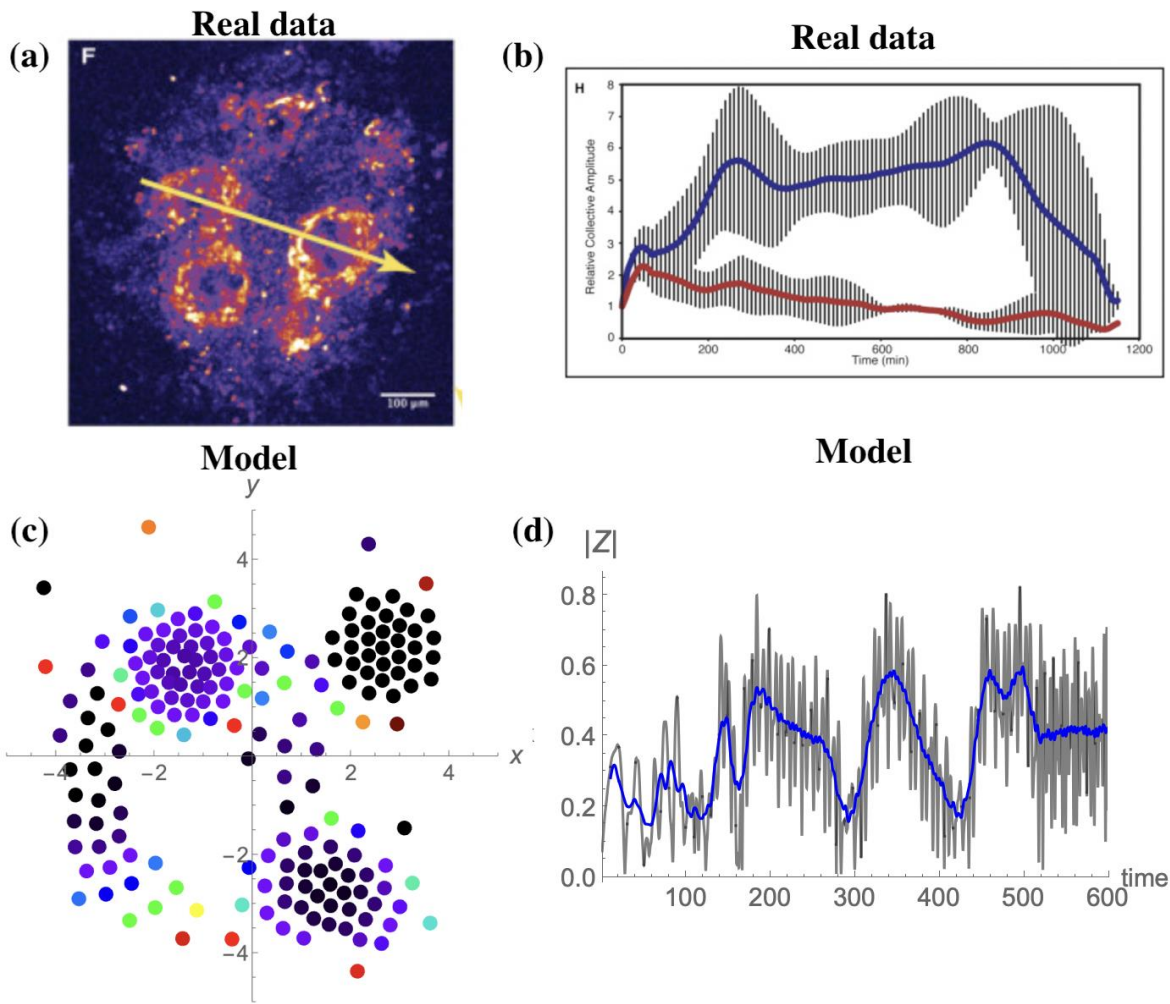

**Supplementary Fig. 61. Local clusters of embryonic genetic oscillators.** **(a)** shows an experimental scenario where multiple disks are formed[40]. **(b)** shows the degree of collective synchrony in the real data[40] (blue curves; ignore the red curve, it corresponds to a random baseline). **(c)** shows the model reproduces the pattern. **(d)** shows our model again reproduces this pattern qualitatively

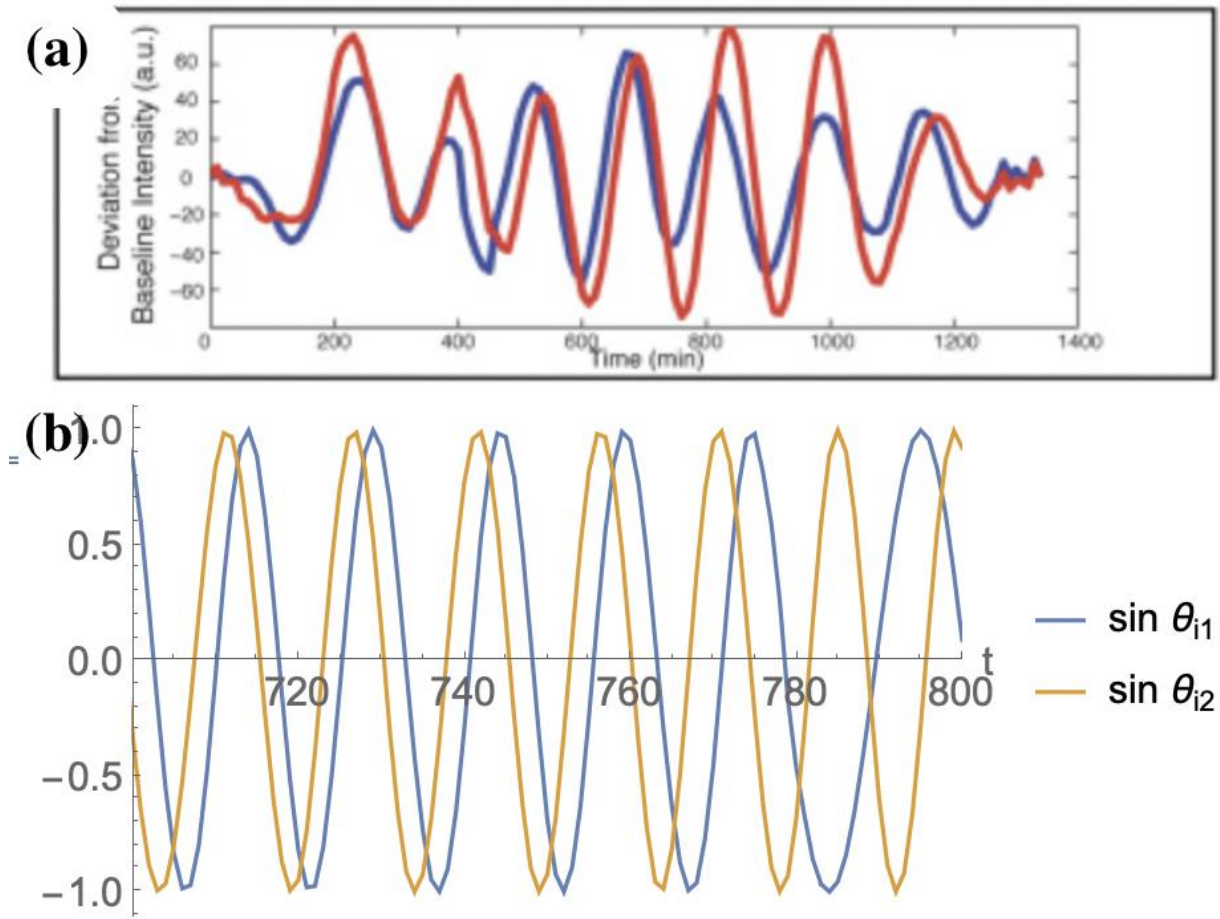

**Supplementary Fig. 62. Synchronization of various clusters.** (**a**) shows that the different foci from Supplementary Fig. 60 are in sync; the time series for two example oscillator are plotted which are phase locked[40]. (**b**) shows the same curves as produced by the model.

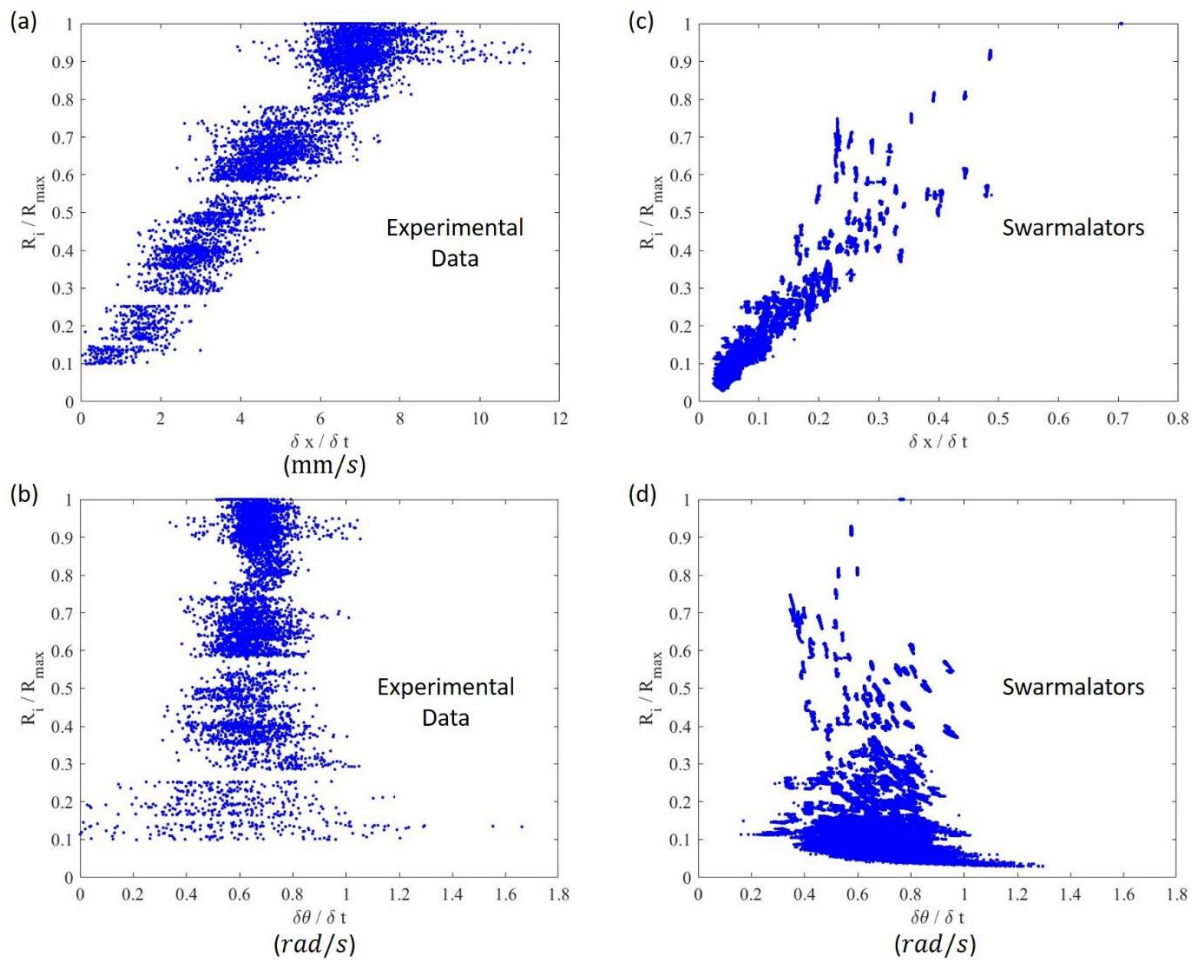

**Supplementary Fig. 63. Comparison to experiments with micro-disks.** Data from the micro-disk experiments conducted in Gardi et al.[50] was used to plot the **(a)** instantaneous speed and the **(b)** instantaneous angular velocity about the collective centroid vs. the normalized distance from collective center. Similar scatter plots **(c-d)** emerge for the swarmalators when the agents' natural frequencies are chosen from an exponential distribution between 0 and 1.

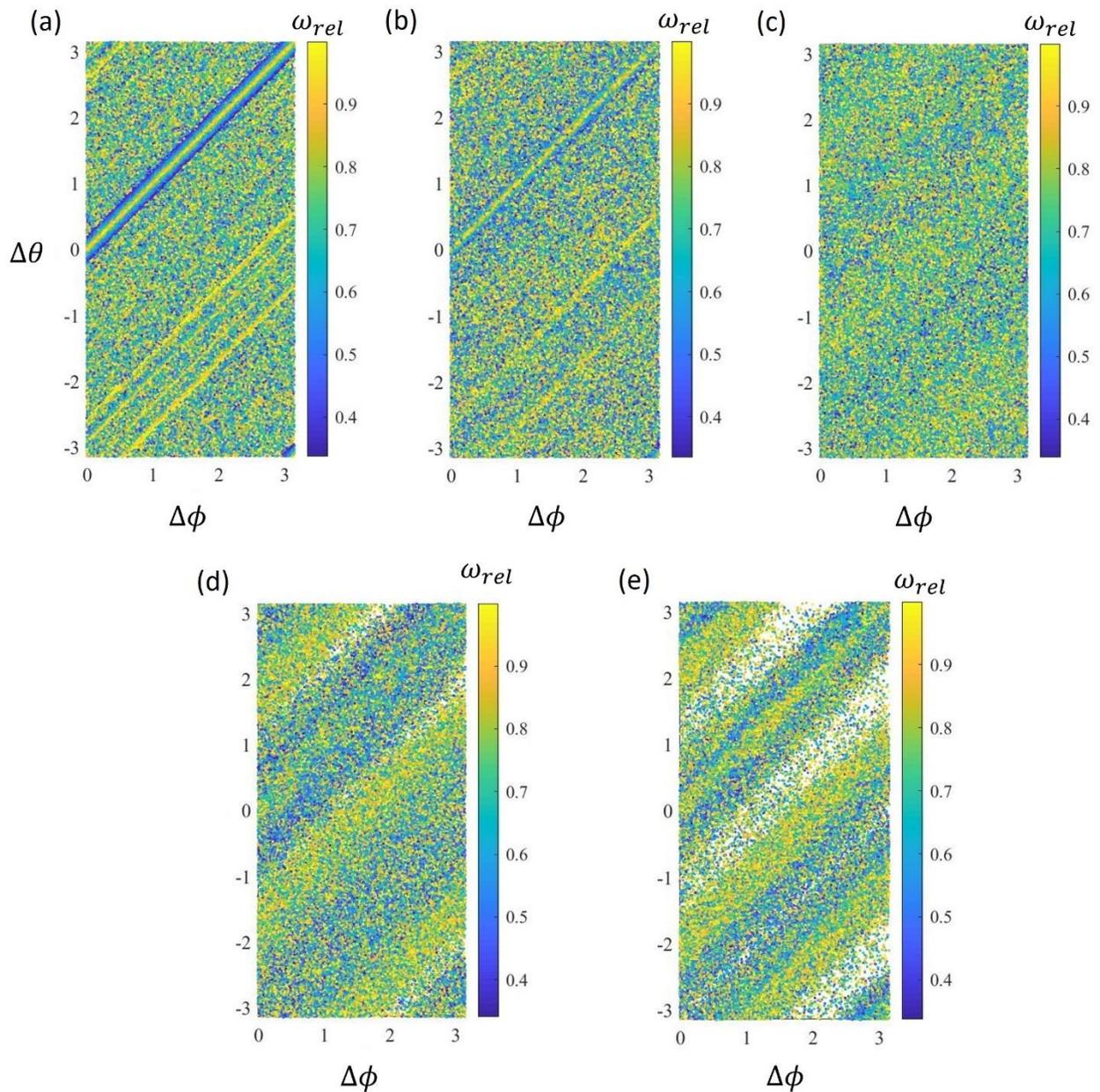

**Supplementary Fig. 64. Phase difference vs. spatial orientation difference.** Data from the micro-disk experiments conducted in Gardi et al.[50] was used to plot the **(a)** instantaneous speed and the **(b)** instantaneous angular velocity about the collective centroid vs. the normalized distance from collective center. Similar scatter plots **(c-d)** emerge for the swarmalators when the agents' natural frequencies are chosen from an exponential distribution between 0 and 1.

## Supplementary Movies

**Supplementary Movie 1. Non-Chiral Swarmalators with No Frequency Coupling.** A collage of movies of non-chiral swarmalators with integer values of $K$ and $J$. The movies corresponding to $F1$ are shown first, followed by $F2$, then $F3$, and finally $F4$.

**Supplementary Movie 2. Splintered Phase Waves.** Collectives with two, three, four, and five natural frequency groups are shown forming splintered phase waves when $K = -0.1, J = 1$.

**Supplementary Movie 3. Natural Frequency Group Separation.** Collectives with three, four, five, and nine natural frequency groups are shown going from a disordered state to exhibiting multiple bouncing clusters when $K = 1, J = 1$.

**Supplementary Movie 4. Concentric Phase Self-Organization.** A movie of a plot of agents' distance from the collective centroid as a function of phase is shown for when there is no chirality, no frequency coupling, $F4$, $K = 2$, and various values of negative $J$.

**Supplementary Movie 5. Non-Chiral Swarmalators with Frequency Coupling.** A collage of movies of non-chiral swarmalators with frequency coupling and integer values of $K$ and $J$ are shown going from a disordered state to their final behavior. The movies corresponding to $F2$ are shown first, followed by $F4$.

**Supplementary Movie 6. Revolving swarmalators with No Frequency Coupling.** A collage of collectives with chirality, no natural frequency-dependent coupling, and integer values of $K$ and $J$ are shown going from a disordered state to their final behavior. The movies corresponding to $F1$ are shown first, followed by $F2$, then $F3$ and finally $F4$.

**Supplementary Movie 7. Revolving Swarmaltors with Frequency Coupling.** A collage of collectives with chirality, frequency coupling, and integer values of $K$ and $J$ are shown going from a disordered state to their final behavior. The movies corresponding to $F2$ and followed by $F4$.

**Supplementary Movie 8. Global coupling, non-chiral, and noise.** Globally coupling non-chiral swarmalators with increasing levels of noise which take the collective further away from a sync state.

**Supplementary Movie 9. Global coupling, non-identical, and non-chiral.** Globally coupling non-chiral swarmalators with increasing spreads of natural frequencies across the collective.

**Supplementary Movie 10. Local coupling, identical, and non-chiral.** Local coupling collectives with different values for radius of interaction with other agents.

**Supplementary Movie 11. Global coupling, non-identical, and non-chiral vortices with noise.** Globally coupling non-chiral swarmalators that form vortices and have increasing spreads of natural frequency and noise.

**Supplementary Movie 12. Global coupling, identical, and non-chiral vortices with noise.** Locally coupling identical and non-chiral swarmalators with various radii of interaction forming vortices.

**Supplementary Movie 13. Local coupling, non-chiral, and no frequency coupling.** A collage of collective with different $\sigma$ are shown for collectives exhibiting no chirality and no natural frequency-dependent coupling. The movies corresponding to $F1$ are shown first, followed by $F2$, then $F3$, and $F4$.

**Supplementary Movie 14. Local coupling, chiral, and no frequency coupling.** A collage of collectives with different $\sigma$ are shown for collectives exhibiting chirality and no frequency coupling. The movies corresponding to $F1$ are shown first, followed by $F2$, then $F3$, and $F4$.

**Supplementary Movie 15. Local coupling, chiral, and frequency coupling.** A collage of locally coupling collectives with different $\sigma$ are shown for collectives exhibiting chirality and frequency coupling. The movies corresponding to $F2$ and followed by $F4$.

**Supplementary Movie 16. Slime Mold.** Locally coupling non-chiral swarmalators that vary the $K$ and $J$ parameters and switch between different states that resemble the various life stages of social slime mold.

**Supplementary Movie 17. Embryonic Genetic Oscillators.** Local clusters are shown to form with radial phase organization; qualitatively similar to the behavior present in embryonic genetic oscillators in the literature.